\documentclass[sigconf, nonacm]{acmart} 
\usepackage{tabularx}
\usepackage{multirow}
\usepackage{threeparttable}
\usepackage{subcaption}
\usepackage{makecell}
\usepackage{float}
\usepackage{tikz}
\usepackage{colortbl}

\newenvironment{icompact}{
    \begin{list}{$\bullet$}{
        \parsep 0pt plus 1pt            
        \partopsep 0pt plus 1pt         
        \topsep 1pt plus 0pt minus 0pt  
        \itemsep 0pt plus 0pt           
        \parskip 1pt plus 1pt           
        \leftmargin 0.13in              
        \labelwidth 0.13in
    }}
{\normalsize\end{list}}

\AtBeginDocument{%
  }

\setcopyright{acmlicensed} 
\copyrightyear{2018} 
\acmYear{2018} 
\acmDOI{XXXXXXX.XXXXXXX} 
\acmConference[Conference acronym 'XX]{Make sure to enter the correct
  conference title from your rights confirmation email}{June 03--05,
  2018}{Woodstock, NY}  
\acmISBN{978-1-4503-XXXX-X/2018/06}  

\begin{document}


\title{Navigating the Latent Manifold: Proactive Concept Drift Adaptation for Resilient NIDS}


\author{Chao Zha}
\orcid{0009-0004-6611-2328}
\email{zhachao@zju.edu.cn}
\affiliation{%
  \institution{Zhejiang University}
  \city{Hangzhou}
  \country{China}
}
\affiliation{%
  \institution{Institute of Computing Technology, Chinese Academy of Sciences}
  \city{Beijing}
  \country{China}
}

\author{Zifeng Kang}
\orcid{0000-0003-0812-0173}
\email{zifengkang@bupt.edu.cn}
\affiliation{%
  \institution{Beijing University of Posts and Telecommunications}
  \city{Beijing}
  \country{China}
}

\author{Tian Liu}
\orcid{0000-0002-0001-8898}
\email{liutian@zaas.ac.cn}
\affiliation{%
  \institution{Institute of Agricultural Equipment, Zhejiang Academy of Agricultural Sciences}
  \city{Hangzhou}
  \country{China}
}

\author{Dakun Shen}
\orcid{0000-0003-2392-8998}
\email{dakun@zju.edu.cn}
\affiliation{%
  \institution{Zhejiang University}
  \city{Hangzhou}
  \country{China}
}

\author{Ruyun Zhang}
\authornote{Corresponding author.}
\orcid{0000-0002-2969-817X}
\email{zcor2021@gmail.com}
\affiliation{%
  \institution{Shanghai AI Laboratory}
  \city{Shanghai}
  \country{China}
}

\renewcommand{\shortauthors}{Trovato et al.}

\begin{abstract} 
Network intrusion detection systems (NIDS) are critical for cybersecurity, safeguarding services and data from potential attacks. However, existing AI-based NIDS often assume static data distributions and fail to handle concept drift, leading to degraded performance and increased false positives in dynamic network environments. To address this issue, we propose DriftXpert\footnotemark[2], a novel NIDS for drift-adaptive detection. Specifically, we propose a decoupled two-stage offline adaptive framework. In Phase 1, we introduce an unsupervised anomaly metric based on latent manifold deviation. By performing outlier analysis within the latent space, the framework achieves high-sensitivity detection of network traffic concept drift. In Phase 2, to mitigate catastrophic forgetting under non-stationary distributions, we design a representation consistency alignment strategy. This strategy constrains the feature mapping between the legacy model and the drifted distribution, ensuring the model captures emerging attack characteristics while retaining discriminative power over known patterns. Furthermore, we incorporate cross-epoch neuron weight aggregation and selective freezing mechanisms to enable fine-grained knowledge transfer in the parameter space, effectively balancing model plasticity and stability. Extensive experiments on public datasets demonstrate that DriftXpert effectively adapts to drifted data without catastrophic forgetting. Furthermore, real-world evaluations on enterprise network further confirm its robustness and practical applicability, contributing to improved security protection for millions of users.
\footnotetext[2]{Github repository: \url{https://github.com/cc-sec-hub/DriftXpert}}
\end{abstract}




\maketitle

\section{Introduction}
\label{sec_introduction}

Network intrusion detection systems are essential for modern network security, protecting environments and providing reliable digital spaces. NIDS monitors system traffic or activity in real time, detecting threats such as malware, DDoS attacks, and data breaches. Traditional anomaly or signature-based methods struggle with unknown threats \cite{dong2016comparison, garcia2009anomaly, ji2016multi, li2019designing, meng2014efm}, while AI-based approaches have shown superior performance \cite{garaffa2021reinforcement, han2022survey, wu2021deep, zha2026flowxpert, yang2025mm4flow, zha2026normality, zha2026sparse}. Classical machine learning methods, such as clustering, decision trees, and random forests, have been applied to classify malicious traffic \cite{taylor2016appscanner, van2020flowprint, fu2021realtime, fu2024flow, fu2023point, kilincer2021machine, yang2021mth}, while reconstruction mechanisms and transfer learning help detect unknown attacks \cite{kim2018zero, mirsky2018kitsune, sameera2020deep, yang2021conditional, yilmaz2021transfer, zha2024skt}.

\begin{figure}[htbp]
\centering
\subfloat[Local drift.]{\includegraphics[width=0.2\textwidth]{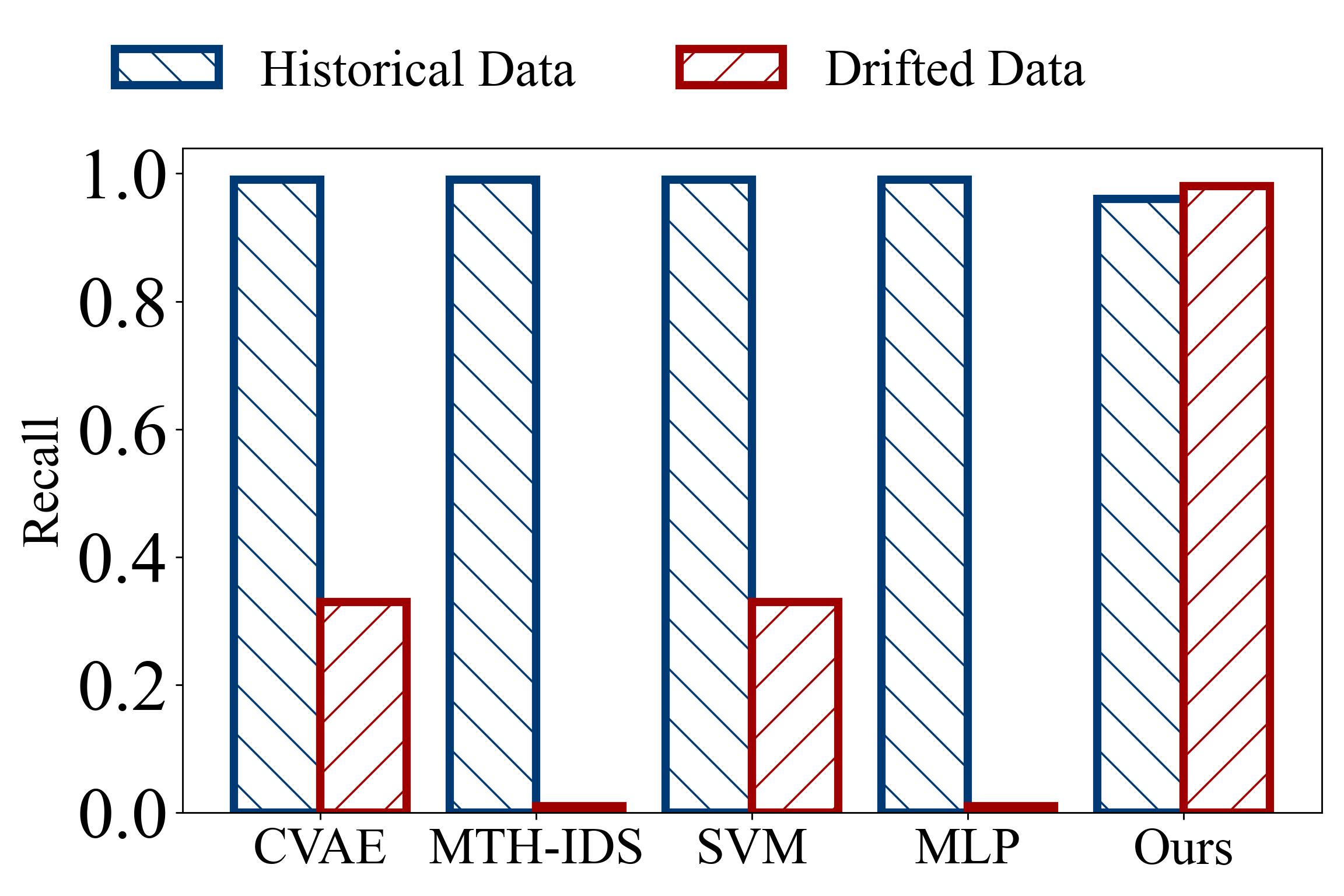}
\label{fig_1_1a}}
\hspace{0.05\textwidth}
\subfloat[Global drift.]{\includegraphics[width=0.2\textwidth]{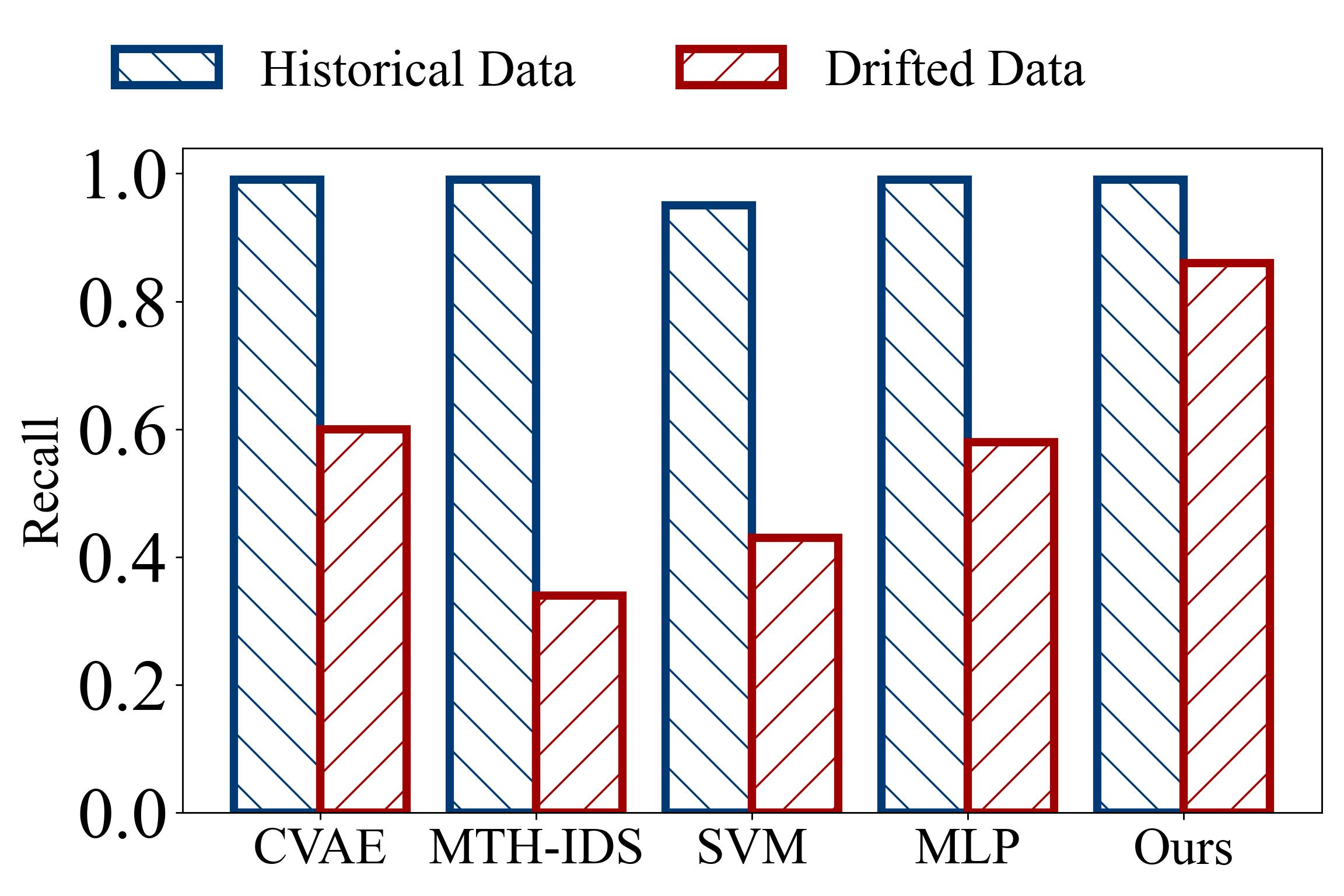}
\label{fig_1_1b}}
\caption{Evaluation of concept drift. We select four different existing methods, train them on the historical data, and then test them on the drifted data on the CICIDS-2017 dataset.}
\label{fig_1_1}
\end{figure}

Despite strong performance in traditional classification tasks \cite{he2015delving, li2022mvitv2, rao2021global}, AI algorithms often underperform in real-world NIDS due to unique domain challenges \cite{sommer2010outside}. Security research should emphasize understanding the nature of security problem over merely improving benchmark scores. Many AI-NIDS studies rely on static traffic assumptions \cite{wang2020cloud, yang2021conditional, yang2021mth}, which are suitable for controlled experiments but do not reflect practical scenarios \cite{fogla2006polymorphic, qing2025training}. \textbf{In reality, network traffic evolves over time with changing statistical properties and inter-feature relationships, which is called concept drift} \cite{gama2014survey}. Concept drift severely degrades detection performance, affecting the model’s ability to recognize emerging threats and evolving patterns \cite{bayram2022concept}. Furthermore, concept drift can bias decision making \cite{korycki2021concept}, increasing false positives and false negatives. To assess the impact of concept drift, we conducted experiments on the CICIDS-2017 dataset, as presented in Figure \ref{fig_1_1}.

Addressing concept drift should include two key components: drift detection and model adaptation. Model adaption should enable the updated model to fit new distributions while mitigating catastrophic forgetting, which is critical as historical patterns often reoccur and forgetting may increase false positives. Drift detection acts as the trigger for model adaption, ensuring that adaptation is performed only after drift is identified. Concept drift in NIDS arises from both temporal and spatial factors \cite{zha2025nids}. Temporally, evolving services and emerging traffic patterns induce distribution shifts, while storing historical data is often impractical due to resource constraints. Spatially, deploying models across different environments further exacerbates drift, while reusing or sharing historical network traffic for adaptation is often infeasible, as such data may contain sensitive user and organizational information and introduce risks of privacy leakage and unauthorized exposure. Consequently, \textbf{the drift detection and the model adaption should be studied under the assumption of no access to historical data.} On the other hand, as real-world network traffic is massive, rendering manual labeling costly and impractical. Moreover, labeling inherently requires data analysis, during which drift can already be observed. Therefore, \textbf{unsupervised drift detection is both necessary and inevitable in practical scenarios.}

However, existing drift detection methods typically rely on labeled drift data or directly incorporate historical data. For example, statistical approaches \cite{costa2018drift, dos2016fast, liu2018accumulating, jain2022k} based on KL divergence or local density changes inherently assume such access, violating the aforementioned practical constraints and limiting their applicability in real-world settings. Moreover, hyperparameter threshold selection remains challenging and often requires careful tuning. From the perspective of model adaptation, existing mask-perturbation-based approaches heavily rely on historical data \cite{yang2024recda}, limiting their practicality in real-world settings. Meanwhile, model update methods \cite{zha2025nids, 11044615, han2023anomaly, wang2022enidrift, liu2020diverse} that do not require historical data (e.g., constrained fine-tuning and generative replay) suffer from unstable training dynamics and suboptimal performance or introduce additional security risks, such as data poisoning.

\textbf{Our work.} We propose \textit{\textbf{DriftXpert}}, a unified and problem-driven framework for autonomous NIDS adaptation under evolving traffic distributions. Unlike prior work that separately studies drift detection, incremental learning, or knowledge preservation, DriftXpert tightly integrates drift awareness with model evolution. Specifically, we introduce two clustering-based drift indicators, CI and OR, with theoretical analysis to enable label-free drift detection and use detected drift as an adaptation criterion to trigger model updates only under distribution shifts. To address catastrophic forgetting, we further design representation consistency alignment, together with weight aggregation and selective freezing, to balance adaptation and knowledge preservation without requiring historical traffic replay or storage. Extensive experiments on public benchmarks and real-world enterprise traffic demonstrate the effectiveness of DriftXpert in long-term adaptive NIDS deployment.

\textbf{Contributions.} This study makes the following contributions.
\begin{itemize}
    \item We propose a unified drift-aware NIDS adaptation framework that integrates unsupervised drift detection and autonomous model evolution, rather than treating them as independent problems.
    
    \item We propose an unsupervised drift detection mechanism based on latent representation deviation, where two clustering-based indicators, \textit{Confidence Interval} and \textit{Outlier Ratio}, are introduced with theoretical analysis to enable effective and interpretable concept drift identification.

    \item We design a drift-driven model adaptation framework that performs representation consistency alignment under evolving distributions, together with weight aggregation and selective freezing, to balance adaptation and knowledge preservation without relying on historical traffic replay.
\end{itemize}

\section{Overview}
\label{sec_overview}

\subsection{Background}
\label{sec_overview_background}
Concept drift in NIDS refers to the evolution of the statistical properties of network traffic over time, leading to a mismatch between training and test data. Most NIDS models assume a stationary distribution, which rarely holds in real-world environments due to changing user behaviors, service updates, and evolving attack strategies. Consequently, learned decision boundaries become outdated, degrading model performance and increasing false positives or false negatives. A common manifestation of concept drift is shifts in benign traffic distributions, where even subtle changes can significantly affect prediction results by altering feature distributions or inter-feature relationships, often leading to a sharp increase in false positives in practical deployments. Handling concept drift typically involves two components: drift detection and model adaptation. Drift detection identifies distribution changes and triggers model updates. Model adaptation then updates the model to fit new data while preserving historical knowledge. Therefore, an effective approach must balance adaptability and stability.

Specifically, concept drift is defined as a change in the joint probability distribution of the samples $x$ (from the feature space $X$) and labels $y$ (from the label space $Y$), denoted $P(x \in X, y \in Y) = P(x, y)$. As in $P(x,y) = P(x) \cdot P(y|x)$, two drift paradigms can be defined as follows.

\noindent \textbf{\textit{DEFINITION 1}} (LOCAL DRIFT). The local drift is the change of $P(x\in X_L, y) = P(x) \cdot P(y|x \in X_L)$, $X_L$ represents the subspace of characteristics where the drift occurs (e.g., a specific attack).
\begin{equation}
    \begin{aligned}
        P_{\text{his}}(y|x) & \ne P_{\text{drift}}(y|x), \forall x \in X_L \\
        P_{\text{his}}(y|x) & = P_{\text{drift}}(y|x), \forall x \in X
    \end{aligned}
\end{equation}
where $P_{\text{his}}$ and $P_{\text{drift}}(y|x)$ denote the conditional probability distributions in the historical and drifted distributions, respectively.

\noindent \textbf{\textit{DEFINITION 2}} (GLOBAL DRIFT). The global drift is the change of $P(x\in X, y) = P(x) \cdot P(y|x \in X)$, denoted $P_{\text{his}}(x, y) \ne P_{\text{drift}}(x,y)$, $\forall x \in X$. There can be two sources of drift:
\begin{icompact}
    \item \textit{Global Concept Drift:} 
    $P_{\text{his}}(y|x) \ne P_{\text{drift}}(y|x), \forall x \in X$ (e.g., novel attack emerges, the original decision boundary to classify attacks becomes invalid, requiring a relearning of attack patterns.)

    \item \textit{Global Covariate Shift:}
    $P_{\text{his}}(x) \ne P_{\text{drift}}(x), \forall x \in X$ (e.g., the statistical characteristics of traffic undergo a global change, leading to a distributional difference between historical and drifted data.)
\end{icompact}

\subsection{Challenges}
\label{sec_overview_challenges}
\subsubsection{\textbf{C1: Limited Capability for Unsupervised Drift Detection}} The occurrence of concept drift indicates that the model requires updating to adapt to new network environments. Supervised drift detection methods depend on extensive manual inspection of prediction results, making them costly and impractical. Consequently, unsupervised approaches offer a more viable solution in real-world settings \cite{gemaque2020overview}. Detecting concept drift serves as a critical initial step for model updates. This should be considered a qualitative issue that can be addressed through quantitative analysis. Furthermore, concept drift detection does not need to be performed in real-time; instead, it should involve periodic monitoring of current network traffic conditions to identify potential drift.

\subsubsection{\textbf{C2: Inadequate Adaptation to Drifted Data}} When concept drift is detected, model updating becomes necessary to adapt to new data. Retraining is often considered the ideal approach, as it typically achieves optimal performance when real data are available. However, in resource-constrained or privacy-sensitive scenarios, storing or accessing historical data is often impractical. Fine-tuning \cite{guo2019spottune} provides an alternative, enabling model updates without relying on historical data. Although it can handle local drift effectively, it may struggle under global drift and lead to performance degradation. Therefore, designing efficient model update methods without historical data remains a key challenge for applying AI to NIDS.

\subsubsection{\textbf{C3: Catastrophic Forgetting of Historical Data}} Model updates can lead to a critical issue: achieving accurate detection of recent data at the cost of significantly reduced detection performance in historical data, a phenomenon known as \textit{catastrophic forgetting} \cite{kemker2018measuring, kirkpatrick2017overcoming}. Although concept drift occurs within the current network traffic space, historical data does not completely vanish. Within a certain timeframe, it remains essential to retain the ability to detect historical data. Balancing improved detection performance for drifted data while preserving detection capabilities for historical data without access to historical datasets represents another key challenge in applying AI to NIDS.

\subsection{Threat Model}
\label{sec_threat_model}
Our threat model assumes a dynamic network environment in which data drift naturally occurs (excluding adversarial drift), and adversaries may exploit such distribution shifts to degrade a deployed NIDS. We consider a setting where the model is trained on historical data, while incoming traffic gradually deviates from the original distribution due to evolving benign behaviors or adaptive attack strategies. If this drift is not properly addressed, the model may produce incorrect predictions, leading to increased false positives or missed detections.

We define two types of risks in this setting. First, undetected drift refers to concept drift that occurs without being identified, leading directly to performance degradation. Second, catastrophic forgetting arises when model updates fail to preserve historical knowledge, potentially amplifying detection errors. When both occur simultaneously, it constitutes a critical failure scenario, where the system loses both adaptability and stability.

\subsubsection{\textbf{In-scope Threats}} We consider three in-scope threats of this study as follows.
\begin{icompact}
    \item \textit{Evasion via distribution shift.} An adversary may craft traffic patterns that mimic benign distribution changes, causing the model to misclassify malicious traffic as benign.

    \item \textit{False positive amplification.} Drift in benign traffic may significantly increase false alarms, degrading system usability in real deployments.

    \item \textit{Forgetting-induced vulnerability}. Improper model updates may erase previously learned attack patterns, allowing known attacks to bypass detection.
\end{icompact}

Note that effective drift detection is indispensable in this context. On the one hand, a system may not currently exhibit significant drift but could become vulnerable as traffic evolves over time. On the other hand, an adversary may implicitly exploit existing drift patterns to bypass detection. Therefore, robust NIDS requires both reliable drift detection and stable model adaptation mechanisms.

\section{Design}
\label{sec_design}
In this section, we delineate the underlying motivation behind the design of DriftXpert and provide a comprehensive exposition of its methodological details.

\subsection{Design Motivation}
\label{sec_design_motivation}

\subsubsection{\textbf{Problem Statement}}
\label{sec_design_problem_statement}
The application of AI to NIDS scenarios can be conceptualized as a classification task. Let $p(X)$ represent the predicted distribution of the model, $q(X)$ denote the true label distribution, and optimize using the loss of cross entropy, which can be expressed as $H(p,q) = H(p) + D_{KL}(p || q)$, where $H(p,q)$ represents the cross-entropy, $H(p)$ denotes the entropy of $p(X)$, $D_{KL}(p || q)$ refers to the Kullback-Leibler (KL) divergence.
\begin{figure}[htbp]
\centering
\subfloat[$p(X)$.]{\includegraphics[width=0.24\textwidth]{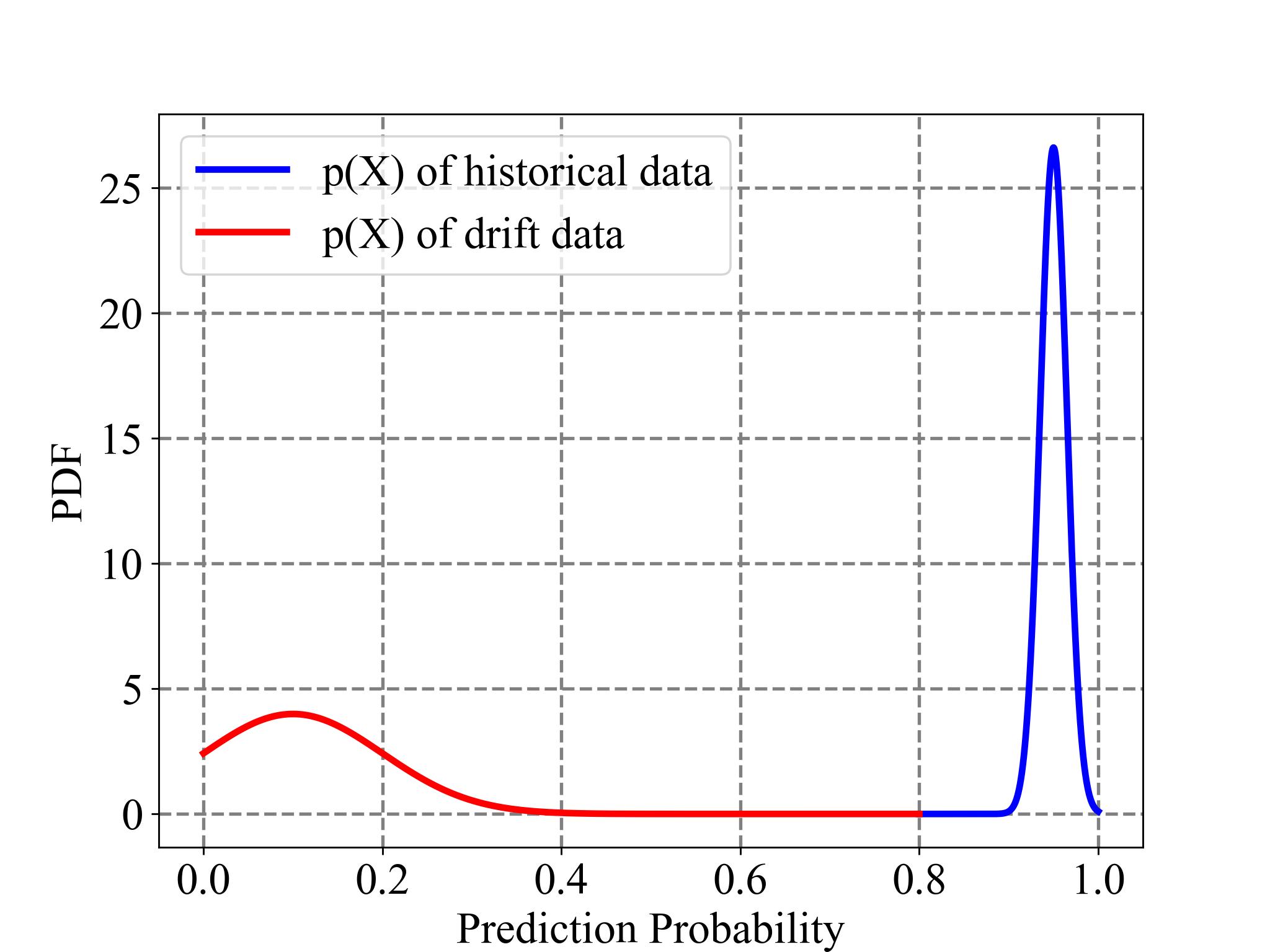}
\label{fig_variations_of_latent_variables_a}}
\subfloat[$z$.]{\includegraphics[width=0.24\textwidth]{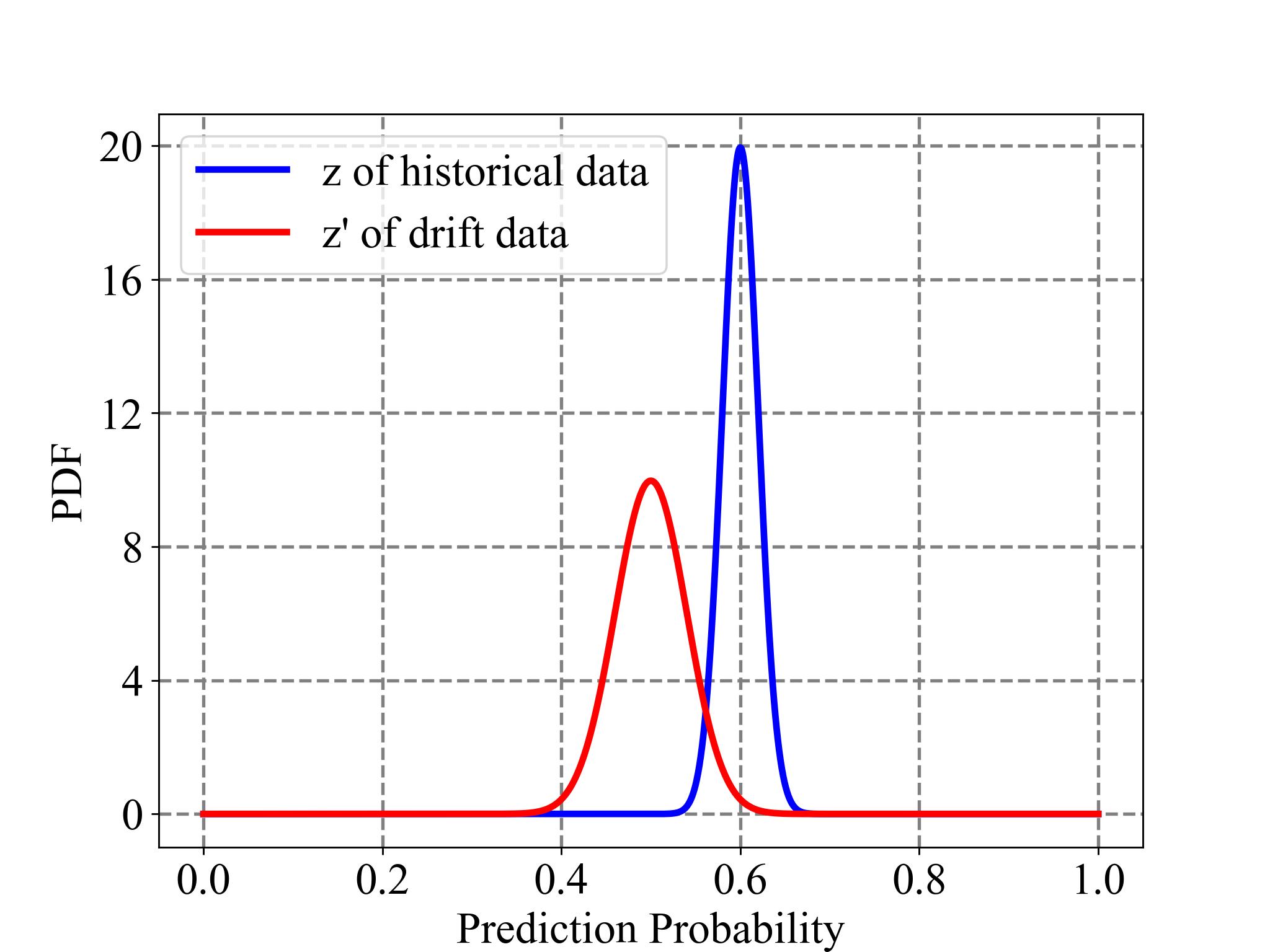}
\label{fig_variations_of_latent_variables_b}}
\caption{Trends in the variations of predicted probability distribution and latent variables.}
\label{fig_variations_of_latent_variables}
\end{figure}

The predict probability density function $p(X)$ can be expressed as a Gaussian distribution $\mathcal{G}(\mu_p, \sigma_p)$ (more detailed theoretical analysis can be found in Appendix \ref{appendix_theoretical_analysis_of_predict_distribution}). If the intrusion detection task is considered a binary classification problem, the positive label distribution $q(X)$ should be treated as a set $q (X) = \{1,... , 1\}$. Based on above classification optimization objective, where $D_{KL}$ represents the KL divergence \cite{van2014renyi} between $p(X)$ and $q(X)$, which quantifies the closeness between the two distributions, and $p(X)$ follows a Gaussian distribution $\mathcal{G}$, the optimal result is that the predictive distribution of the classifier is given by:
\begin{equation}
\label{eq_history_predict_distribution}
    p(X) = \frac{1}{\sqrt{2 \pi \sigma_p^2}} \exp{-\frac{1}{2 \sigma_p^2} }(X - \mu_p)^2, ~ \mu_p \rightarrow 1, \sigma_p \rightarrow 0.
\end{equation}

We further consider the impact of concept drift, where the prediction probability for positive labels decreases, leading to misclassification as the negative class. This results in an erroneous classification outcome. In this case, the predicted probability distribution $p'(X)$ for the drifted data can be expressed as:
\begin{equation}
\label{eq_drifted_predict_distribution}
    p'(X)= \frac{1}{\sqrt{2 \pi \sigma_{p'}^2}} \exp{-\frac{1}{2 \sigma_{p'}^2} }(X-\mu_{p'})^2, ~ \mu_{p'} \rightarrow 0, ~ \sigma_{p'} \rightarrow 1.
\end{equation}

Compared to Equation \eqref{eq_history_predict_distribution}, Equation \eqref{eq_drifted_predict_distribution} illustrates that the impact of concept drift manifests itself as a degradation in predictive performance, with changes in the probability distribution depicted in Figure \ref{fig_variations_of_latent_variables_a}. Based on this analysis, our goal is to attribute and interpret the effect of concept drift on AI models for intrusion detection, identifying patterns of drift and the corresponding feature changes in the model. However, the inherent nonlinearity and opacity of deep learning models make them challenging to interpret, a well-known difficulty in the academic community.

To address this, we shift our focus to reverse attribution - inferring possible changes in latent variables by analyzing variations in prediction outcomes. The model architecture widely used in the current study consists of an encoder with multiple fully connected layers and a classifier with a single fully connected layer \cite{lecun2015deep, rumelhart1986learning}. By examining the changes in the predicted probability distributions, our aim is to perform inverse causal inference on the behavior of the latent variables $z$ produced by the encoder. Since $z$ and $p(X)$ exhibit a linear relationship, this approach significantly reduces the complexity of the attribution process.

The predicted probability and the latent variables are connected through a single fully connected layer, and their relationship can be expressed as:
\begin{equation}
\begin{aligned}
    p(X) &= W \cdot z(X) + b, \\
    p'(X) &= W \cdot z'(X) + b,
\end{aligned}
\end{equation}
where $p(X)$ and $z(X)$ represent the predicted probability and latent variables of historical data, and $p'(X)$ and $z'(X)$ represent the predicted probability and latent variables of the drifted data.

Since the relationship between the predicted probability $p(X)$ and the latent variables $z$ is linear, combining Equation \eqref{eq_history_predict_distribution} and Equation \eqref{eq_drifted_predict_distribution}, we can infer that the variation pattern of $z$ and $z'$ should be approximately similar to that of $p$ and $p'$. The pattern of variation of latent variables $z$ is also depicted graphically in Figure \ref{fig_variations_of_latent_variables_b}. From the perspective of the encoder, the drifted data is not included in its training process. As the encoder extracts features from historical data to achieve classification, this results in differences in the distribution of latent variables, as illustrated in Figure \ref{fig_variations_of_latent_variables_b}. Similarly, when a new model is trained using drifted data to identify new samples, the historical data may exhibit drift relative to the recently collected data. The attribution results remain consistent with the aforementioned analysis.

\subsubsection{\textbf{Heuristic Insights}} 
\label{sec_design_insights}
The analytical results mentioned above provide crucial insight that informs the architectural design of DriftXpert. Specifically, these findings lead to the following design principles:
\begin{icompact}
    \item As illustrated in Figure \ref{fig_variations_of_latent_variables_b}, the evolution of drifted data distributions within the latent space exhibits strong congruence with the fluctuations in predictive probability distributions. Since the representation process in the latent space is label-independent, this distributional heterogeneity provides a pivotal entry point: we can achieve high-sensitivity, unsupervised drift detection by monitoring distribution shifts directly within the latent representation space.

    \item Figure \ref{fig_variations_of_latent_variables_b} suggests that aligning the latent distributions of drifted and historical data via distribution overlap or subspace embedding enables the model to generalize across temporal domains. Since the output layer is typically linear, if a weighted combination of drifted samples yields a prediction close to 1, historical features in the same subspace will similarly produce near-unity outputs, providing a mathematical guarantee of consistent detection during model evolution.
\end{icompact}

\subsection{System Architecture}
\label{sec_system_architecture}
We illustrate the proposed DriftXpert in Figure \ref{fig_driftxpert_framework}, a lightweight and robust real-time NIDS. The system is structured into the following components:
\begin{figure*}[htbp]
    \centering
    \includegraphics[width=0.95\textwidth]{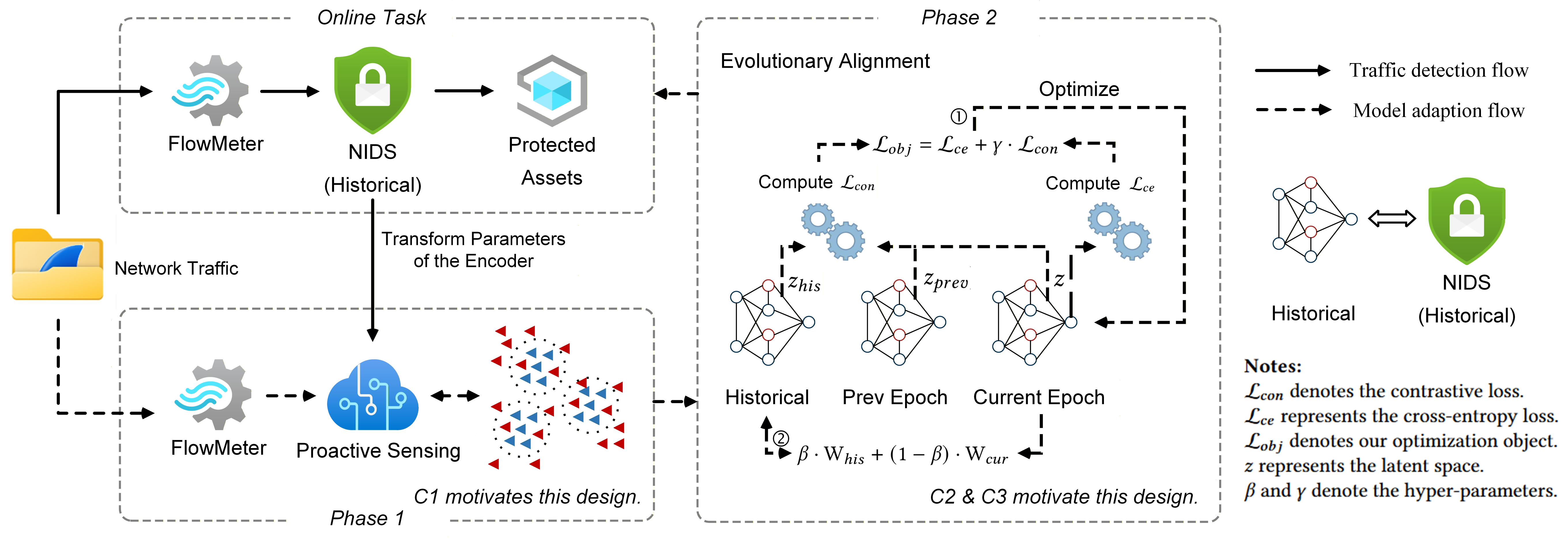}
    \caption{The overall architecture of DriftXpert. The system operates across three main components: an Online Task for real-time intrusion detection using a lightweight encoder-classifier structure. Phase 1 for continuous drift detection and deep clustering within traffic windows, and Phase 2 for model adaptation. Phase 2 employs a dual-objective optimization (\textcircled{\small 1}) and model aggregation (\textcircled{\small 2}) to integrate new knowledge while mitigating catastrophic forgetting.}
    \label{fig_driftxpert_framework} 
\end{figure*}

\begin{icompact}
    \item \textit{Online Task (Intrusion Detection).} Given the resource constraints and the real-time requirements, we designed a lightweight model structure. The structure consists of a multi-layer fully connected stacked encoder and a classifier made of a single fully connected layer. The classifier outputs a binary result: \textit{Benign} or \textit{Attack}.
    
    \item \textit{Phase 1 (Proactive Sensing).} As shown in Figure \ref{fig_driftxpert_framework}, phase 1 continuously monitors traffic within a fixed window at regular intervals (C1). Once drift is detected, the system triggers a model update to mitigate its impact. This process runs in the offline mode without real-time constraints, as detailed in $\S$ \ref{sec_drift_detection}.
    
    \item \textit{Phase 2 (Evolutionary Alignment).} As shown in Figure \ref{fig_driftxpert_framework}, upon detecting concept drift, phase 2 employs representation alignment, cross-epoch weight aggregation, and selective freezing to facilitate new knowledge acquisition (C2) while mitigating catastrophic forgetting (C3). This task also runs in the offline mode without real-time requirements, as detailed in $\S$ \ref{sec_model_adaptation}.
\end{icompact}

From a holistic perspective, DriftXpert employs two phases to accomplish model updates: drift detection continuously monitors network traffic for data drift, and once detected, triggers model adaptation. Through model alignment, a new NIDS model is trained and replaces the historical one, achieving adaptability while preserving robustness to historical data.

\subsection{Phase 1: Proactive Sensing}
\label{sec_drift_detection}
As discussed in $\S$ \ref{sec_design_insights}, using the consistency between latent variable changes and predicted probability changes (see $\S$ \ref{sec_design_problem_statement} and Figure \ref{fig_variations_of_latent_variables}), we propose detecting drift through latent space dynamics. Specifically, we develop a deep clustering-based detection framework \cite{caron2018deep, chang2017deep, zha2025nids} as shown in Figure \ref{fig_driftxpert_framework}. By reusing the historical model's encoder and performing outlier analysis on latent variables, this model achieves robust unsupervised drift detection.

To construct the deep clustering model, we first freeze all encoder parameters and then perform clustering in the latent variable space $Z$. Various approaches can be used to implement the clustering model \cite{xu2015comprehensive}, we choose the K-Means algorithm \cite{liu2023transforming}. This algorithm iteratively optimizes an objective function by assigning data points to $K$ clusters, minimizing the sum of squared distances between the data points and their respective cluster centers. The clustering model can be represented as $f_C: z \rightarrow [0, 1]^K$, $C$ denotes the cluster centers, and $Y=\{1, 2, ..., K\}$ represents the set of labels. Let $D(z, C_z)=||z-C_z||^2$ denote the distance of latent variables $z$ to their corresponding cluster centers $C_z$. Based on $D(z, C_z)$, we define two critical metrics, the confidence interval and the outlier ratio, to detect concept drift.

\noindent \textit{\textbf{Confidence Interval ($CI$).}} $CI$ defines a reliable distance range. When the distance $D(z, C_z)$ is less than $CI$, the data point is considered reliable, indicating that no drift has occurred. In contrast, when $D(z, C_z)$ exceeds $CI$, the data point is classified as an outlier, suggesting that drift has taken place. The mathematical definition of $CI$ is as follows:
\begin{equation}
\label{eq_ci}
    CI = \mathbb{E}_{z \sim Z_{h}}[D(z, C_z)] + \alpha \cdot \mathbb{S}_{z \sim Z_{h}}[D(z, C_z)],
\end{equation}
where $Z_h$ represents the latent variable space of historical data, $\mathbb{E}_{z \sim Z_{h}}$ and $\mathbb{S}_{z \sim Z_{h}}$ denote the expectation and standard deviation of the distances from each sample point in $Z_h$ to $C$, and $\alpha$ denotes an adjustable hyperparameter. 
    
\noindent \textit{\textbf{Outlier Ratio ($OR$).}} $OR$ is defined as the ratio of outliers in $Z_r$ within the recent data window to outliers in $Z_h$ within the historical data window, under the same $CI$. The mathematical definition of $OR$ is as follows:
\begin{equation}
    OR =  \frac{\frac{1}{N_{Z_r}}\sum_{i=0}^{N_{Z_r}} \mathbb{I}\{D(z, C_z) \notin [0, CI]\}}
               {\frac{1}{N_{Z_h}}\sum_{j=0}^{N_{Z_h}} \mathbb{I}\{D(z, C_z) \notin [0, CI]\}},
\end{equation}
where $N_{Z_r}$ and $N_{Z_h}$ denote the number of samples in $Z_r$ and $Z_h$, respectively. $I\{A\} = 1$ if and only if condition $A$ is true. When the denominator is 0, we set it to 1 for numerical stability. This only decreases the calculated $OR$ and does not affect drift detection, since a smaller $OR$ already satisfies the drift criterion under our detection rule.

Once $OR$ is calculated, we can determine whether drift has occurred by comparing it with a threshold $T$. The threshold $T$ is not treated as a model hyperparameter requiring sensitivity analysis. Since $OR=1$ naturally indicates equal outlier ratios in the recent and historical windows, $OR>1$ already provides evidence of drift. The choice of $T$ only determines the detector's operating sensitivity and should therefore be specified according to the application's tolerance for drift and the relative costs of false alarms and missed detections. This module also involves two key hyperparameters, $K$ and $\alpha$. Since the encoder is optimized for the classification task, we recommend that $K$ be within the range $[L, 3L]$, where $L$ represents the total number of classification labels. A more detailed theoretical analysis of $\alpha$ will be provided in Appendix \ref{appendix_hyperparameter_alpha_analysis}. Moreover, our method operates in the latent variable rather than in the original data, where the encoder incorporates knowledge from historical and drift data. This enables the drift detection model to be updated using only drift data, preventing catastrophic forgetting.

\subsection{Phase 2: Evolutionary Alignment}
\label{sec_model_adaptation}
As shown in Figure \ref{fig_variations_of_latent_variables_b}, latent variable variations exhibit trends consistent with predicted probability distributions. Consistent with $\S$ \ref{sec_design_insights}, our method is based on latent space synchronization. By embedding historical latent space into the drifted subspace, DriftXpert enables adaptive learning while preserving prior knowledge. Based on this, we propose the following design strategies.

\noindent \textbf{\textit{Model Alignment.}} Figure \ref{fig_driftxpert_framework} illustrates our proposed model update process. To align the distribution of latent variables of the drifted data with that of the historical data, we employ contrastive learning \cite{khosla2020supervised, li2021model}. We maintain three models: the current epoch model, the previous epoch model, and the historical model, producing latent variables $z$, $z_{prev}$, and $z_{his}$, respectively. The goal is to align the current latent variables with the historical ones while distancing them from the previous epoch, formalized through the following contrastive loss.
\begin{equation}
\label{con_loss}
    \mathcal{L}_{con} = -\log{
        \frac{ \exp{(\text{sim}(z, z_{his}) / \mathcal{T}}) }
        { \exp{(\text{sim}(z, z_{prev}) / \mathcal{T}}) + \exp{(\text{sim}(z, z_{his}) / \mathcal{T}}) }
        },
\end{equation}
where $\text{sim}(a,b)$ represents the cosine similarity between two vectors $a$ and $b$, and $\mathcal{T}$ denotes the temperature coefficient.

\noindent \textbf{\textit{Model Aggregation.}} Our design philosophy aims to approximate the latent variable distribution of historical data to that of drifted data, thereby addressing the catastrophic forgetting problem while simultaneously learning new knowledge. However, during model updates, the training data consists primarily of the most recent data. As a result, the output layer focuses not only on the important features of the latent space corresponding to historical data, but also on those of the recent data. To mitigate this, we aggregate the historical model with the updated model, allowing the model to converge more effectively and minimize the loss of knowledge due to forgetting. The aggregation process of the models is as follows.
\begin{equation}
    \text{W}_{his} = \beta \cdot \text{W}_{his} + (1-\beta) \cdot \text{W}_{cur},
\end{equation}
where $\text{W}_{his}$ and $\text{W}_{cur}$ represent the weights of the historical model and the model of the current epoch, respectively. $\beta$ is a hyperparameter.

\noindent \textbf{\textit{Input-Layer Weight Freezing.}} To mitigate catastrophic forgetting, we employ a parameter-freezing strategy that keeps the weights of the initial encoder layers unchanged during adaptation. Since early layers typically encode general and transferable features, freezing them helps preserve previously learned representations. Meanwhile, the higher layers remain trainable to capture new patterns, enabling effective adaptation while reducing the risk of forgetting.

\noindent \textbf{\textit{Optimizing Objective.}} Beyond aligning the latent space, classification on drifted data is performed using cross-entropy loss \cite{mao2023cross} to ensure convergence. The loss function is defined as follows.
\begin{equation}
\label{ce_loss}
    \mathcal{L}_{ce} = -\sum_{i=1}^{n} q(X) \log{p(X)},
\end{equation}
where $q(X)$ represents the true label distribution and $p(x)$ denotes the predicted label distribution.

Finally, we optimize the model update objective using two loss functions (Equation \eqref{con_loss} and Equation \eqref{ce_loss}. Specifically, our goal is to learn new knowledge while ensuring that the latent variable distribution of the drifted data approximates the latent variable distribution of the historical model, thus preventing catastrophic forgetting. Our optimization objective can be expressed as follows.
\begin{equation}
    \mathcal{L}_{obj} = \mathcal{L}_{ce} + \gamma \cdot \mathcal{L}_{con}, 
\end{equation}
where $\gamma$ is a hyperparameter.

\section{Experimental Setup}
\label{sec_experimental_setup}

\subsection{Implementation}
\label{sec_implementation}

\subsubsection{\textbf{Environment}} We deployed DriftXpert on PC with the following main configuration: 13th Gen Intel® Core™ i5-13500 CPU, Windows 11, 16GB DDR5 5200Hz RAM, and 1TB SSD. Our prototype is implemented in Python, the phase 1 is built using scikit-learn \cite{pedregosa2011scikit}, and the phase 2 is built using PyTorch \cite{paszke2019pytorch}. 

\subsubsection{\textbf{Public Datasets}} 
We used two widely public datasets, CICIDS-2017 \cite{CICIDS2017, sharafaldin2018toward} and CICIDS-2018 \cite{CICIDS2018, sharafaldin2018toward}, to simulate real-world scenarios. Using CICFlowMeter \cite{lashkari2017characterization}, we extracted 80 features, and retained 50 after evaluation, removing around 30 for the following reasons: i) Features causing "cheat effects" (e.g., 5-tuples). ii) Features with excessive outliers that could mislead results. iii) Zero-dominant binary features that are confirmed to have negligible impact on performance through experiments. Furthermore, we also addressed common issues, such as duplicate packets and labeling errors \cite{engelen2021troubleshooting}. Our evaluation encompasses two data drift paradigms, local drift and global drift, to rigorously assess the system's adaptability across diverse distributional shift scenarios. Local drift affects only specific regions of the data, while others remain stable, typically caused by distribution shifts in certain subsets. In contrast, global drift alters the entire distribution, often due to environmental changes or the emergence of new attacks. 

We partitioned the data sets to simulate global and local drift scenarios (more details on Table \ref{tab_dataset_settings} in Appendix \ref{subsec_drift_scenario_settings}). Since drift is distribution-related rather than time-related, we applied distribution-based rather than time-based partitioning.
\begin{icompact}
    \item For local drift, we partition the data with the same label into historical and drift data by distribution-based sampling.
    \item For global drift, several new attack traffic differing from historical data is considered drift data.
\end{icompact}

Historical and current data were divided into training and test sets, and the experiments used only the training set to keep the test set unseen. 

\subsubsection{\textbf{Real-world Datasets}} 
We collected real-world network traffic over several consecutive months in early 2026 from the external network of a large-scale enterprise. True labels are generated by the existing security system and then manually verified by domain experts to ensure high data quality and minimal labeling noise. The dataset includes diverse and representative attack types, such as command injection, directory traversal, and SQL injection, ensuring strong realism and coverage. For evaluation, the data are partitioned monthly: earlier traffic is treated as historical data, while later traffic is regarded as newly observed data. This setup enables systematic assessment of performance degradation under dynamic environments and the effectiveness of the proposed update mechanism. Due to the continuous evolution of real-world systems, the two stages exhibit not only significant shifts in benign traffic distributions but also previously unseen attack patterns, forming a more challenging concept drift scenario. Moreover, the enterprise serves millions of users, resulting in highly concurrent and complex traffic patterns. Consequently, the dataset faithfully reflects real operational environments, providing a solid foundation to evaluate the robustness, adaptability, and deployability.

\subsubsection{\textbf{Baselines}} We selected five methods for comparison to validate the effectiveness of our method, as detailed below.
\begin{icompact}
    \item \textit{Mlp-FineTuning (M.F.)}. No parameter freezing, and the new model is fine-tuned over 5 epochs using the drifted dataset.
    \item \textit{Mlp-LayerFreezed-FT (M.LF.FT)}. The first encoder layer is frozen, and the new model is fine-tuned for 5 epochs.
    \item \textit{OWAD @NDSS23}. Han et al. \cite{han2023anomaly} proposed weighting model parameters to restrict updates on important ones, preserving old knowledge while adapting new distributions.
    \item \textit{A-NIDS @TIFS25}. Zha et al. \cite{zha2025nids} proposed a method that uses CTGAN to replay historical data, combining it with the latest data to update the model.
    \item \textit{SSF @INFOCOM25}. Zhang et al. \cite{11044615} proposed SSF, an incremental method that captures concept drift by selecting representative new samples and discarding outdated ones.
\end{icompact}

\subsubsection{\textbf{Metrics}} We used five metrics to evaluate DriftXpert, including five commonly used metrics: \textit{Accuracy (AC)}, \textit{Precision}, \textit{Recall}, \textit{F1-Score}, and \textit{True Positive Rate (TPR)} \cite{japkowicz2011evaluating}.

\section{Benchmark Evaluation on Public Datasets}
\label{evaluation_on_public_datasets}
In this section, we evaluate the performance of DriftXpert through comparative experiments on two publicly available datasets, complemented by a series of analytical studies, including ablation experiments and hyperparameter sensitivity analyses.

\subsection{Research Questions}
\label{sec_public_rq}
Our research questions (RQs) in this section are as follows.
\begin{icompact}
    \item RQ1: How effective is DriftXpert in accurately identifying concept drift within dynamic network traffic streams?
    
    \item RQ2: What extent does DriftXpert outperform SOTA methodologies in terms of model adaptation efficiency?

    \item RQ3: What are the individual contributions of the key components within DriftXpert to the overall detection performance?

    \item RQ4: Can DriftXpert effectively align latent representation spaces across different temporal stages?

    \item RQ5: How sensitive is the performance of DriftXpert to variations in key hyperparameters, such as $\alpha$, $\beta$, $\gamma$?
\end{icompact}

\subsection{RQ1: Proactive Sensing Results by DriftXpert}
\label{sec_public_rq1}
To evaluate the efficacy of the drift detection module, we conducted a suite of evaluation experiments in four distinct scenarios derived from two benchmark datasets. We quantified the presence of outliers for both historical and drifted telemetry within each cluster and calculated the corresponding outlier ratio. The empirical results are detailed in Table \ref{tab_local_drift_detection_results} and Table \ref{tab_global_drift_detection_results}. Experimental results indicate that outlier divergence is more pronounced within isolated data clusters. Across four distinct scenarios, our proposed deep clustering-based drift detection methodology effectively identifies concept drift. Crucially, as the primary objective is to serve as a high-fidelity trigger for phase 2, these findings empirically corroborate the module's functional integrity. Furthermore, this component incorporates two critical hyperparameters: the cluster count $K$ and the outlier threshold coefficient $\alpha$. While $K$ can be precisely estimated leveraging historical data and clustering heuristics, the sensitivity of $\alpha$ is systematically analyzed in $\S$ \ref{sec_public_rq5}, with its optimal value derivation and corresponding mathematical simulations provided in Appendix \ref{appendix_hyperparameter_alpha_analysis}.
\begin{table}[htbp]
    \centering
    \caption{Local Drift Detection Results on Public Datasets.}
    \begin{tabularx}{0.49\textwidth}{
        >{\centering\arraybackslash}p{0.08\textwidth} | 
        >{\centering\arraybackslash}p{0.08\textwidth} 
        >{\centering\arraybackslash}X
        >{\centering\arraybackslash}X
        >{\centering\arraybackslash}X 
        >{\centering\arraybackslash}p{0.04\textwidth} 
        }
        \toprule
        \multirow{2}{*}{Datasets} & \multicolumn{5}{c}{\textbf{\textit{Outliers / Samples of i-th Cluster}}} \\
        \cmidrule(lr){2-6}
        & \textit{1} & \textit{2} & \textit{3} & \textit{4} & \textit{5} \\
        \midrule
        2017-Set1 & 686/4,277 & 84/551 & 99/692 & 15/259 & 0/1 \\
        2017-Set2 & 15,727/21,331 & 860/860 & 766/3,195 & 356/3,528 & 1/2 \\
        OR & 4.60 & 6.60 & 1.68 & 1.74 & - \\

        \midrule
        2018-Set1 & 22/2,049 & 19/3,791 & 5/988 & 1/277 & 0/629 \\
        2018-Set2 & 773/12,928 & 23/23 & 16/15,269 & 10/3,100 & 1/3,106 \\
        OR & 5.57 & - & - & - & - \\
        \bottomrule
    \end{tabularx}
    \begin{tablenotes}
        \vspace{0.5ex} 
        \footnotesize \item[*] - OR is not reported for clusters with negligible outlier counts due to lack of representative significance.
    \end{tablenotes}
    \label{tab_local_drift_detection_results}
\end{table}

\begin{table}[htbp]
    \centering
    \caption{Global Drift Detection Results on Public Datasets.}
    \begin{tabularx}{0.49\textwidth}{
        >{\centering\arraybackslash}p{0.07\textwidth} | 
        >{\centering\arraybackslash}p{0.07\textwidth}
        >{\centering\arraybackslash}p{0.08\textwidth}
        >{\centering\arraybackslash}p{0.06\textwidth}
        >{\centering\arraybackslash}X 
        >{\centering\arraybackslash}p{0.04\textwidth} 
        }
        \toprule
        \multirow{2}{*}{Datasets} & \multicolumn{5}{c}{\textbf{\textit{Outliers / Samples of i-th Cluster}}} \\
        \cmidrule(lr){2-6}
        & \textit{1} & \textit{2} & \textit{3} & \textit{4} & \textit{5} \\
        \midrule
        2017-Set1 & 169/6,936 & 1/1,948 & 4/407 & 12/1,044 & 0/1,670 \\
        2017-Set2 & 91,90/45,676 & 50/7,187 & 35/178 & 0/2,238 & 0/741 \\
        OR & 8.26 & 13.55 & - & - & - \\

        \midrule
        2018-Set1 & 37/7,095 & 1/2,696 & 100/3,717 & 1/97 & 0/1,486 \\
        2018-Set2 & 4,988/8,184 & 4,298/38,978 & 843/13,486 & 9/1,325 & 0/1,466 \\
        OR & 116.87 & 297.28 & 2.32 & - & - \\
        \bottomrule
    \end{tabularx}
    \label{tab_global_drift_detection_results}
\end{table}

\subsection{RQ2: Performance of Evolutionary Alignment vs. Baselines}
\label{sec_public_rq2}
To evaluate our proposed method, we conducted comparative experiments with two fine-tuning-based methods and three SOTA methods \cite{han2023anomaly, zha2025nids, 11044615} under two different drift scenarios in the CICIDS-2017 and CICIDS-2018 dataset. As shown in Table \ref{tab_cmp_in_cicids17_local_drift}, Table \ref{tab_cmp_in_cicids17_global_drift}, Table \ref{tab_cmp_in_cicids18_local_drift} and Table \ref{tab_cmp_in_cicids18_global_drift}, \textit{Model-FineTuning} achieves high recall under local drift on CICIDS-2017 (99.86\% benign, 86.22\% attack) without forgetting, but its attack recall drops sharply to 10.37\% under global drift. \textit{Model-LayerFreezed-FT} improves attack recall by 14\% under local drift and increases it to 24.40\% under global drift, but remains overall inferior. In CICIDS-2018, both fine-tuning methods adapt well to recent data under local drift (recall $>$98\%) but suffer severe catastrophic forgetting, with only slight mitigation from layer freezing; similar trends persist under global drift. Among SOTA methods, OWAD shows limited adaptability (64.30\% recall for drifted anomalies in CICIDS-2017), A-NIDS improves detection via generative replay but still lags behind, and SSF preserves historical knowledge yet fails on drifted anomalies (30.41\% recall), further degrading under global drift. On CICIDS-2018, OWAD struggles with both forgetting and adaptation, A-NIDS sacrifices historical knowledge for recent performance, and SSF exhibits the opposite behavior. Overall, \textit{DriftXpert} consistently achieves a better balance between adaptability and stability, outperforming all baselines across datasets.
\begin{table*}[htbp]
    \centering
    \caption{Comparative Experiments (Benign\%/ Malicious\%) in the CICIDS-2017 Dataset on the Local Drift Scenario.}
    \begin{tabularx}{1.0\textwidth}{>{\centering\arraybackslash}p{0.18\textwidth}|
    >{\centering\arraybackslash}X >{\centering\arraybackslash}X >{\centering\arraybackslash}X | 
    >{\centering\arraybackslash}X >{\centering\arraybackslash}X >{\centering\arraybackslash}X}
        \toprule
        \multirow{2}{*}{\textbf{\textit{Methods}}} & \multicolumn{3}{c|}{\textbf{\textit{Set 1 (historical data)}}} & \multicolumn{3}{c}{\textbf{\textit{Set 2 (recent data)}}} \\
        \cmidrule(lr){2-4} \cmidrule(lr){5-7}
         & \textit{Precision.} & \textit{Recall.} & \textit{F1-Score.} & \textit{Precision.} & \textit{Recall.} & \textit{F1-Score.} \\
        \midrule
        Model-FineTuning & 98.67\%/ 98.31\% & 99.86\%/ 86.22\% & 99.26\%/ 91.87\% & 99.66\%/ 98.22\% & 99.83\%/ 96.51\% & 99.74\%/ 97.36\% \\
        Model-LayerFreezed-FT & 99.99\%/ 94.87\% & 99.47\%/ 99.99\% & 99.74\%/ 97.37\% & 98.87\%/ 96.27\% & 99.68\%/ 87.98\% & 99.27\%/ 91.94\% \\
        OWAD @NDSS23 & 99.91\%/ 33.31\% & 82.27\%/ 99.16\% & 90.23\%/ 49.47\% & 96.01\%/ 26.10\% & 82.50\%/ 64.30\% & 88.74\%/ 37.13\% \\
        A-NIDS @TIFS25 & 99.97\%/ 98.78\% & 99.89\%/ 99.73\% & 99.93\%/ 99.25\% & 99.16\%/ 97.04\% & 99.73\%/ 91.14\% & 99.45\%/ 94.00\% \\
        SSF @INFOCOM25 & 99.77\%/ 99.40\% & 99.94\%/ 97.65\% & 99.86\%/ 98.52\% & 93.32\%/ 81.91\% & 99.31\%/ 30.41\% & 96.22\%/ 44.35\% \\
        \rowcolor{gray!20}
        DriftXpert (\textbf{Ours}) & 99.87\%/ 97.09\% & 99.71\%/ 98.72\% & 99.79\%/ 97.90\% & 99.70\%/ 98.88\% & 99.89\%/ 97.06\% & 99.79\%/ 97.96\% \\
        \bottomrule
    \end{tabularx}
    \label{tab_cmp_in_cicids17_local_drift}
\end{table*}

\begin{table*}[htbp]
    \centering
    \caption{Comparative Experiments (Benign\%/Malicious\%) in the CICIDS-2017 Dataset on the Global Drift Scenario.}
    \begin{tabularx}{1.0\textwidth}{>{\centering\arraybackslash}p{0.18\textwidth}|
    >{\centering\arraybackslash}X >{\centering\arraybackslash}X >{\centering\arraybackslash}X | 
    >{\centering\arraybackslash}X >{\centering\arraybackslash}X >{\centering\arraybackslash}X}
        \toprule
        \multirow{2}{*}{\textbf{\textit{Methods}}} & \multicolumn{3}{c|}{\textbf{\textit{Set 1 (historical data)}}} & \multicolumn{3}{c}{\textbf{\textit{Set 2 (recent data)}}} \\
        \cmidrule(lr){2-4} \cmidrule(lr){5-7}
         & \textit{Precision.} & \textit{Recall.} & \textit{F1-Score.} & \textit{Precision.} & \textit{Recall.} & \textit{F1-Score.} \\
        \midrule
        Model-FineTuning & 50.13\%/ 75.39\% & 96.38\%/ 10.37\% & 65.95\%/ 18.23\% & 98.40\%/ 95.06\% & 95.34\%/ 98.30\% & 96.85\%/ 96.65\% \\
        Model-LayerFreezed-FT & 51.60\%/ 65.45\% & 86.22\%/ 24.40\% & 64.56\%/ 35.55\% & 97.12\%/ 93.72\% & 94.05\%/ 96.96\% & 95.56\%/ 95.31\% \\
        OWAD @NDSS23 & 82.10\%/ 97.02\% & 97.33\%/ 80.42 & 89.07\%/ 87.94\% & 51.31\%/ 47.86\% & 63.58\%/ 35.66\% & 56.79\%/ 40.87\% \\
        A-NIDS @TIFS25 & 72.11\%/ 96.21\% & 97.33\%/ 64.29\% & 82.85\%/ 77.07\% & 98.56\%/ 98.61\% & 98.74\%/ 98.42\% & 98.65\%/ 98.52\% \\
        SSF @INFOCOM25 & 98.89\%/ 91.86\% & 90.81\%/ 99.02\% & 94.67\%/ 95.31\% & 60.94\%/ 56.32\% & 60.32\%/ 56.96\% & 60.63\%/ 56.64\% \\
        \rowcolor{gray!20}
        DriftXpert (\textbf{Ours}) & 78.84\%/ 97.22\% & 97.70\%/ 75.35\% & 87.20\%/ 84.90\% & 98.89\%/ 97.64\% & 97.81\%/ 98.80\% & 98.35\%/ 98.22\% \\
        \bottomrule
    \end{tabularx}
    \label{tab_cmp_in_cicids17_global_drift}
\end{table*}

\begin{table*}[htbp]
    \centering
    \caption{Comparative Experiments (Benign\%/Malicious\%) in the CICIDS-2018 Dataset on the Local Drift Scenario.}
    \begin{tabularx}{1.0\textwidth}{>{\centering\arraybackslash}p{0.18\textwidth}|
    >{\centering\arraybackslash}X >{\centering\arraybackslash}X >{\centering\arraybackslash}X | 
    >{\centering\arraybackslash}X >{\centering\arraybackslash}X >{\centering\arraybackslash}X}
        \toprule
        \multirow{2}{*}{\textbf{\textit{Methods}}} & \multicolumn{3}{c|}{\textbf{\textit{Set 1 (historical data)}}} & \multicolumn{3}{c}{\textbf{\textit{Set 2 (recent data)}}} \\
        \cmidrule(lr){2-4} \cmidrule(lr){5-7}
         & \textit{Precision.} & \textit{Recall.} & \textit{F1-Score.} & \textit{Precision.} & \textit{Recall.} & \textit{F1-Score.} \\
        \midrule
        Model-FineTuning & 50.70\%/ 0.01\% & 98.21\%/ 0.01\% & 66.88\%/ 0.01\% & 99.95\%/ 97.91\% & 98.44\%/ 99.93\% & 99.19\%/ 98.91\% \\
        Model-LayerFreezed-FT & 50.69\%/ 0.01\% & 98.17\%/ 0.01\% & 66.86\%/ 0.01\% & 99.97\%/ 97.50\% & 98.15\%/ 99.97\% & 99.05\%/ 98.72\% \\
        OWAD @NDSS23 & 50.96\%/ 0.01\% & 98.24\%/ 0.01\% & 67.11\%/ 0.01\% & 56.51\%/ 11.27\% & 97.86\%/ 0.36\% & 71.65\%/ 0.70\% \\
        A-NIDS @TIFS25 & 51.48\%/ 48.48\% & 98.87\%/ 1.14\% & 67.98\%/ 2.24\% & 99.97\%/ 97.77\% & 98.29\%/ 99.98\% & 99.12\%/ 98.85\% \\
        SSF @INFOCOM25 & 99.99\%/ 99.99\% & 99.99\%/ 99.99\% & 99.99\%/ 99.99\% & 99.97\%/ 0.00\% & 98.29\%/ 0.00\% & 73.10\%/ 0.00\% \\
        \rowcolor{gray!20}
        DriftXpert (\textbf{Ours}) & 87.47\%/ 97.70\% & 98.08\%/ 85.29\% & 92.47\%/ 91.07\% & 99.99\%/ 97.54\% &  98.08\%/ 99.99\% &  99.03\%/ 98.75\% \\
        \bottomrule
    \end{tabularx}
    \label{tab_cmp_in_cicids18_local_drift}
\end{table*}

\begin{table*}[htbp]
    \centering
    \caption{Comparative Experiments (Benign\%/Malicious\%) in the CICIDS-2018 Dataset on the Global Drift Scenario.}
    \begin{tabularx}{1.0\textwidth}{>{\centering\arraybackslash}p{0.18\textwidth}|
    >{\centering\arraybackslash}X >{\centering\arraybackslash}X >{\centering\arraybackslash}X | 
    >{\centering\arraybackslash}X >{\centering\arraybackslash}X >{\centering\arraybackslash}X}
        \toprule
        \multirow{2}{*}{\textbf{\textit{Methods}}} & \multicolumn{3}{c|}{\textbf{\textit{Set 1 (historical data)}}} & \multicolumn{3}{c}{\textbf{\textit{Set 2 (recent data)}}} \\
        \cmidrule(lr){2-4} \cmidrule(lr){5-7}
         & \textit{Precision.} & \textit{Recall.} & \textit{F1-Score.} & \textit{Precision.} & \textit{Recall.} & \textit{F1-Score.} \\
        \midrule
        Model-FineTuning & 45.68\%/ 83.89\%  & 97.58\%/ 9.82\% & 62.23\%/ 17.58\% & 96.78\%/ 98.84\% & 98.95\%/ 96.44\% & 97.85\%/ 97.63\% \\
        Model-LayerFreezed-FT & 81.08\%/ 83.94\% & 78.91\%/ 85.69\% & 79.98\%/ 84.80\% & 96.31\%/ 92.18\% & 92.52\%/ 96.13\% & 94.38\%/ 94.11\% \\
        OWAD @NDSS23 & 43.84\%/ 51.90\% & 98.34\%/ 1.40\% & 60.64\%/ 2.73\% & 51.57\%/ 28.02\% & 97.35\%/ 1.12\% & 67.43\%/ 2.15\% \\
        A-NIDS @TIFS25 & 70.29\%/ 92.28\% & 92.34\%/ 70.11\% & 79.82\%/ 79.68\% & 98.21\%/ 99.60\% & 99.64\%/ 98.02\% & 98.92\%/ 98.80\% \\
        SSF @INFOCOM25 & 98.73\%/ 92.69\% & 91.59\%/ 98.91\% & 95.02\%/ 95.70\% & 62.51\%/ 58.60\% & 61.52\%/ 59.62\% & 62.01\%/ 59.10\% \\
        \rowcolor{gray!20}
        DriftXpert (\textbf{Ours}) & 79.98\%/ 99.07\% & 99.03\%/ 80.73\% & 88.49\%/ 88.97\% & 97.11\%/ 99.11\% & 99.19\%/ 96.83\% & 98.14\%/ 97.96\% \\
        \bottomrule
    \end{tabularx}
    \label{tab_cmp_in_cicids18_global_drift}
\end{table*}

\begin{figure}[htbp]
\centering
\subfloat[CICIDS-2017 (Local Drift).]{\includegraphics[width=0.24\textwidth]{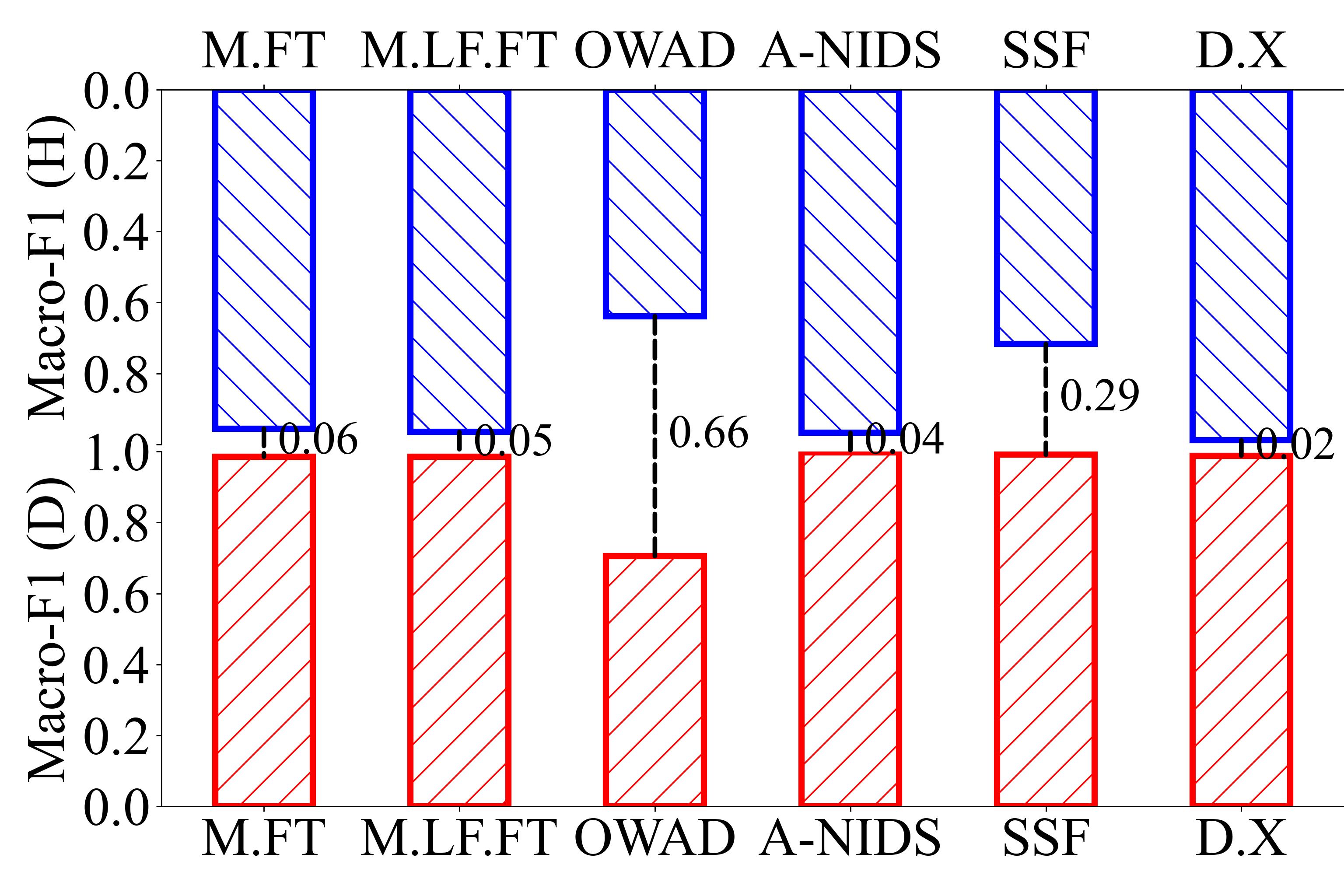}
\label{fig_cmp_macro_f1_cicids17_local}}
\subfloat[CICIDS-2017 (Global Drift).]{\includegraphics[width=0.24\textwidth]{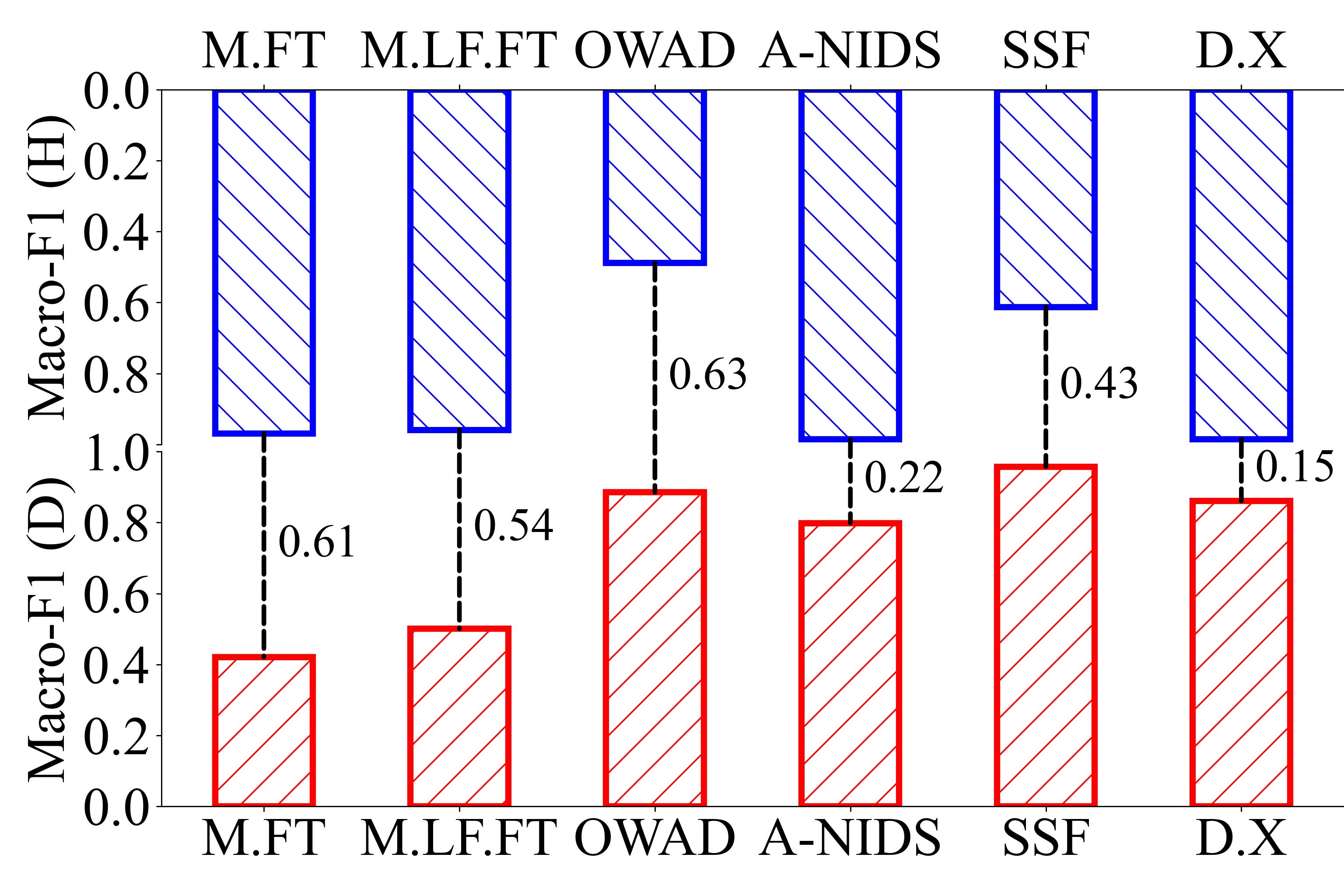}
\label{fig_cmp_macro_f1_cicids17_global}}

\subfloat[CICIDS-2018 (Local Drift).]{\includegraphics[width=0.24\textwidth]{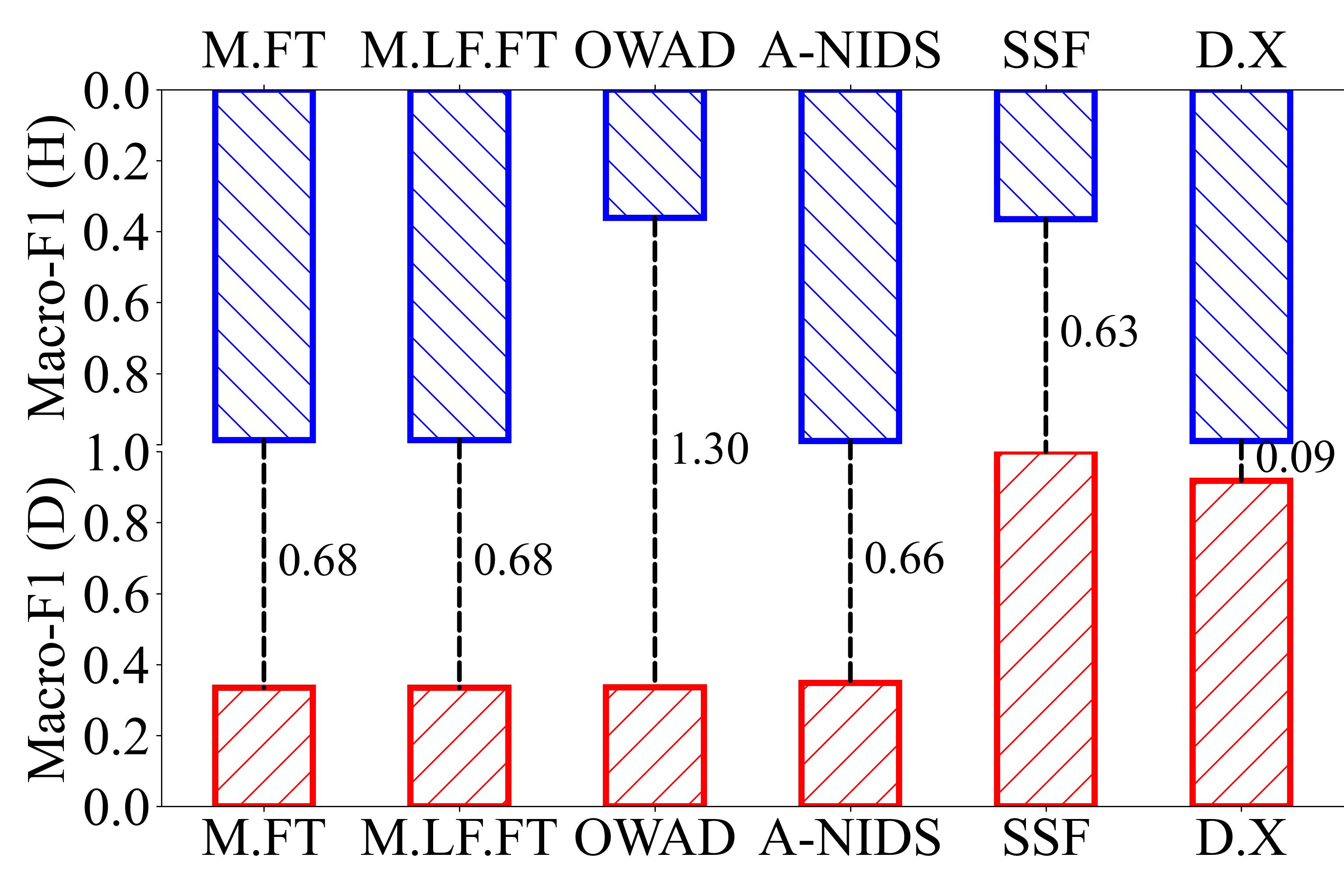}
\label{fig_cmp_macro_f1_cicids18_local}}
\subfloat[CICIDS-2018 (Global Drift).]{\includegraphics[width=0.24\textwidth]{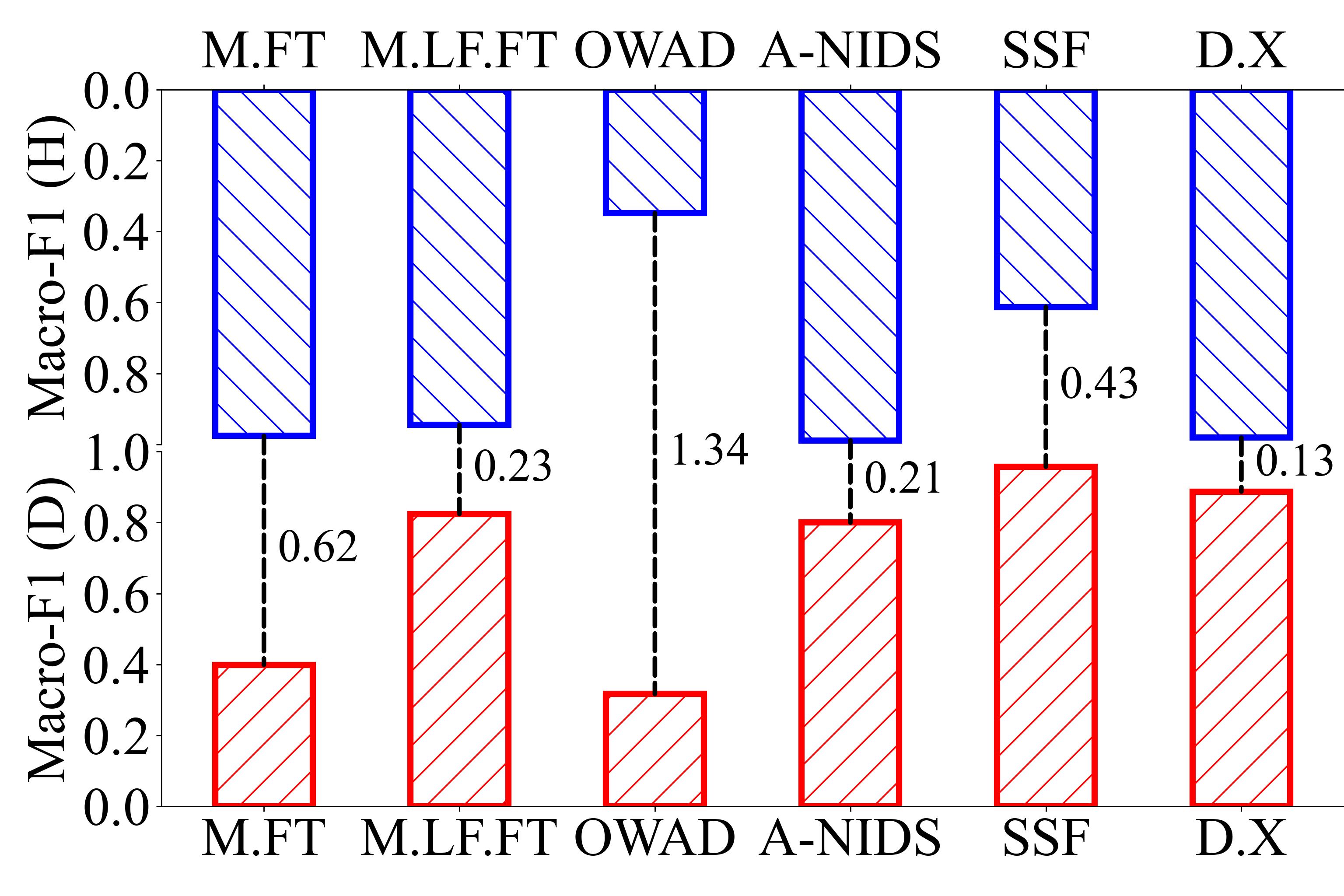}
\label{fig_cmp_macro_f1_cicids18_global}}
\caption{Comparative Results (Macro-F1) vs. Baselines.}
\label{fig_cmp_macro_f1}
\end{figure}

To further address the severe class-imbalance issue, we additionally report the Macro-F1 score on both the historical and drifted datasets. As illustrated in Figure \ref{fig_cmp_macro_f1}, most baseline methods exhibit a pronounced adaptation–retention gap, indicating an inherent trade-off between learning new traffic patterns and preserving historical knowledge. Conventional fine-tuning methods fail to adequately adapt to the drifted distribution, whereas SSF achieves stronger adaptation at the cost of substantial degradation on historical data, revealing severe catastrophic forgetting. In contrast, DriftXpert consistently achieves the smallest adaptation–retention gap across all four experimental settings, demonstrating its ability to effectively accommodate concept drift while maintaining balanced recognition performance on previously learned traffic.

\subsection{RQ3: Ablation Study}
\label{sec_public_rq3}
We conducted ablation experiments in the CICIDS-2017 \cite{CICIDS2017, sharafaldin2018toward} and CICIDS-2018 \cite{CICIDS2018, sharafaldin2018toward}, and presented in Table \ref{tab_ablation_in_cicids17_local_drift}, Table \ref{tab_ablation_in_cicids17_global_drift}, Table \ref{tab_ablation_in_cicids18_local_drift} and Table \ref{tab_ablation_in_cicids18_global_drift}. Specifically, the following nomenclature defines the ablation variants, delineating the correspondence between the model configurations, their sub-modules, and the data scopes.
\begin{icompact}
    \item \textit{D.X-With-H}. Consistent with the DriftXpert model architecture, the training dataset used is Set1 (historical data).
    \item \textit{D.X-With-D}. Consistent with the DriftXpert model architecture, the training dataset used is Set2 (drifted data).
    \item \textit{D.X-Only-C}. Consistent with the DriftXpert model architecture, only contrastive learning is retained as the adaptation strategy.
    \item \textit{D.X-EnFre}. All encoder parameters are frozen, the new model is trained in Set2 with contrastive loss and aggregation.
    \item \textit{D.X-NoFre}. Without parameters freezing, the new model is trained in Set2 using contrastive loss with aggregation.
    \item \textit{D.X-NoAgg}. The first layer parameters are frozen, the new model is trained in Set2 with contrastive loss, without aggregation.
\end{icompact}

In the ablation study, D.X-Static-H and D.X-Static-R exhibit high intra-distribution accuracy but suffer from catastrophic forgetting or poor adaptability. D.X-Only-C demonstrates strong adaptability to new knowledge, however, its ability to mitigate catastrophic forgetting remains limited. D.X-EncFreezed partially preserves historical knowledge by freezing the encoder, yet its adaptability remains limited. Further comparisons with D.X-NoFreezed and D.X-NoAggregation confirm that first-layer freezing is essential for feature retention, while the aggregation strategy is vital for drifted traffic detection. Consequently, DriftXpert optimizes the trade-off between adaptability and stability. It maintains approximately 98\% precision and recall on drifted data and high historical recall on the CICIDS-2017 dataset, effectively mitigating catastrophic forgetting.
\begin{table}[htbp]
    \centering
    \caption{Ablation Studies (Benign\%/Malicious\%) in the CICIDS-2017 Dataset on the Local Drift Scenario.}
    \begin{tabularx}{0.49\textwidth}{
        >{\centering\arraybackslash}p{0.09\textwidth} |
        >{\centering\arraybackslash}X >{\centering\arraybackslash}X | 
        >{\centering\arraybackslash}X >{\centering\arraybackslash}X
    }
        \toprule
        \multirow{2}{*}{\textbf{\textit{Methods}}} & \multicolumn{2}{c|}{\textbf{\textit{Set 1 (historical data)}}} & \multicolumn{2}{c}{\textbf{\textit{Set 2 (recent data)}}} \\
        \cmidrule(lr){2-3} \cmidrule(lr){4-5}
        & \textit{Recall.} & \textit{F1-Score.} & \textit{Recall.} & \textit{F1-Score.} \\
        \midrule
        D.X-With-H & 99.99/99.99 & 99.99/99.99 & 99.97/29.27 & 96.64/45.18 \\
        D.X-With-D & 99.99/0.01 & 95.34/0.01 & 99.98/97.95 & 99.89/98.87 \\
        D.X-Only-C & 99.89/72.48 & 99.91/83.50 & 99.98/98.58 & 99.91/99.19 \\
        D.X-EnFre & 99.97/99.69 & 99.97/99.69 & 99.90/32.06 & 96.68/48.21 \\
        D.X-NoFre & 99.92/80.15 & 99.00/88.60 & 99.90/98.69 & 99.89/98.88 \\
        D.X-NoAgg & 98.24/99.99 & 99.11/91.72 & 99.79/96.04 & 99.70/96.95 \\
        \rowcolor{gray!20}
        DriftXpert & 99.71/98.72 & 99.79/97.90 & 99.89/97.06 & 99.79/97.96 \\
        \bottomrule
    \end{tabularx}
    \label{tab_ablation_in_cicids17_local_drift}
\end{table}

\begin{table}[htbp]
    \centering
    \caption{Ablation Studies (Benign\%/Malicious\%) in the CICIDS-2017 Dataset on the Global Drift Scenario.}
    \begin{tabularx}{0.49\textwidth}{
        >{\centering\arraybackslash}p{0.09\textwidth}|
        >{\centering\arraybackslash}X >{\centering\arraybackslash}X |
        >{\centering\arraybackslash}X >{\centering\arraybackslash}X 
    }
        \toprule
        \multirow{2}{*}{\textbf{\textit{Methods}}} & \multicolumn{2}{c|}{\textbf{\textit{Set 1 (historical data)}}} & \multicolumn{2}{c}{\textbf{\textit{Set 2 (recent data)}}} \\
        \cmidrule(lr){2-3} \cmidrule(lr){4-5}
        & \textit{Recall.} & \textit{F1-Score.} & \textit{Recall.} & \textit{F1-Score.} \\
        \midrule
        D.X-With-H & 98.59/99.98 & 99.28/99.34 & 79.40/40.34 & 67.73/49.60 \\
        D.X-With-D & 99.83/0.57 & 65.21/1.14 & 98.53/97.83 & 98.29/98.09 \\
        D.X-Only-C & 92.13/19.27 & 66.16/30.44 & 99.29/99.24 & 99.26/99.20 \\
        D.X-EnFre & 76.21/81.62 & 77.62/80.08 & 72.04/78.37 & 74.86/75.61 \\
        D.X-NoFre & 92.49/31.74 & 69.67/45.75 & 99.29/99.17 & 99.27/99.19 \\
        D.X-NoAgg & 97.90/69.35 & 84.87/80.96 & 97.96/98.81 & 98.44/98.27 \\
        \rowcolor{gray!20}
        DriftXpert & 97.70/75.35 & 87.20/84.90 & 97.81/98.80 & 98.35/98.22 \\
        \bottomrule
    \end{tabularx}
    \label{tab_ablation_in_cicids17_global_drift}
\end{table}

\subsection{RQ4: Alignment Analysis of Latent Space}
\label{sec_public_rq4}
\begin{figure}[htbp]
\centering
\subfloat[D.X-Static-H (Local Drift).]{\includegraphics[width=0.24\textwidth]{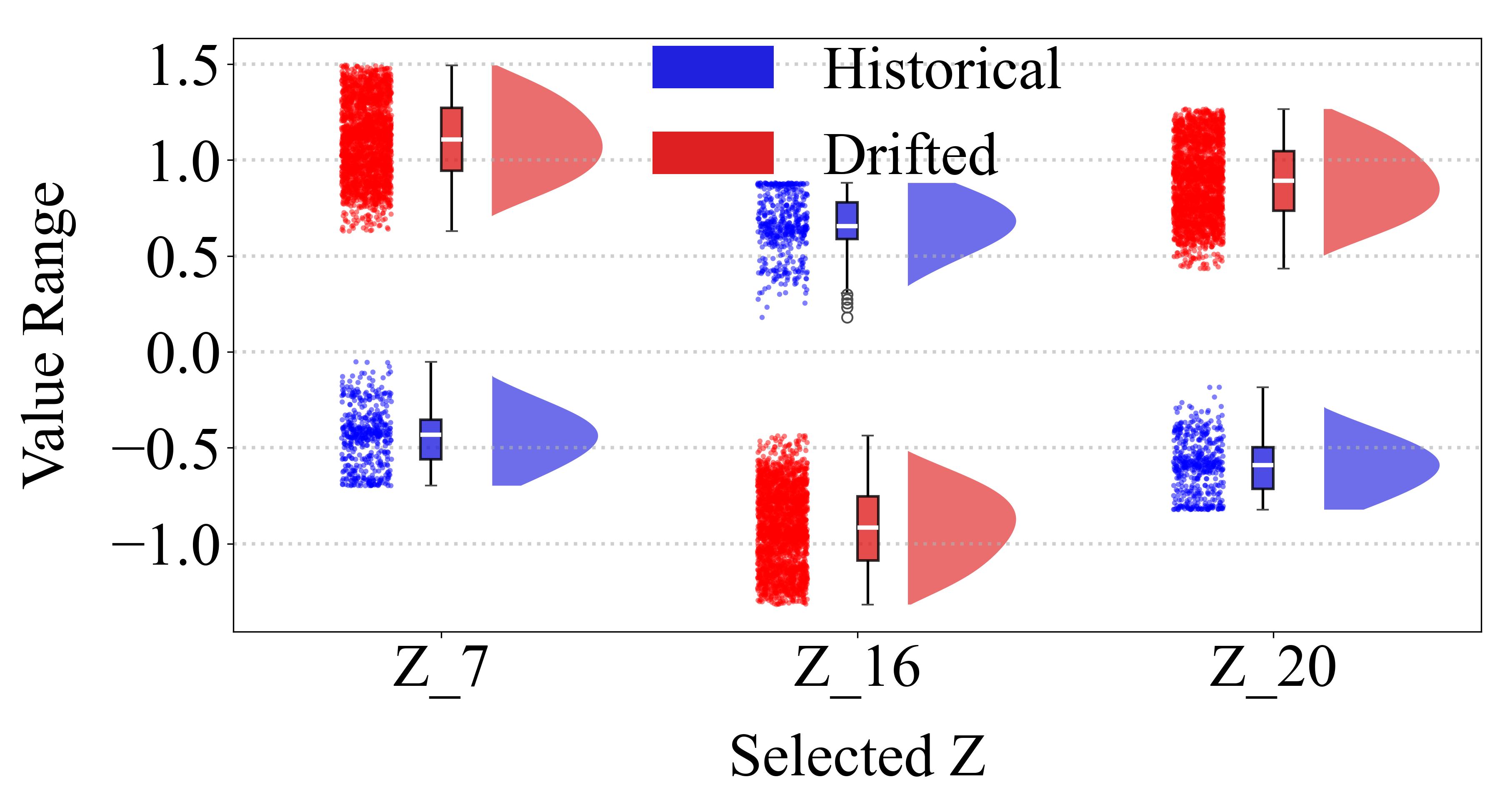}
\label{fig_alignment_analysis_cicids17_a}}
\subfloat[DriftXpert (Local Drift).]{\includegraphics[width=0.24\textwidth]{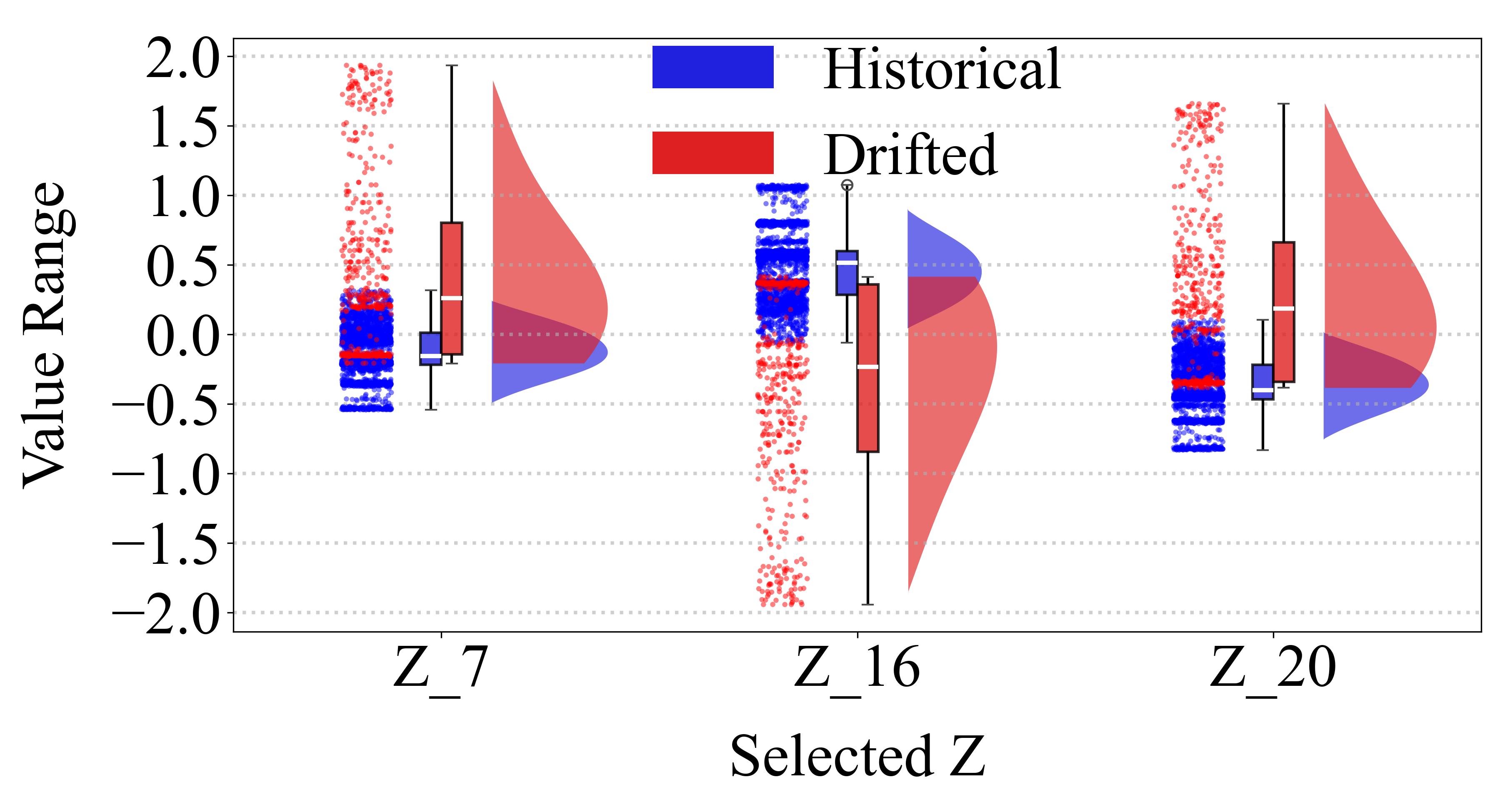}
\label{fig_alignment_analysis_cicids17_b}}

\subfloat[D.X-Static-H (Global Drift).]{\includegraphics[width=0.24\textwidth]{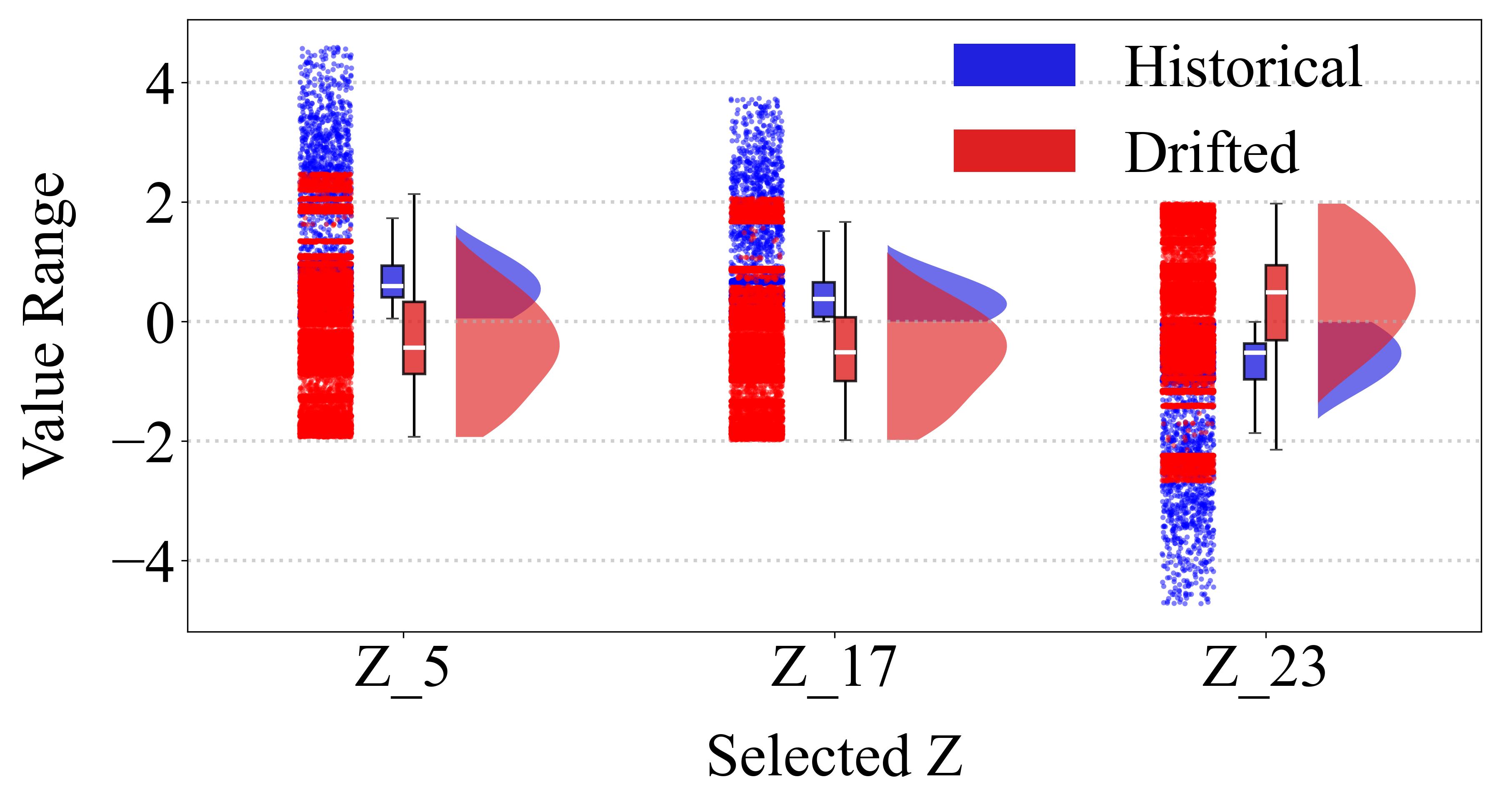}
\label{fig_alignment_analysis_cicids17_c}}
\subfloat[DriftXpert (Global Drift).]{\includegraphics[width=0.24\textwidth]{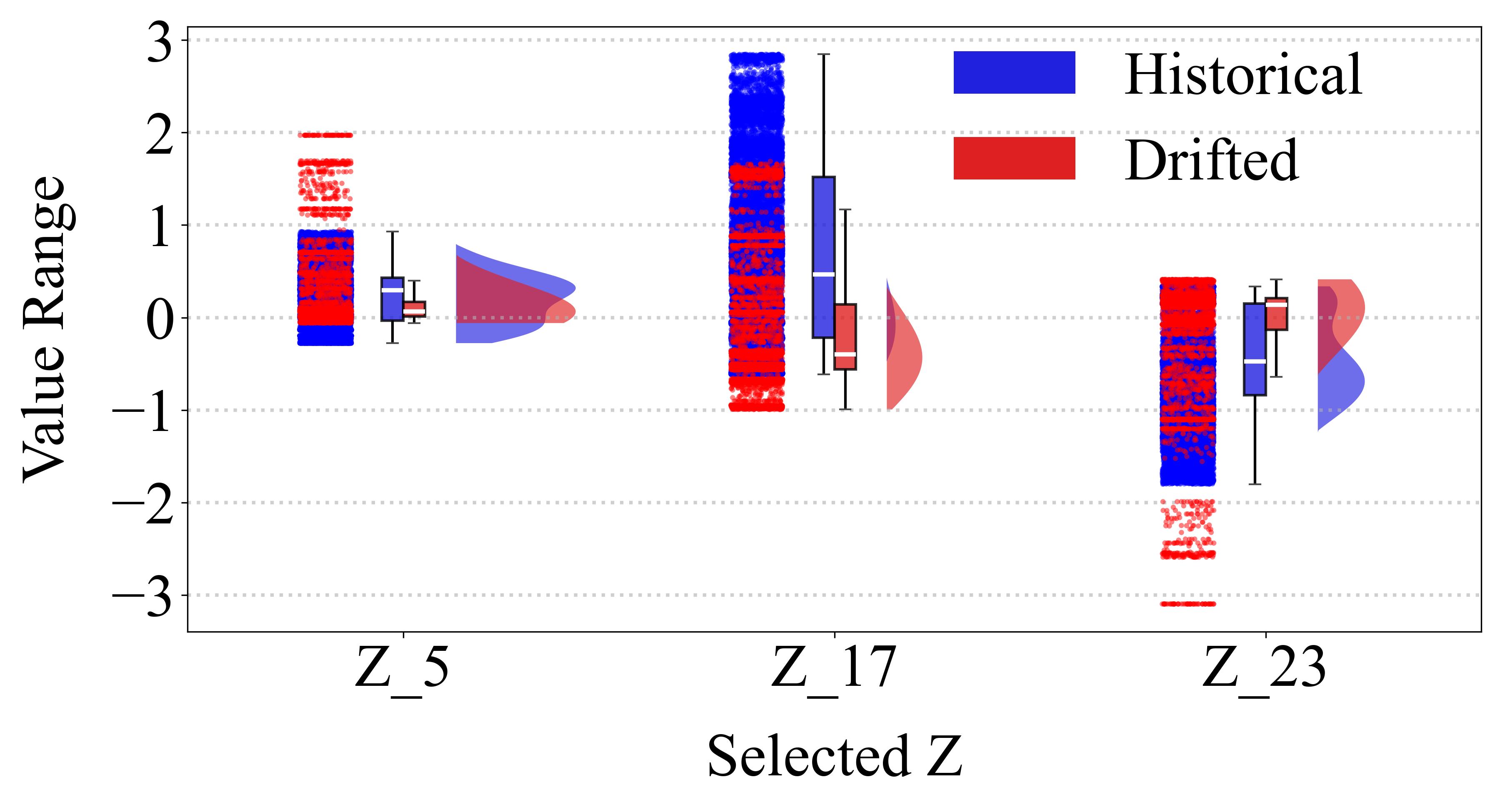}
\label{fig_alignment_analysis_cicids17_d}}
\caption{Alignment Analysis on the CICIDS-2017 Dataset.}
\label{fig_alignment_analysis_cicids17}
\end{figure}

\begin{figure}[htbp]
\centering
\subfloat[D.X-Static-H (Local Drift).]{\includegraphics[width=0.24\textwidth]{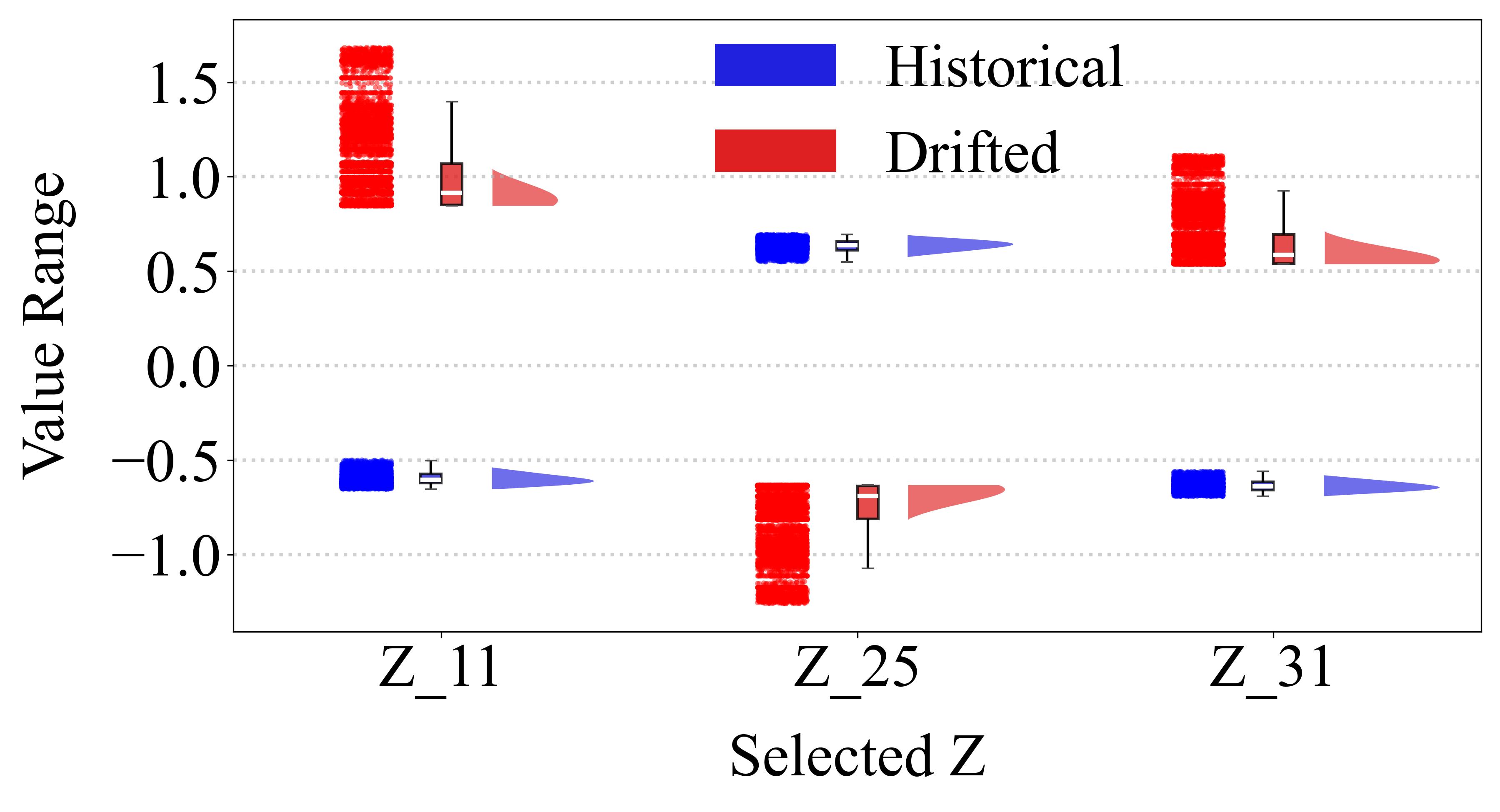}
\label{fig_alignment_analysis_cicids18_a}}
\subfloat[DriftXpert (Local Drift).]{\includegraphics[width=0.24\textwidth]{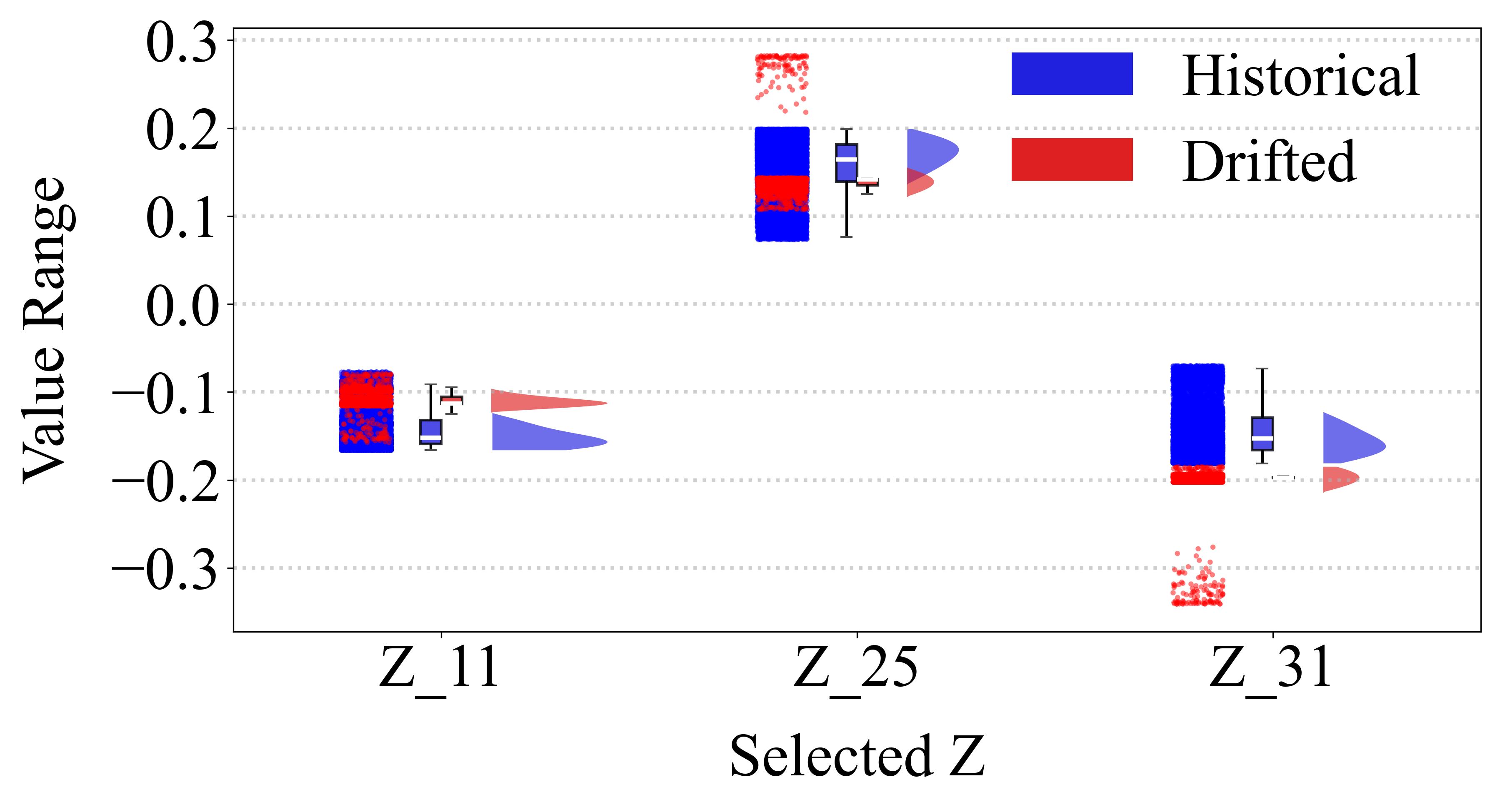}
\label{fig_alignment_analysis_cicids18_b}}

\subfloat[D.X-Static-H (Global Drift).]{\includegraphics[width=0.24\textwidth]{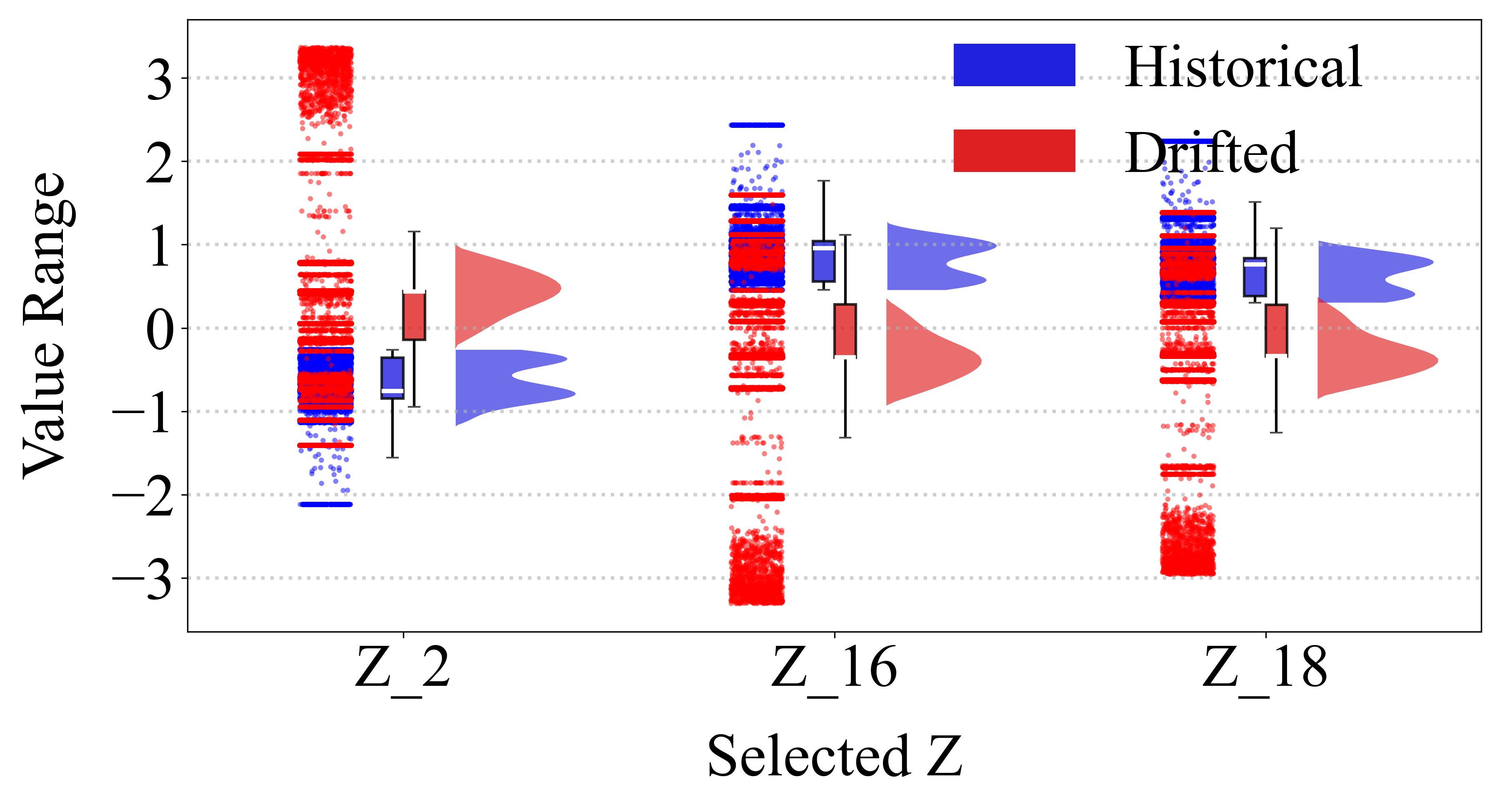}
\label{fig_alignment_analysis_cicids18_c}}
\subfloat[DriftXpert (Global Drift).]{\includegraphics[width=0.24\textwidth]{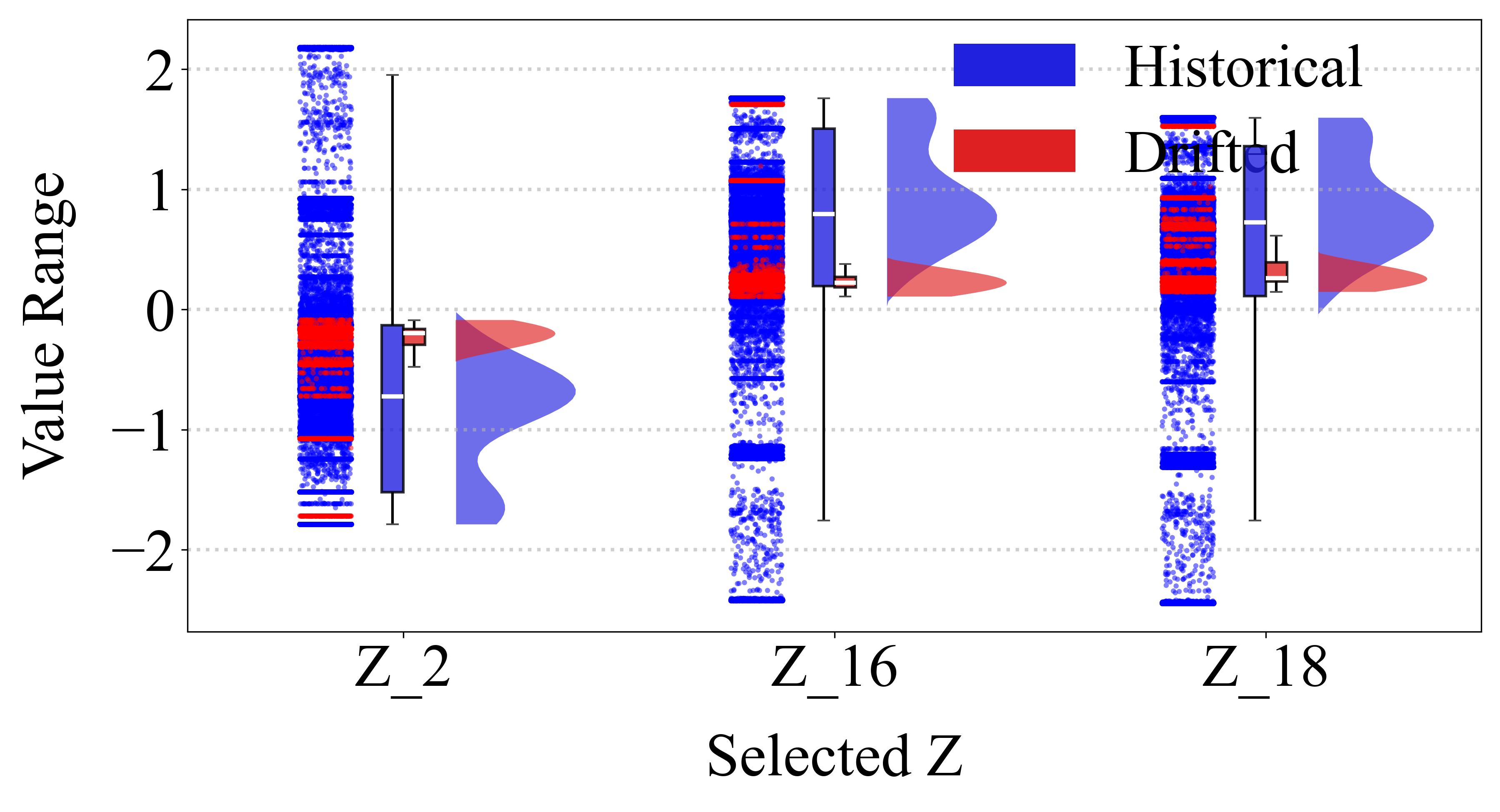}
\label{fig_alignment_analysis_cicids18_d}}
\caption{Alignment Analysis on the CICIDS-2018 Dataset.}
\label{fig_alignment_analysis_cicids18}
\end{figure}

To quantify the contribution of the alignment mechanism during phase 2, we performed a quantitative analysis and visualized the latent space distributions of both drifted and historical traffic. By comparing these distributions across the historical and updated model states, we aim to elucidate the pivotal role of this mechanism in mitigating distribution shifts. The results are illustrated in Figure \ref{fig_alignment_analysis_cicids17} and Figure \ref{fig_alignment_analysis_cicids18}. In the latent space of the historical model, the distributions of historical data and drifted data exhibit a certain degree of separation, which is particularly pronounced in the local drift scenario. In contrast, although the separation under global drift is less evident than that of local drift, observable distribution differences can still be identified from the box plots. After model adaption, the historical and drifted data demonstrate a more consistent alignment. From the perspective of density distributions, the proportion of aligned samples between the two data types can be observed more intuitively. It should be noted that the two distributions are unlikely to fully overlap in visualization, however, achieving effective alignment within certain subspaces is sufficient to support the adaptive capacity of the model.

\begin{table}[htbp]
    \centering
    \caption{Ablation Studies (Benign\%/Malicious\%) in the CICIDS-2018 Dataset on the Local Drift Scenario.}
    \begin{tabularx}{0.49\textwidth}{
        >{\centering\arraybackslash}p{0.09\textwidth}|
        >{\centering\arraybackslash}X >{\centering\arraybackslash}X | 
        >{\centering\arraybackslash}X >{\centering\arraybackslash}X
    }
        \toprule
        \multirow{2}{*}{\textbf{\textit{Methods}}} & \multicolumn{2}{c|}{\textbf{\textit{Set 1 (historical data)}}} & \multicolumn{2}{c}{\textbf{\textit{Set 2 (recent data)}}} \\
        \cmidrule(lr){2-3} \cmidrule(lr){4-5}
        & \textit{Recall.} & \textit{F1-Score.} & \textit{Recall.} & \textit{F1-Score.} \\
        \midrule
        D.X-With-H & 99.52/99.97 & 99.75/99.74 & 99.15/0.01 & 72.58/$~$0.01 \\
        D.X-With-D & 99.94/$~$0.01 & 67.66/$~$0.01 & 99.97/99.99 & 99.99/99.98 \\
        D.X-Only-C & 98.56/0.01 & 67.04/0.01 & 98.37/99.97 & 99.16/98.88 \\
        D.X-EnFre & 91.83/$~$0.01 & 63.92/$~$0.01 & 90.51/98.10 & 94.26/92.71 \\
        D.X-NoFre & 98.56/$~$0.01 & 67.04/$~$0.01 & 98.48/99.99 & 99.23/98.99 \\
        D.X-NoAgg & 98.08/85.56 & 92.59/91.23 & 97.60/99.97 & 98.77/98.42 \\
        \rowcolor{gray!20}
        DriftXpert & 98.08/85.29 & 92.47/91.07 & 98.08/99.99 & 99.03/98.75 \\
        \bottomrule
    \end{tabularx}
    \label{tab_ablation_in_cicids18_local_drift}
\end{table}

\begin{table}[htbp]
    \centering
    \caption{Ablation Studies (Benign\%/Malicious\%) in the CICIDS-2018 Dataset on the Global Drift Scenario.}
    \begin{tabularx}{0.49\textwidth}{
        >{\centering\arraybackslash}p{0.09\textwidth}|
        >{\centering\arraybackslash}X >{\centering\arraybackslash}X | 
        >{\centering\arraybackslash}X >{\centering\arraybackslash}X
    }
        \toprule
        \multirow{2}{*}{\textbf{\textit{Methods}}} & \multicolumn{2}{c|}{\textbf{\textit{Set 1 (historical data)}}} & \multicolumn{2}{c}{\textbf{\textit{Set 2 (recent data)}}} \\
        \cmidrule(lr){2-3} \cmidrule(lr){4-5}
        & \textit{Recall.} & \textit{F1-Score.} & \textit{Recall.} & \textit{F1-Score.} \\
        \midrule
        D.X-With-H & 99.82/99.80 & 99.78/99.83  & 94.43/25.77 & 71.84/39.10 \\
        D.X-With-D & 61.97/72.00 & 62.60/71.44 & 99.88/97.33 & 98.71/98.58 \\
        D.X-Only-C & 92.58/37.63 & 67.87/52.49 & 99.94/97.42 & 98.82/98.66 \\
        D.X-EnFre & $~~$4.71/99.99 & $~~$8.99/72.97 & 64.28/73.11 & 68.21/68.71 \\
        D.X-NoFre & 97.04/25.70 & 66.32/40.15 & 99.76/97.65 & 98.81/98.68 \\
        D.X-NoAgg & 99.71/69.01 & 83.24/81.55 & 99.26/96.69 & 98.12/97.92 \\
        \rowcolor{gray!20}
        DriftXpert & 99.03/80.73 & 88.49/88.97 & 99.19/96.83 & 98.14/97.96 \\
        \bottomrule
    \end{tabularx}
    \label{tab_ablation_in_cicids18_global_drift}
\end{table}

\subsection{RQ5: Sensitivity Analysis of Hyperparameter}
\label{sec_public_rq5}
In this section, we conducted a comprehensive sensitivity analysis on three key hyperparameters, $\alpha$, $\beta$, and $\gamma$, to evaluate the model's performance stability and robustness. Regarding the hyperparameter $K$, we omit a separate sensitivity evaluation as its optimal value can be empirically estimated from the clustering profiles of historical data, providing a principled basis for its selection.

\subsubsection{\textbf{Sensitivity Analysis of $\alpha$}} 
\begin{figure}[htbp]
\centering
\subfloat[Local Drift (CICIDS-2017).]{\includegraphics[width=0.24\textwidth]{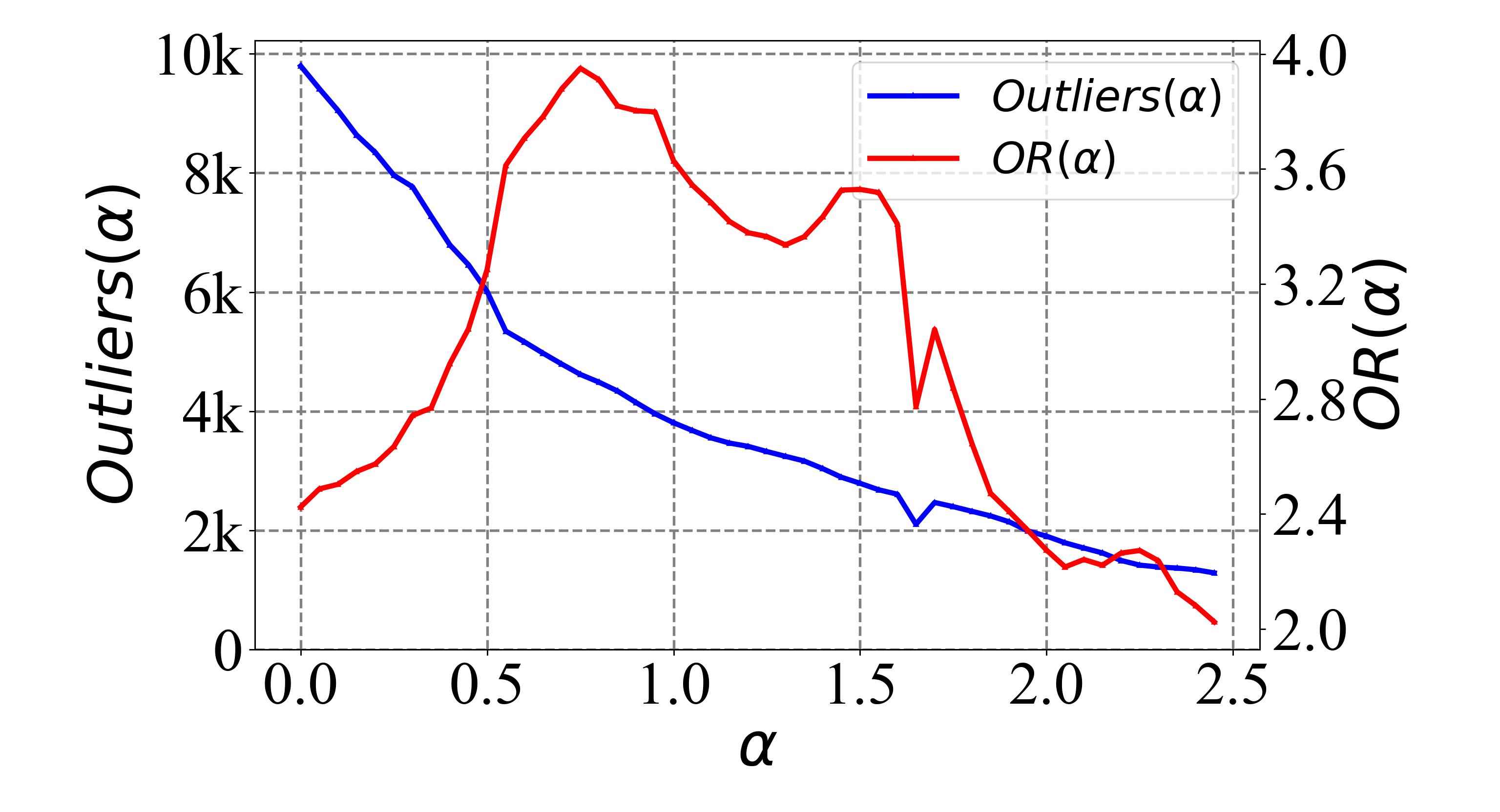}
\label{fig_sensitivity_analysis_of_alpha_a}}
\subfloat[Global Drift (CICIDS-2017).]{\includegraphics[width=0.24\textwidth]{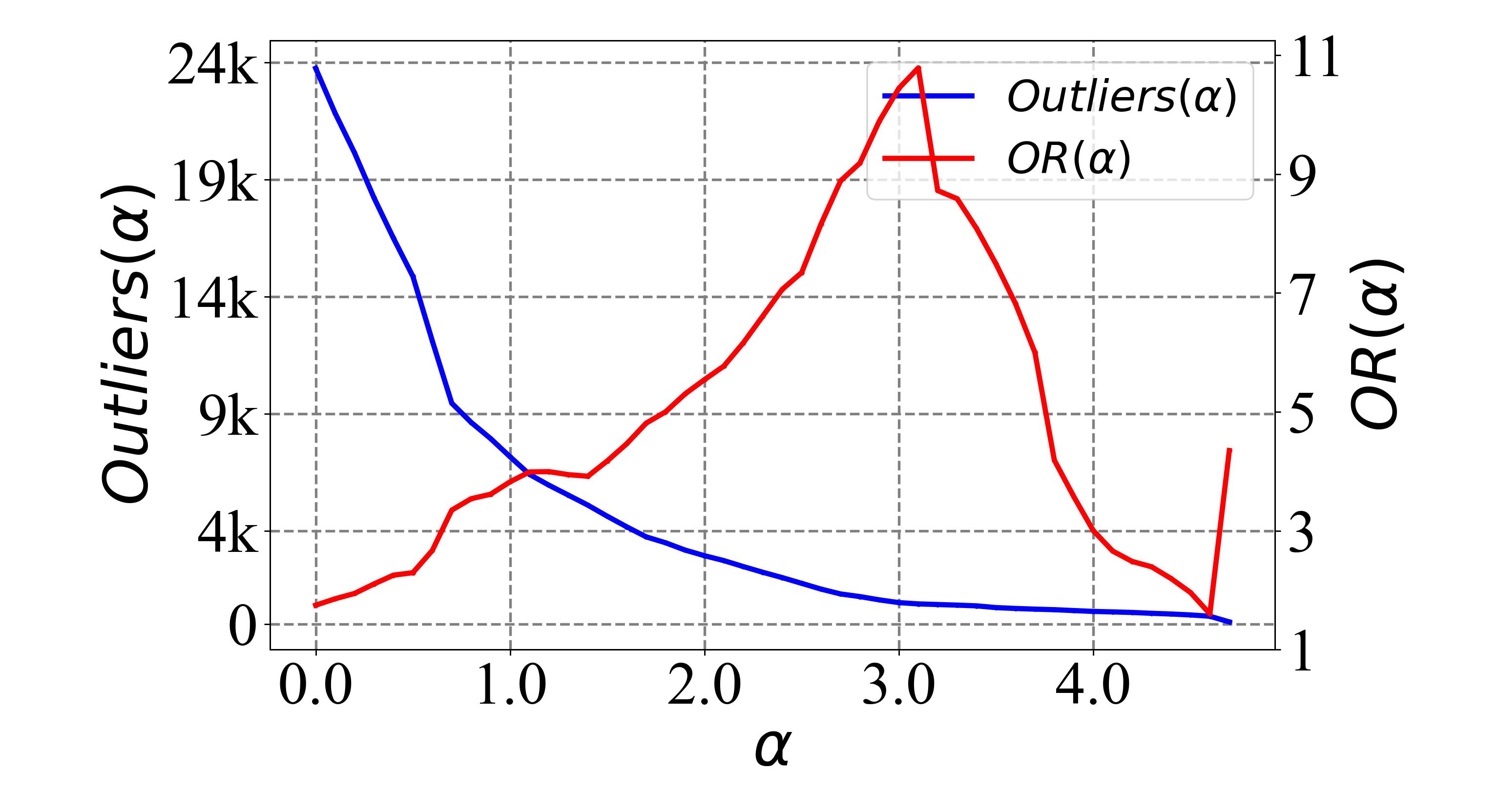}
\label{fig_sensitivity_analysis_of_alpha_b}}

\subfloat[Local Drift (CICIDS-2018).]{\includegraphics[width=0.24\textwidth]{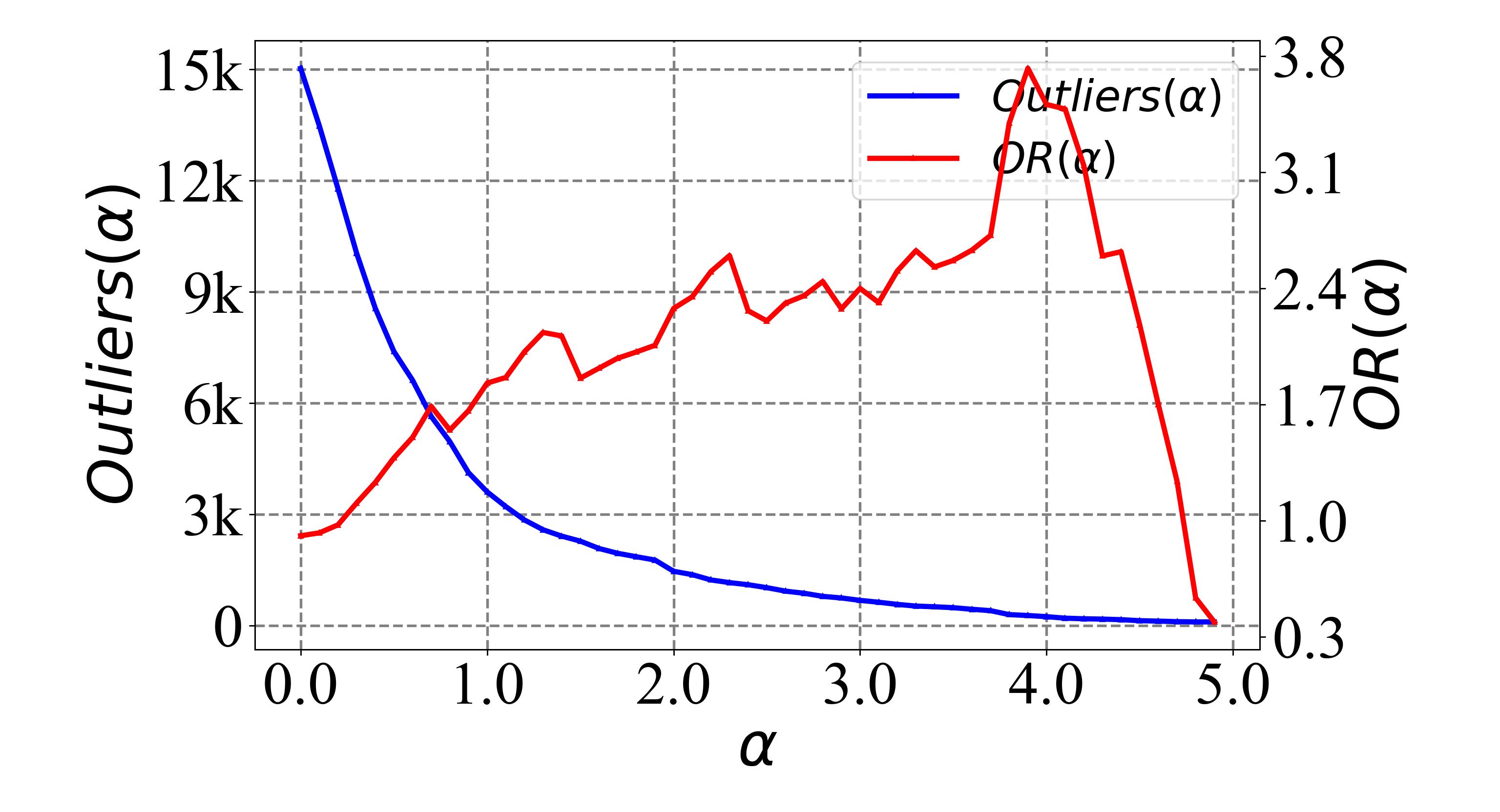}
\label{fig_sensitivity_analysis_of_alpha_c}}
\subfloat[Global Drift (CICIDS-2018).]{\includegraphics[width=0.24\textwidth]{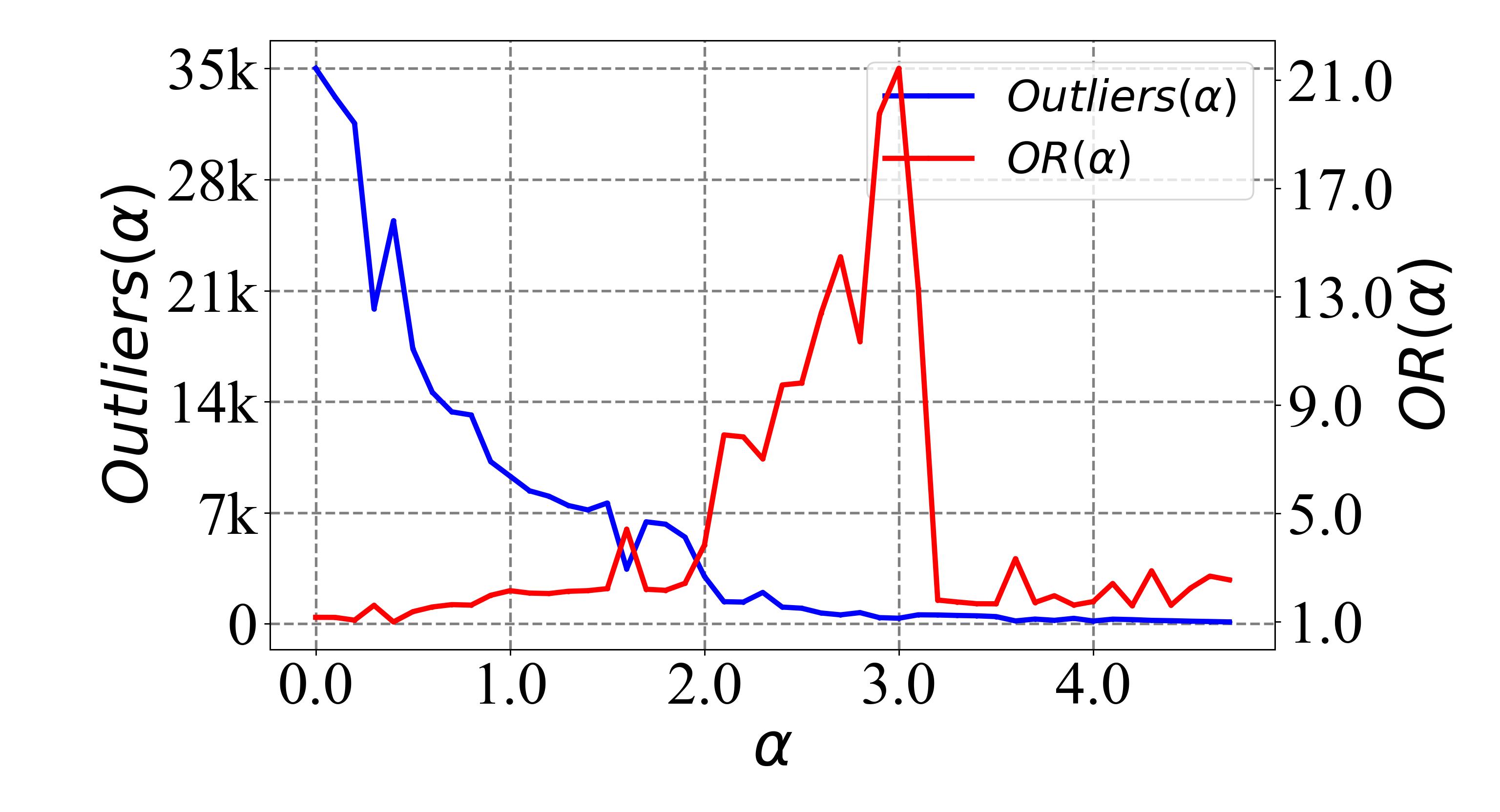}
\label{fig_sensitivity_analysis_of_alpha_d}}
\caption{Sensitivity Analysis of Hyperparameter $\alpha$.}
\label{fig_sensitivity_analysis_of_alpha}
\end{figure}

We conducted sensitivity analysis on $\alpha$ using CICIDS-2017 \cite{CICIDS2017, sharafaldin2018toward} and CICIDS-2018 \cite{CICIDS2018, sharafaldin2018toward}. The clustering model uses 5 clusters, and a minimum number of outliers is required to ensure a reliable OR, since drift detection is qualitative and OR alone is insufficient. As $\alpha$ increases from 0, historical outliers decrease while the OR first increases and then decreases, as shown in Figure \ref{fig_sensitivity_analysis_of_alpha}. The maximum occurs when the derivative of outliers with respect to $\alpha$ is zero, consistent with Appendix \ref{appendix_hyperparameter_alpha_analysis}, indicating the inflection point as an appropriate choice of $\alpha$. 

\subsubsection{\textbf{Sensitivity Analysis of $\beta$}} 
\begin{figure}[htbp]
\centering
\subfloat[Local Drift (Historical).]{\includegraphics[width=0.24\textwidth]{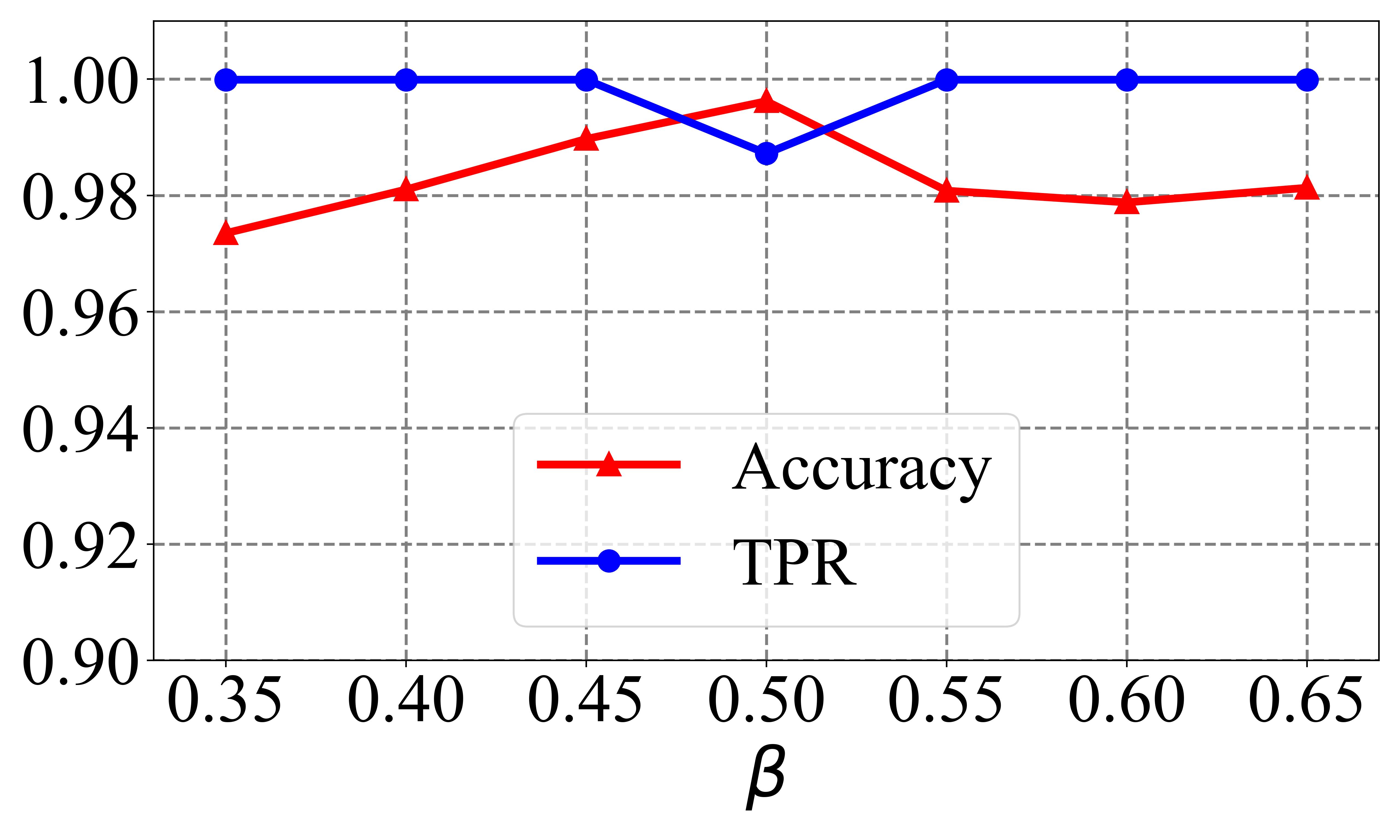}
\label{fig_sensitivity_analysis_of_beta_a}}
\subfloat[Local Drift (Drifted).]{\includegraphics[width=0.24\textwidth]{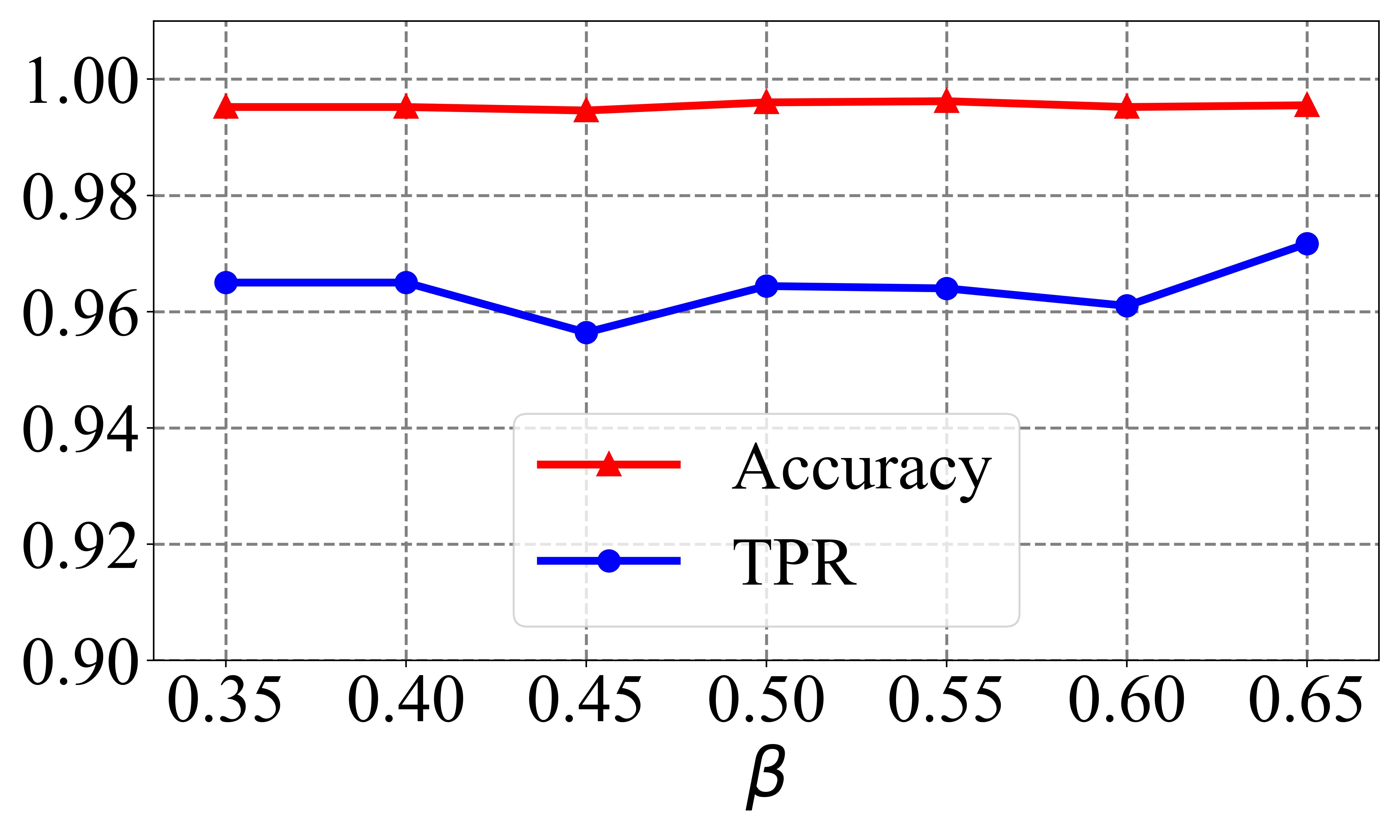}
\label{fig_sensitivity_analysis_of_beta_b}}

\subfloat[Global Drift (Historical).]{\includegraphics[width=0.24\textwidth]{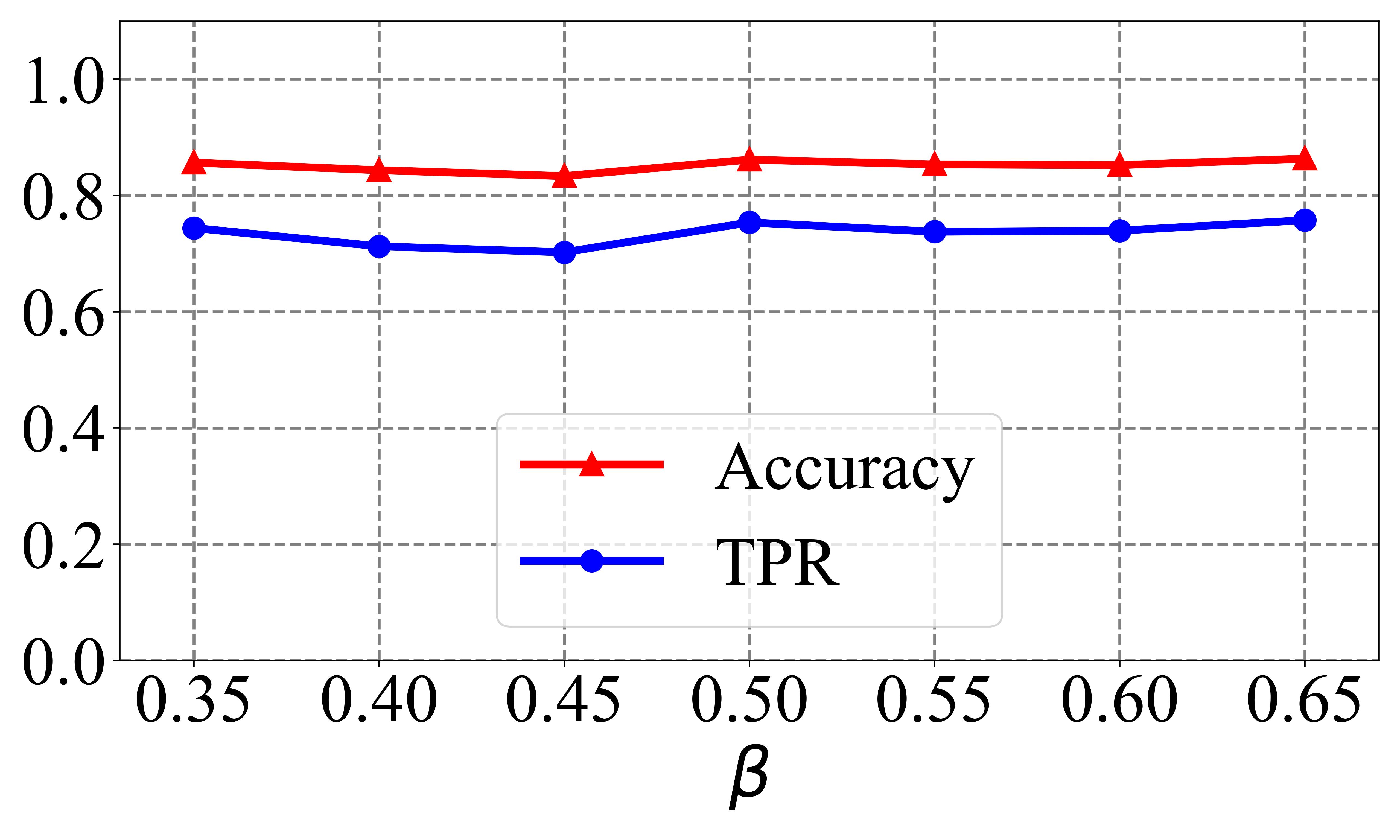}
\label{fig_sensitivity_analysis_of_beta_c}}
\subfloat[Global Drift (Drifted).]{\includegraphics[width=0.24\textwidth]{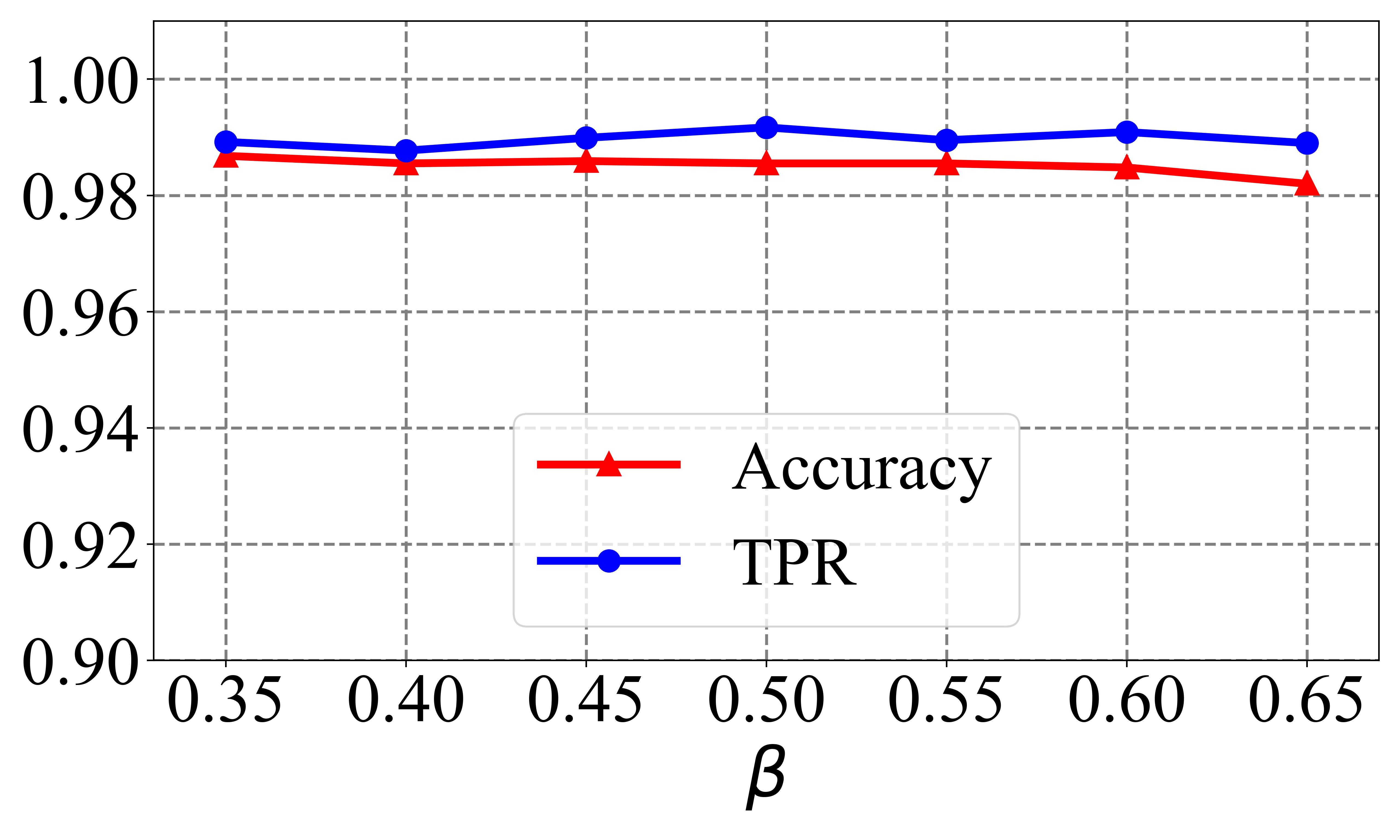}
\label{fig_sensitivity_analysis_of_beta_d}}
\caption{Sensitivity Analysis of Hyperparameter $\beta$ on the CICIDS-2017 Dataset.}
\label{fig_sensitivity_analysis_of_beta}
\end{figure}

The hyperparameter $\beta$ is used to control the aggregation weight between the historical model and the updated model, where $\beta = 0.5$ indicates equal contributions from both. Based on this, we conducted a sensitivity analysis by selecting seven values within the range of $0.35$ to $0.65$, as shown in Figure \ref{fig_sensitivity_analysis_of_beta}. The experimental results demonstrate that, under different values, the model performance remains relatively stable in both local and global drift scenarios. This indicates that setting $\beta$ around $0.5$ can effectively ensure the robustness of the model.

\subsubsection{\textbf{Sensitivity Analysis of $\gamma$}} 
\begin{figure}[htbp]
\centering
\subfloat[Local Drift (Historical).]{\includegraphics[width=0.24\textwidth]{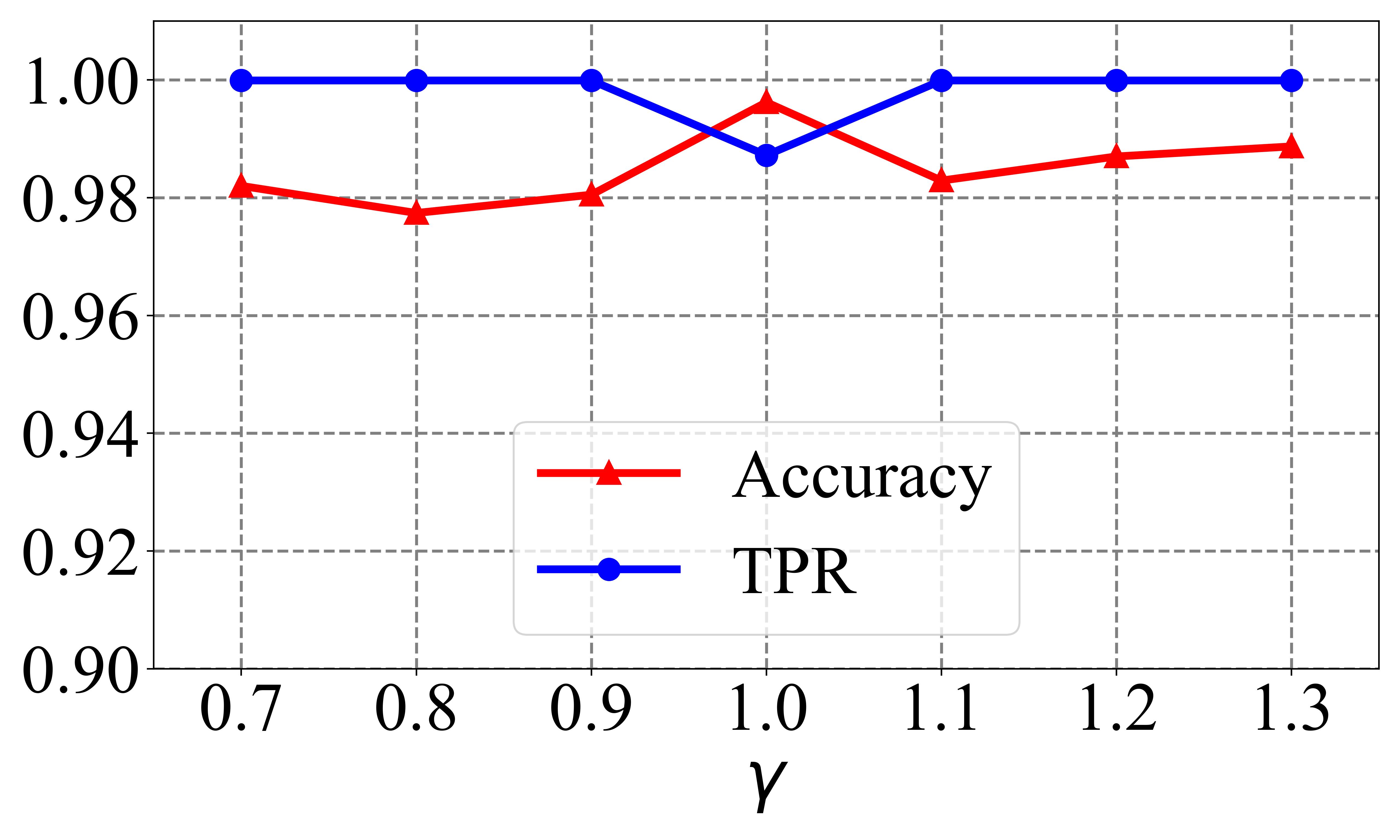}
\label{fig_sensitivity_analysis_of_gamma_a}}
\subfloat[Local Drift (Drifted).]{\includegraphics[width=0.24\textwidth]{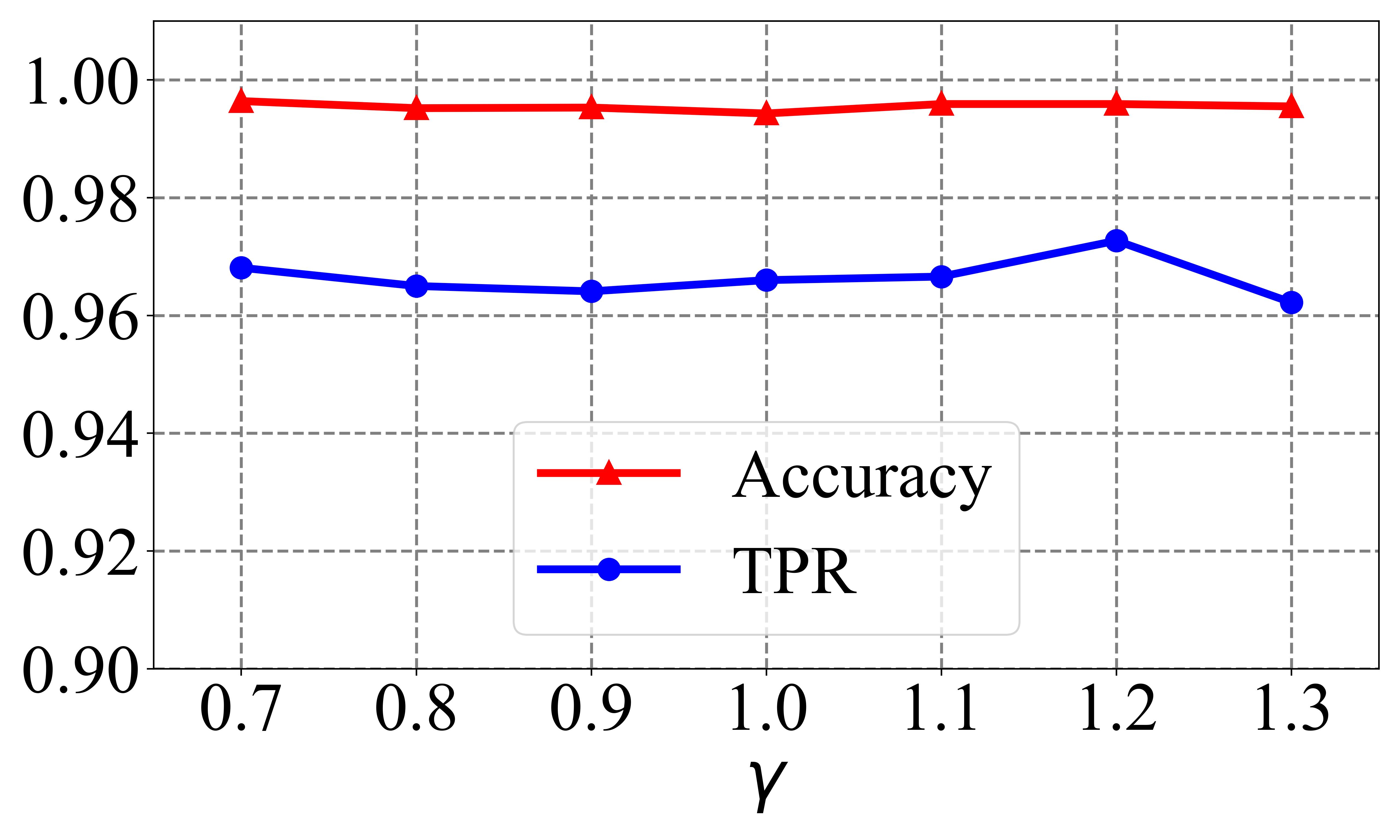}
\label{fig_sensitivity_analysis_of_gamma_b}}

\subfloat[Global Drift (Historical).]{\includegraphics[width=0.24\textwidth]{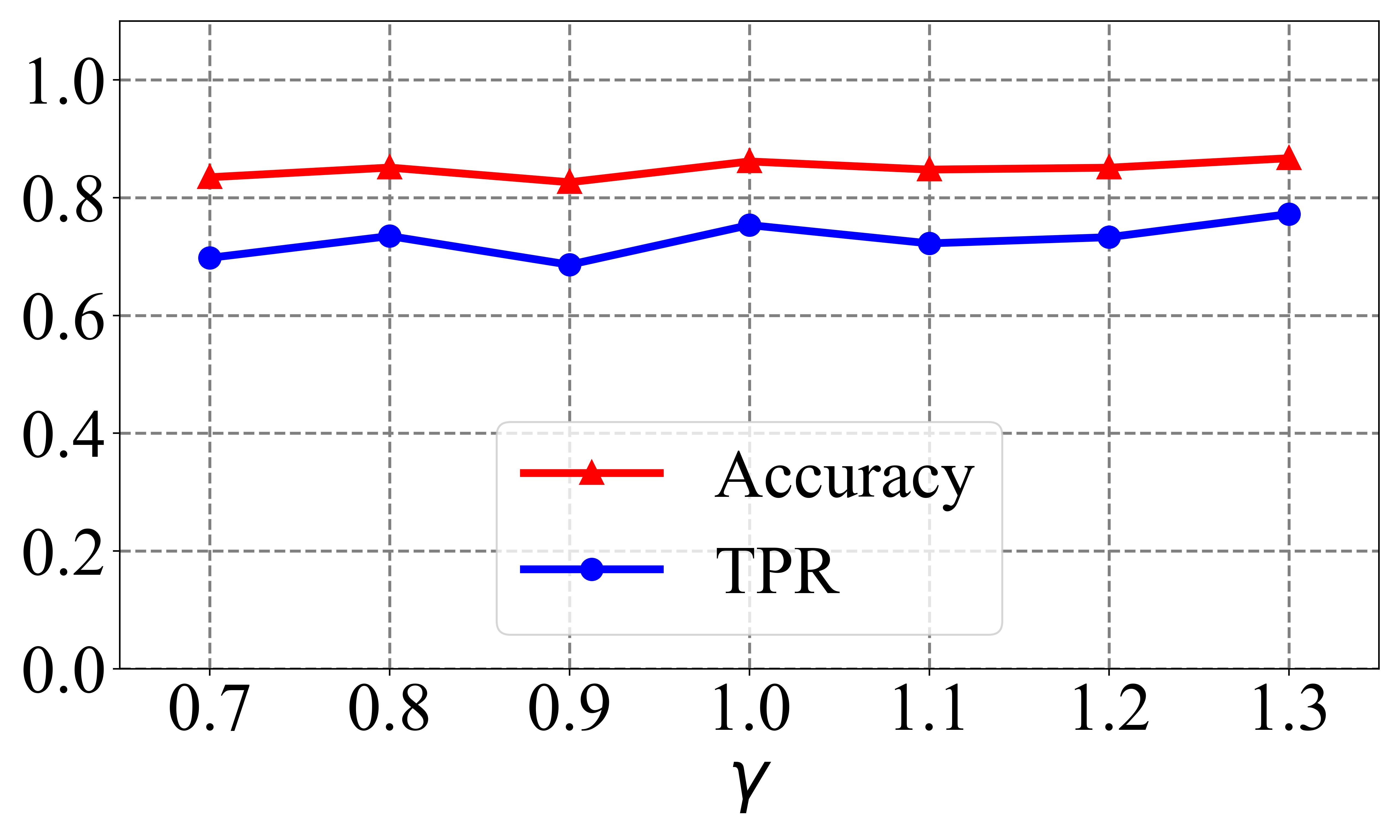}
\label{fig_sensitivity_analysis_of_gamma_c}}
\subfloat[Global Drift (Drifted).]{\includegraphics[width=0.24\textwidth]{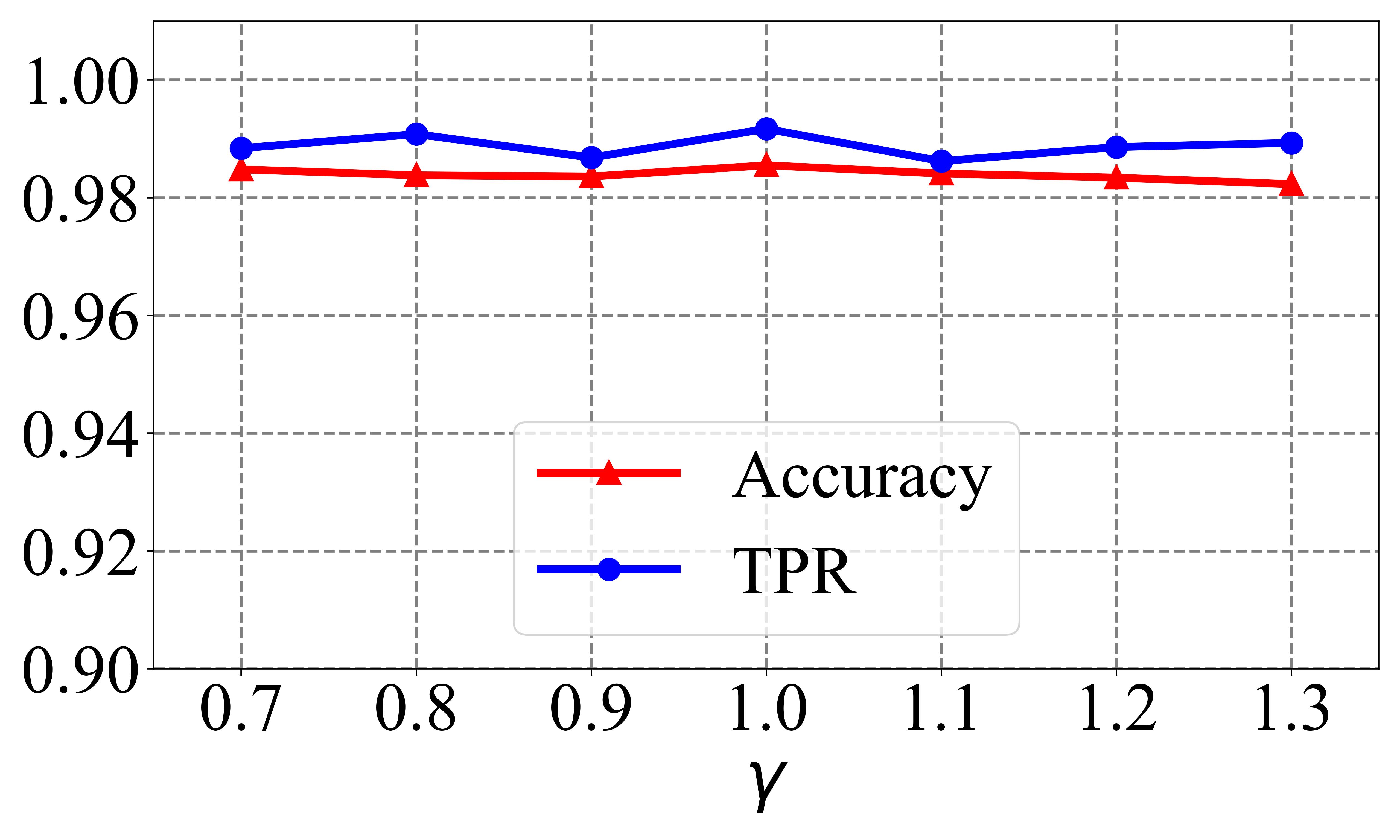}
\label{fig_sensitivity_analysis_of_gamma_d}}
\caption{Sensitivity Analysis of Hyperparameter $\gamma$ on the CICIDS-2017 Dataset.}
\label{fig_sensitivity_analysis_of_gamma}
\end{figure}

The hyperparameter $\gamma$ controls the weight of the alignment regularization term. We performed a sensitivity analysis by selecting seven values within the range of $0.7$ to $1.3$, as shown in Figure \ref{fig_sensitivity_analysis_of_gamma}. The experimental results indicate that, under different values, the model performance remains relatively stable in two drift scenarios. This suggests that setting $\gamma$ around $1.0$ effectively ensure the robustness of the model.

\section{Real-world Test on Enterprise Network}
\label{sec_real_world_test}
For real-world validation, we deployed DriftXpert within a major corporation's infrastructure, utilizing empirical datasets from their external-facing services for training and performance monitoring.

\subsection{Research Questions}
\label{sec_real_world_rq}
Our RQs in this section are as follows.
\begin{icompact}
    \item RQ1: Does DriftXpert maintain its efficacy in identifying drift when deployed within real-world enterprise infrastructures?
    \item RQ2: Can DriftXpert demonstrate robust adaptability to drifted telemetry while concurrently mitigating catastrophic forgetting?
    \item RQ3: Can DriftXpert achieve robust performance under encrypted traffic conditions, while still outperforming baseline methods?
\end{icompact}

\subsection{RQ1: Proactive Sensing in the Real-world Environment}
\label{sec_real_world_rq1}
\begin{figure}[htbp]
\centering
\includegraphics[width=0.35\textwidth]{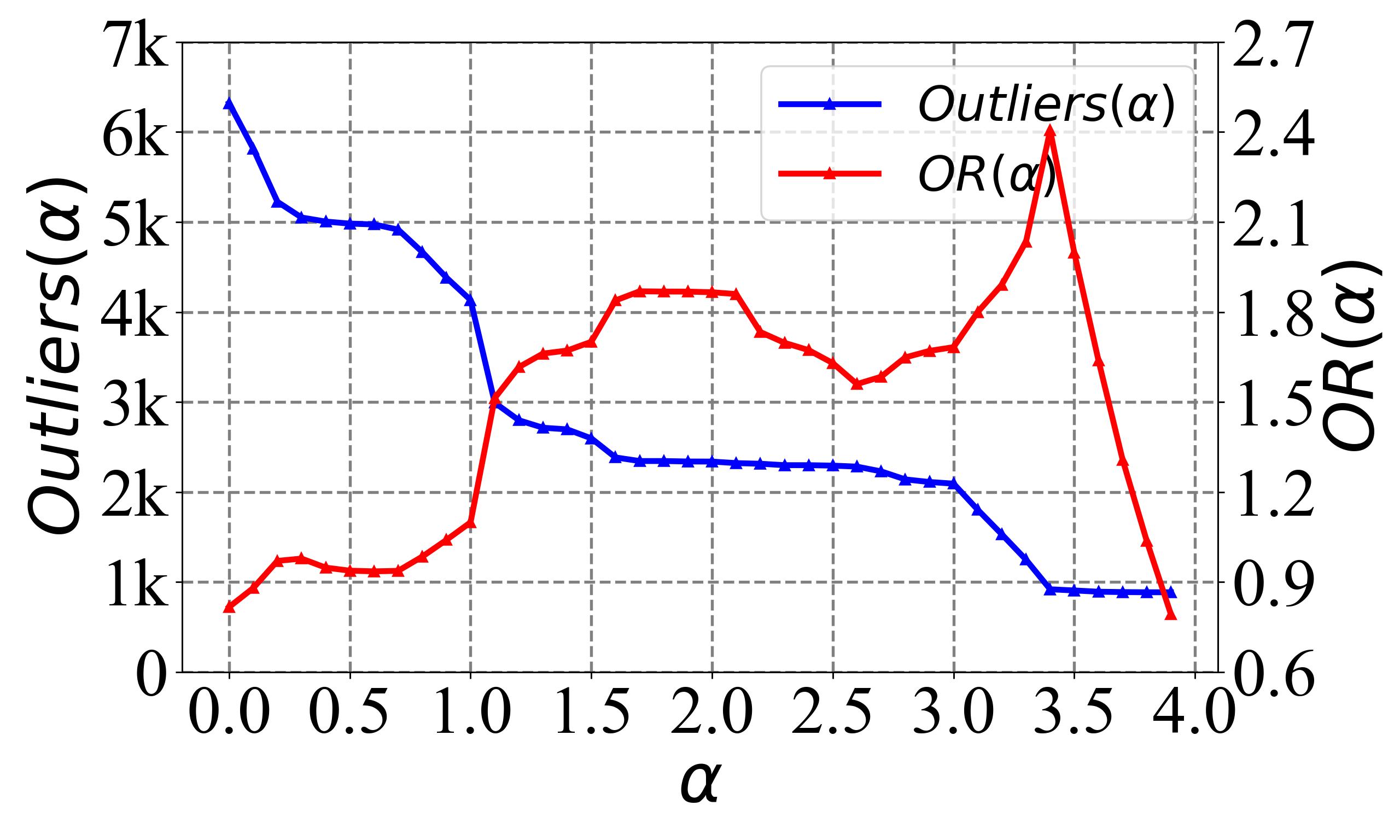}
\caption{Proactive Sensing in the Real-world Environment.}
\label{fig_drift_detection_on_real_world}
\end{figure}

Initially, we conducted drift detection experiments within an enterprise network environment, with the results summarized in Figure \ref{fig_drift_detection_on_real_world}. The experimental curves indicate that as $\alpha$ increases, the outlier count in historical telemetry progressively declines. The optimal configuration is identified at the inflection point ($\alpha \approx 3.5$), where the outlier ratio reaches its peak. This selection process is entirely deterministic, relying solely on historical data for parameter calibration. With $\alpha$ fixed at 3.5, we further analyzed the outlier distributions across individual clusters. Specifically, in the cluster exhibiting the most significant drift, the outlier count for drifted telemetry reached 1,926 out of 10,326 samples, compared to only 39 out of 4,265 for historical data. This yields a localized OR of 20.4, providing strong empirical evidence of localized distribution shifts and highlighting the imperative for adaptive model updates.

\subsection{RQ2: Evolutionary Alignment in the Real-world Environment vs. Baselines}
\label{sec_real_world_rq2}
In our real-world enterprise network evaluation, we benchmarked DriftXpert against three SOTA baselines, including OWAD \cite{han2023anomaly}, A-NIDS \cite{zha2025nids}, and SSF \cite{11044615}, as illustrated in Table \ref{tab_real_world}. The results demonstrate that DriftXpert achieves a superior F1-Score of 99.43\% and 97.10\% on the drifted data set, significantly outperforming all baselines and manifesting an exceptional capacity for adaptation of new knowledge. Crucially, while existing models such as OWAD suffer from acute performance degradation due to catastrophic forgetting (with an F1-Score that plummets to 15.22\%), DriftXpert maintains a robust recall of 96.12\% on historical data. This underscores our model’s ability to concurrently capture emerging threat patterns and preserve foundational knowledge, providing a decisive advantage in continuous learning for dynamic environments.
\begin{table*}[htbp]
    \centering
    \caption{Real-world Test on the Enterprise Network (Benign\% / Malicious\%).}
    \begin{tabularx}{1.0\textwidth}{>{\centering\arraybackslash}p{0.16\textwidth}|
    >{\centering\arraybackslash}X >{\centering\arraybackslash}X >{\centering\arraybackslash}X | 
    >{\centering\arraybackslash}X >{\centering\arraybackslash}X >{\centering\arraybackslash}X}
        \toprule
        \multirow{2}{*}{\textbf{\textit{Methods}}} & \multicolumn{3}{c|}{\textbf{\textit{Set 1 (historical data)}}} & \multicolumn{3}{c}{\textbf{\textit{Set 2 (recent data)}}} \\
        \cmidrule(lr){2-4} \cmidrule(lr){5-7}
         & \textit{Precision.} & \textit{Recall.} & \textit{F1-Score.} & \textit{Precision.} & \textit{Recall.} & \textit{F1-Score.} \\
        \midrule
        D.X-With-H & 98.33\%/ \textbf{90.49\%} & \textbf{96.75\%}/ 94.96\% & \textbf{97.53\%}/ \textbf{92.67\%} & 97.80\%/ 91.98\% & 98.43\%/ 89.06\% & 98.11\%/ 90.49\% \\
        D.X-With-D & 93.11\%/ 48.37\% & 70.88\%/ 83.88\% & 80.49\%/ 61.36\% & 99.29\%/ 95.94\% & 99.21\%/ 96.35\% & 99.25\%/ 96.15\% \\
        OWAD @NDSS23 & 75.59\%/ 26.74\% & 90.48\%/ 10.63\% & 82.37\%/ 15.22\% & 83.12\%/ 13.89\% & 93.06\%/ 5.59\% & 87.81\%/ 7.97\% \\
        A-NIDS @TIFS25 & 97.41\%/ 48.34\% & 67.43\%/ 94.44\% & 79.69\%/ 63.95\% & 99.37\%/ 94.80\% & 98.94\%/ 96.88\% & 99.15\%/ 95.83\% \\
        SSF @INFOCOM25 & 90.72\%/ 75.65\% & 92.96\%/ 69.69\% & 91.82\%/ 72.55\% & 92.80\%/ 88.18\% & 98.38\%/ 61.26\% & 95.51\%/ 72.29\% \\
        DriftXpert (\textbf{Ours}) & \textbf{98.45\%}/ 61.16\% & 80.14\%/ \textbf{96.12\%} & 88.36\%/ 74.75\% & \textbf{99.64\%}/ \textbf{96.08\%} & \textbf{99.22\%}/ \textbf{98.14\%} & \textbf{99.43\%}/ \textbf{97.10\%} \\
        \bottomrule
    \end{tabularx}
    \label{tab_real_world}
\end{table*}

\subsection{RQ3: Evaluation on Encrypted Environment vs. Baselines}
\label{subsec_real_world_encrypted}
\begin{table}[htbp]
    \centering
    \caption{Evaluation on Encrypted Environment.}
    \begin{tabularx}{0.48\textwidth}{
        >{\centering\arraybackslash}p{0.08\textwidth} |
        >{\centering\arraybackslash}p{0.14\textwidth} |
        >{\centering\arraybackslash}X >{\centering\arraybackslash}X 
        >{\centering\arraybackslash}X >{\centering\arraybackslash}X
    }
    \toprule
    \textbf{\textit{Datasets}} & \textbf{\textit{Methods}} & \textbf{\textit{ACC}} & \textbf{\textit{F1}} & \textbf{\textit{TPR}} \\

    \midrule
    \multirow{4}{*}{\textit{Historical}} & OWAD @NDSS23 & 0.29 & 0.39 & 0.24 \\
    & A-NIDS @TIFS25 & 0.87 & 0.93 & 0.93 \\
    & SSF @INFOCOM25 & 0.64 & 0.77 & 0.64 \\
    & DriftXpert (\textbf{Ours}) & \textbf{0.88} & \textbf{0.93} & \textbf{0.94} \\
    \midrule
    \multirow{4}{*}{\textit{Drifted}} & OWAD @NDSS23 & 0.32 & 0.21 & 0.13 \\
    & A-NIDS @TIFS25 & 0.71 & 0.82 & \textbf{0.92} \\
    & SSF @INFOCOM25 & 0.54 & 0.60 & 0.49 \\
    & DriftXpert (\textbf{Ours}) & \textbf{0.85} & \textbf{0.89} & 0.90 \\
    
    \bottomrule
    \end{tabularx}
    \label{tab_encrypted_evaluation}
\end{table}
Given that encrypted payloads eliminate plaintext visibility, assessing detection integrity under encryption is critical for real-world deployment. As shown in Table \ref{tab_encrypted_evaluation}, our comparative analysis reveals that DriftXpert consistently outperforms OWAD, A-NIDS, and SSF in encrypted scenarios. While DriftXpert secures a 0.88 ACC and 0.94 TPR on baseline distributions and preserves an 0.85 ACC post-drift, competing models suffer from severe performance degradation. Specifically, the accuracy of A-NIDS falls to 0.71 after drift, while OWAD and SSF do not provide reliable detection. These findings underscore the efficacy of our latent alignment strategy, which enables DriftXpert to navigate complex encrypted traffic evolution with a superior balance between plasticity and stability compared to main baselines.

\section{Discussion}
\label{sec_discussion}

\subsection{Foundation Models and Drift Adaptation}
\label{subsec_discussion_foundation_models}
As Large Language Models (LLMs) demonstrate remarkable potential in zero-shot task \cite{brown2020language, ouyang2022training, dai2023can, devlin2019bert}, a fundamental question naturally arises: \textit{\textbf{Can the inherent robustness of general-purpose foundation models entirely obviate the need for explicit concept drift adaptation?}} In this section, we provide a critical discussion.

\noindent \textbf{\textit{Protocol v.s. Linguistic Semantics.}} Network traffic consists of unstructured bitstreams governed by rigorous state-machine logic and spatio-temporal dependencies, fundamentally diverging from natural language semantics. Although LLMs capture macroscopic patterns, they lack the perceptual precision to sense fine-grained protocol shifts, such as subtle variations in proprietary industrial protocols. By quantifying latent manifold deviation, DriftXpert models the underlying geometry of traffic representations, capturing essential distribution shifts more acutely than the probabilistic heuristics of LLMs.

\noindent \textbf{\textit{The Plasticity-Stability Dilemma.}} Even foundation models require fine-tuning to maintain precision when encountering new attacks or environmental shifts. However, full-parameter tuning is not only computationally prohibitive but also suffers from catastrophic forgetting, where adapting to new patterns inevitably compromises the discriminative power over legacy attack signatures. DriftXpert addresses this via a decoupled two-stage framework, employing representation consistency alignment and selective neuron freezing. This approach provides a surgical and controllable path for model evolution that is far more precise than vanilla fine-tuning, thereby ensuring the long-term integrity of the defensive system.

\noindent \textbf{\textit{Operational Bottlenecks and Deployment Overhead.}} NIDS are typically deployed on high-throughput backbone links, requiring microsecond-level inference latency. The prohibitive computational overhead and hardware redundancy necessitated by LLMs are impractical for most production environments, such as the power grid infrastructure evaluated in this study. In contrast, DriftXpert maintains a lightweight architecture, significantly lowering deployment thresholds and energy consumption while meeting the stringent real-time requirements of industrial networks. Crucially, our approach is not antithetical to the rise of LLMs. We envision a profound synergistic potential between foundation models and task-specific frameworks like DriftXpert. In future autonomous security ecosystems, LLMs can function as high-level commanders that leverage reasoning capabilities for cross-domain threat modeling and strategic orchestration. Conversely, DriftXpert acts as a front-line scout residing at the high-speed network edge. Its primary role is to perform sub-millisecond real-time sensing and robust distribution alignment. This hierarchical integration ensures a defensive posture that is both strategically comprehensive and operationally agile.

\subsection{Drift Detection and Annotation Efficiency}
\label{subsec_discussion_annotation_efficiency}
We first evaluate the drift detection performance under both local and global drift scenarios using Macro-Precision (Macro-Prc), Macro-Recall (Macro-Rec), and Accuracy, as summarized in Table~\ref{tab_evaluation_on_drift_detection}. The results demonstrate that DriftXpert achieves effective drift identification in both scenarios, with Macro-Recall values of 84.29\% and 81.77\% for local and global drift, respectively, indicating that most drifted samples can be successfully detected. The relatively lower Macro-Precision (50.07\%) under local drift is mainly attributed to the inherent class imbalance, where only a small fraction of samples undergo distribution changes while the majority remain consistent with the historical distribution.
\begin{table}[htbp]
    \centering
    \caption{Evaluation of Drift Detection.}
    \begin{tabularx}{0.48\textwidth}{
        >{\centering\arraybackslash}p{0.12\textwidth} |
        >{\centering\arraybackslash}X >{\centering\arraybackslash}X 
        >{\centering\arraybackslash}X 
    }
    \toprule
    \textbf{\textit{Scenarios}} & \textbf{\textit{Macro-Prc}} & \textbf{\textit{Macro-Rec}} & \textbf{\textit{Accuracy}} \\

    \midrule
    Local Drift & 0.5007 & 0.8429 & 0.8107 \\
    Global Drift & 0.8181 & 0.8177 & 0.7869 \\
    
    \bottomrule
    \end{tabularx}
    \label{tab_evaluation_on_drift_detection}
\end{table}

With accurate drift detection, DriftXpert can substantially reduce the annotation overhead by requiring manual labeling only for detected drift samples, rather than continuously annotating all incoming traffic. For example, in our enterprise network deployment, we collected two months of traffic data and observed that only 1,357 out of 42,875 sampled flows in the second month exhibited distribution shifts compared with the previous month. This observation indicates that, in practical scenarios, drifted samples typically constitute only a small fraction. In real-world networks, traffic evolution is generally gradual, and drifted samples rarely dominate the data stream within a short period unless major events such as large-scale business changes or system migrations occur. Such events are usually foreseeable, allowing operators to proactively adjust data collection and annotation strategies. Therefore, with reliable drift detection, DriftXpert effectively minimizes unnecessary annotation efforts during continuous model maintenance.

\section{Related Work}
\label{sec_related_work}
In this section, we summarize previous related research efforts, with detailed information provided as below.

\noindent \textit{\textbf{Detection Methods of Concept Drift.}} Statistics-based methods are widely used for concept drift detection \cite{gemaque2020overview}. Examples include using density changes in active learning as a drift trigger \cite{costa2018drift}, Treap-based Kolmogorov-Smirnov tests for distribution comparison \cite{dos2016fast}, density estimation for local or global drift \cite{liu2018accumulating}, pseudo-error rates from dynamic classifiers \cite{pinage2020drift}, distance measures like Kullback-Leibler (KL) divergence \cite{jain2022k}, and trainable detectors based on Restricted Boltzmann Machines \cite{korycki2021concept}. However, in NIDS scenarios, since historical data is not retained, direct comparison of historical data with drifted data is not applicable \cite{dos2016fast, liu2018accumulating, jain2022k}. Furthermore, model updates must be considered, as continuous stacking of classifiers can introduce additional problems \cite{pinage2020drift}. Methods based on distance, among others, require specific hyperparameters, and selecting appropriate hyperparameters is a key challenge \cite{korycki2021concept}. In contrast, our method can be implemented without access to historical data and does not rely on labeled data. Moreover, the selection of hyperparameters is supported by solid theoretical foundations and an automated analysis strategy.

\noindent \textit{\textbf{ML-Based Intrusion Detection Methods.}} Machine learning algorithms have been extensively applied to intrusion detection. For example, Yang et al. \cite{yang2021conditional} proposed a conditional probability model for detecting known traffic, complemented by a reconstruction mechanism for zero-day attacks. Fu et al. \cite{fu2021realtime} employed discrete Fourier transform to extract frequency-domain features, combined with clustering models for detection. Yang et al. \cite{yang2021mth} introduced an ensemble tree model for known attack detection alongside a clustering-based approach for zero-day scenarios. Wang et al. \cite{wang2020cloud} developed a cloud-based method integrating a Stacked Contractive Auto-Encoder with Support Vector Machine (SVM). Fu et al. \cite{fu2023point} proposed an unsupervised point cloud analysis method that models traffic feature vectors as high-dimensional points, aggregates them into voxels, and leverages voxel density to reduce false positives. Mirsky et al. \cite{mirsky2018kitsune} combined stacked autoencoders with reconstruction techniques to identify malicious traffic, while King et al. \cite{king2023edgetorrent} utilized temporal graph representations and message-passing neural networks for anomaly detection. Ding et al. \cite{ding2023tmg} further introduced generative adversarial networks to mitigate class imbalance. Despite achieving strong performance on benchmark datasets, these approaches largely overlook concept drift in real-world environments. Consequently, they implicitly assume a static traffic distribution, under which model effectiveness is only temporary. Once drift occurs, performance deteriorates significantly, often resulting in a substantial increase in false positives.

\noindent \textit{\textbf{Adaptation Methods for Concept Drift.}} The objective of model adaptation is to update models upon drift detection, enabling them to accommodate new data while mitigating catastrophic forgetting. Existing approaches can be broadly categorized into ensemble learning \cite{dong2020survey} and incremental learning \cite{masana2022class}. For instance, Wang et al. \cite{wang2022enidrift} proposed constructing new subclassifiers and adjusting their weights to adapt to dynamic data; however, this method lacks explicit drift detection and depends on continuous labeling. Liu et al. \cite{liu2020diverse} introduced an instance-based ensemble approach that dynamically selects classifiers under different drift scenarios. Andresini et al. \cite{andresini2021insomnia} leveraged clustering for drift detection and employed a neighbor-based strategy to generate pseudo-labels for model updates, though these pseudo-labels are susceptible to noise and may induce self-poisoning. Yang et al. \cite{yang2024recda} generated drift-aware perturbation vectors via masking, aligned them with historical representations, and fine-tuned a constrained classifier using selected historical samples; however, this approach relies on access to historical data and lacks a robust drift detection mechanism. Han et al. \cite{han2023anomaly} assigned importance weights to model parameters, restricting updates to critical parameters to preserve prior knowledge while adapting less significant ones to new distributions. Zhang et al. \cite{11044615} proposed SSF (Strategic Selection and Forgetting), an incremental NIDS method that captures drift by selecting representative new samples and discarding outdated ones. Zha et al. \cite{zha2025nids} utilized CTGAN to replay historical data and combine it with recent data for model updates, though such generation inevitably incurs information loss, potentially degrading performance. Lian et al. \cite{lian2025facing} proposed leveraging uncertainty to identify drift-induced anomalies. However, because their method cannot learn new benign patterns, it may generate a large number of false positives during natural drift, such as that caused by business upgrades.

Unlike most prior work that separately focuses on drift detection, incremental learning, or knowledge preservation, we propose a unified and problem-driven adaptation framework rather than a simple combination of existing techniques. Specifically, we introduce two clustering-based drift indicators, CI and OR, with theoretical analysis to validate their effectiveness and guide hyperparameter selection, enabling unsupervised drift detection. The detected drift serves as an adaptation criterion to trigger updates only under significant distribution shifts. Motivated by the analysis in $\S$ \ref{sec_design_motivation}, we further introduce contrastive learning to align representations between historical and evolving traffic distributions. To balance adaptation and catastrophic forgetting, we employ weight aggregation and first-layer freezing to preserve stable knowledge. Unlike existing methods that rely on historical data replay or storage, our framework requires no access to historical traffic and performs label-free drift detection, reducing storage and annotation costs in long-term deployment.

\section{Conclusion}
\label{sec_conclusion}
We propose a robust and lightweight NIDS, DriftXpert. We begin with a backward-attribution analysis of concept drift, common in IDS research, to identify variation patterns in the latent space. Based on this, we introduce a deep clustering method for drift detection and a model contrastive approach to align latent spaces of historical and drifted data, mitigating catastrophic forgetting during updates. We provide a theoretical analysis of drift detection hyperparameters and propose a reliable selection method using image-based simulations. Finally, we validate the performance of drift and intrusion detection in two public datasets, analyze latent space distributions to confirm our attribution, and conduct latency and throughput tests on a low-end PC, showing the suitability of DriftXpert for resource-constrained deployment.



\begin{acks}  
This work was supported by the National Key R\&D Program of China under Grant No. 2023YFB2904000, No. 2023YFB2904001 and No. 2024YFE0203800. 
\end{acks}

\bibliographystyle{ACM-Reference-Format}
\bibliography{reference}

\appendix 

\section{Open Science} 

All artifacts of this work are available at our GitHub repository (https://github.com/cc-sec-hub/DriftXpert), including the complete implementation of our adaptation framework, hyperparameter configurations, and scripts to reproduce the main results. The public benchmarks used in this study are accessible via their respective official websites.

\section{Ethical Considerations} 

This study focuses on enhancing the robustness of NIDS against data drift. Our ethical statements are as follows:
\begin{icompact}
    \item Data Provenance and Privacy: We used public benchmarks and a  dataset provided by an industrial partner. All data was strictly authorized for research and underwent thorough anonymization. No personally identifiable information or sensitive network configurations are included.

    \item No Subject Involvement: This work involves no human or animal subjects. Thus, IRB approval was not required.
\end{icompact}

\section{Generative AI Usage}
Generative AI tools, such as ChatGPT and Gemini, are used only for polishing the language and improving the readability of the manuscript. These tools are not used for data analysis, code implementation, or generating research ideas. All contents are carefully reviewed and verified by the authors to ensure their technical accuracy and academic integrity.

\section{Theoretical Analysis of Predict Distribution}
\label{appendix_theoretical_analysis_of_predict_distribution}
\textbf{\textit{DEFINITION 3.}} \textit{Let $p(X)$ represent the predicted distribution of the model and $q(X)$ denote the true distribution of the labels.}

\textbf{\textit{THEOREM 1.}} \textit{The predict probability density function $p(X)$ can be expressed as:}
\begin{equation}
    p(X) = \frac{1}{\sqrt{2 \pi \sigma^2}} \exp{-\frac{1}{2 \sigma^2} }(X-\mu)^2,
\end{equation}

\textbf{\textit{PROOF.}} The classification results are optimized through convergence using cross-entropy loss, which can be expressed as follows:
\begin{equation}
    H(p,q) = H(p) + D_{KL}(p || q),
\end{equation}
where $H(p,q)$ represents the cross-entropy, $H(p)$ denotes the entropy of $p(X)$, $D_{KL}(p || q)$ refers to the Kullback-Leibler (KL) divergence.

$H(p)$ represents the intrinsic uncertainty of the predicted distribution $p(X)$, expressed as follows:
\begin{equation}
    H(p) = -\int_{-\infty}^{\infty} p(X) \ln{p(X)} \, dX.
\end{equation}

To derive the probability density function of the predicted distribution $p(X)$ from the perspective of maximizing entropy \cite{jaynes1957information, thomas2006elements}, the following constraints must first be satisfied:
\begin{equation}
    \int_{-\infty}^{\infty} p(X) \,dX = 1,
\end{equation}
\begin{equation}
    \int_{-\infty}^{\infty} X p(X) \, dX = \mu,
\end{equation}
\begin{equation}
    \int_{-\infty}^{\infty} (X-\mu)^2 p(X) \, dX = \sigma ^ 2,
\end{equation}
where $\mu$ represents the mean, $\sigma ^ 2$ denotes the variance.

For constrained optimization problems, we solve them using the Lagrangian function. Thus, our objective function can be formulated as:
\begin{align}
    \mathcal{L} &= -\int_{-\infty}^{\infty} p(X) \ln{p(X)} \, dX \notag \\
                &\quad + \lambda_0 \left( \int_{-\infty}^{\infty} p(X) \,dX - 1 \right) \notag \\
                &\quad + \lambda_1 \left( \int_{-\infty}^{\infty} X p(X) \, dX - \mu \right) \notag \\
                &\quad + \lambda_2 \left( \int_{-\infty}^{\infty} (X-\mu)^2 p(X) \, dX - \sigma^2 \right).
\end{align}

By solving the above objective function, we can derive $p(X) = C \exp{(\lambda_1 X + \lambda_2 (X - \mu) ^2)}$. Further analysis leads to the following parameter values: $C = \frac{1}{\sqrt{2 \pi \sigma^2}}$, $\lambda_1 = 0$, $\lambda_2 = -\frac{1}{2 \sigma^2}$. 

Consequently, the predict probability density function $p(X)$ can be expressed as:
\begin{equation}
    p(X) = \frac{1}{\sqrt{2 \pi \sigma^2}} \exp{-\frac{1}{2 \sigma^2} }(X-\mu)^2,
\end{equation}
which follows a Gaussian distribution.

\section{Theoretical Analysis of Hyperparameter $\alpha$ for Drift Detection}
\label{appendix_hyperparameter_alpha_analysis}
The hyperparameters $\alpha$ directly determine the value of $CI$, thus influencing the number of outliers from both historical data and drifted data in the drift detection model. This, in turn, affects the magnitude of $OR$. Following the theoretical analysis in \cite{zha2025nids}, we examine the relationship among $Outliers$, $OR$, and $\alpha$, with the goal of understanding their interdependencies and possible correlations.

\textbf{\textit{ASSUMPTION 1.}} \textit{Assume the distribution of i-th cluster follows the Gaussian distribution, that is,}
\begin{equation}
\label{eq_p_i_z}
    P_i(z) = \frac{1}{\sqrt{2 \pi \sigma_i ^ 2}} \exp{-\frac{1}{2 \sigma_i ^2} }(z-\mu_i)^2,
\end{equation}
\textit{where $\mu_i$ and $\sigma_i$ are the mean and standard deviation of $P_i(z)$, respectively.}

\textbf{\textit{THEOREM 2.}} \textit{When $|x| > V_i (V_i>\mu_i)$, samples
 are outliers, such that the number of outliers is:}
\begin{equation}
\label{eq_theorem_1}
    Outliers = \sum_{i=1}^{K} N_i \cdot \frac{\exp{(-(A\alpha+B)^2)}}{(A\alpha + B)\sqrt{\pi}}.
\end{equation}
\textit{where $K$ is the total number of clusters, $N_i$ is the sample number of i-th cluster, $A$ and $B$ are constants.}

\textit{\textbf{PROOF.}} According to $|x| > V_i (V_i>\mu_i)$, the number of outliers can be estimated by the probability distribution:
\begin{equation}
\label{eq_outliers}
    Outliers = \sum_{i=1}^{K} 2N_i \cdot \int_{V_i}^{+\infty} P_i(z) dz.
\end{equation}

A standard transformation is applied as $t=\frac{z-\mu_i}{\sigma_i}$; then
\begin{equation}
\label{eq_standard_transformation_1}
\begin{split}
    \int_{V_i}^{+\infty} P_i(z) dz &= \int_{\frac{V_i-\mu_i}{\sigma_i}}^{+\infty} \frac{1}{\sqrt{2\pi}} \exp{(-\frac{t^2}{2})}dt \\
    &= 1 - \Phi(\frac{V_i-\mu_i}{\sigma_i}),
\end{split}
\end{equation}
\begin{equation}
\label{eq_standard_transformation_2}
    \Phi(\frac{V_i-\mu_i}{\sigma_i}) = \int_{-\infty}^{\frac{V_i-\mu_i}{\sigma_i}} \frac{1}{\sqrt{2\pi}} \exp{(-\frac{t^2}{2})}dt.
\end{equation}

Based on Plackett's approximation and the Gaussian error function, $\Phi(\frac{z-\mu_i}{\sigma_i})$ is approximated as follows:
\begin{equation}
\label{eq_plackett_approximation}
    \Phi(\frac{V_i-\mu_i}{\sigma_i}) = \frac{1}{2} \left[ 1 + \text{erf}(\frac{V_i-\mu_i}{\sqrt{2} \sigma_i})\right],
\end{equation}
\begin{equation}
\label{eq_erf}
    \text{erf}(\frac{V_i-\mu_i}{\sqrt{2} \sigma_i}) = \frac{2}{\sqrt{\pi}} \int_{0}^{\frac{V_i-\mu_i}{\sigma_i}} \exp{(-t^2)} dt.
\end{equation}

Applying the Taylor series expansion gives the following.
\begin{equation}
\label{eq_taylor_expansion}
    \text{erf}(x) = 1 - \frac{e^{-x^2}}{x \sqrt{\pi}} \left[ 1 - \frac{1}{2x^2} + \frac{3}{4x^4} +O(\frac{1}{x^6})\right].
\end{equation}

Let $V_i=C_i+CI$, $C_i$ be the center of the i-th cluster, according to Equation \eqref{eq_ci}, $\frac{V_i - \mu_i}{\sigma_i} = \frac{C_i + CI -\mu_i}{\sigma_i} = A\alpha + B$. Consequently, combining Equation \eqref{eq_p_i_z}, Equation \eqref{eq_outliers} - Equation \eqref{eq_taylor_expansion}, $Outliers$ can be represented as:
\begin{equation}
\label{eq_outliers_result}
    Outliers = \sum_{i=1}^{K} N_i \cdot \frac{\exp{(-(A\alpha+B)^2)}}{(A\alpha + B)\sqrt{\pi}}.
\end{equation}

Considering the relationship between $Outliers$ and $\alpha$, the connection between $OR$ and $\alpha$ becomes clearer. To simplify the analysis, two scenarios are considered to illustrate their relationship.
\begin{itemize}
    \item As $\alpha$ gradually increases from 0, the drifted data distribution completely consists of outliers, whereas the number of outliers in the historical data distribution gradually decreases. The outlier ratio $OR_1$ can be expressed as:
    \begin{equation}
    \label{eq_or_1}
        OR_1 = \frac{\frac{1}{N_{Z_r}} \cdot N_{Z_r}}{N_{Z_h} \cdot Outliers(\alpha)}.
    \end{equation}
    \item As $\alpha$ continues to increase, there are no outliers in the historical data (to ensure that the denominator is not zero, we set it to $1$). Meanwhile, the number of outliers in the drifted data distribution gradually decreases, following a trend similar to that of $Outliers(\alpha)$. The outlier ratio $OR_2$ can be expressed as:
    \begin{equation}
    \label{eq_or_2}
        OR_2 = \frac{\frac{1}{N_{Z_r}} \cdot Outliers(\alpha)}{\frac{1}{N_{Z_h}} \cdot 1}.
    \end{equation}
\end{itemize}

\begin{figure}[htbp]
    \centering
    \includegraphics[width=0.3\textwidth]{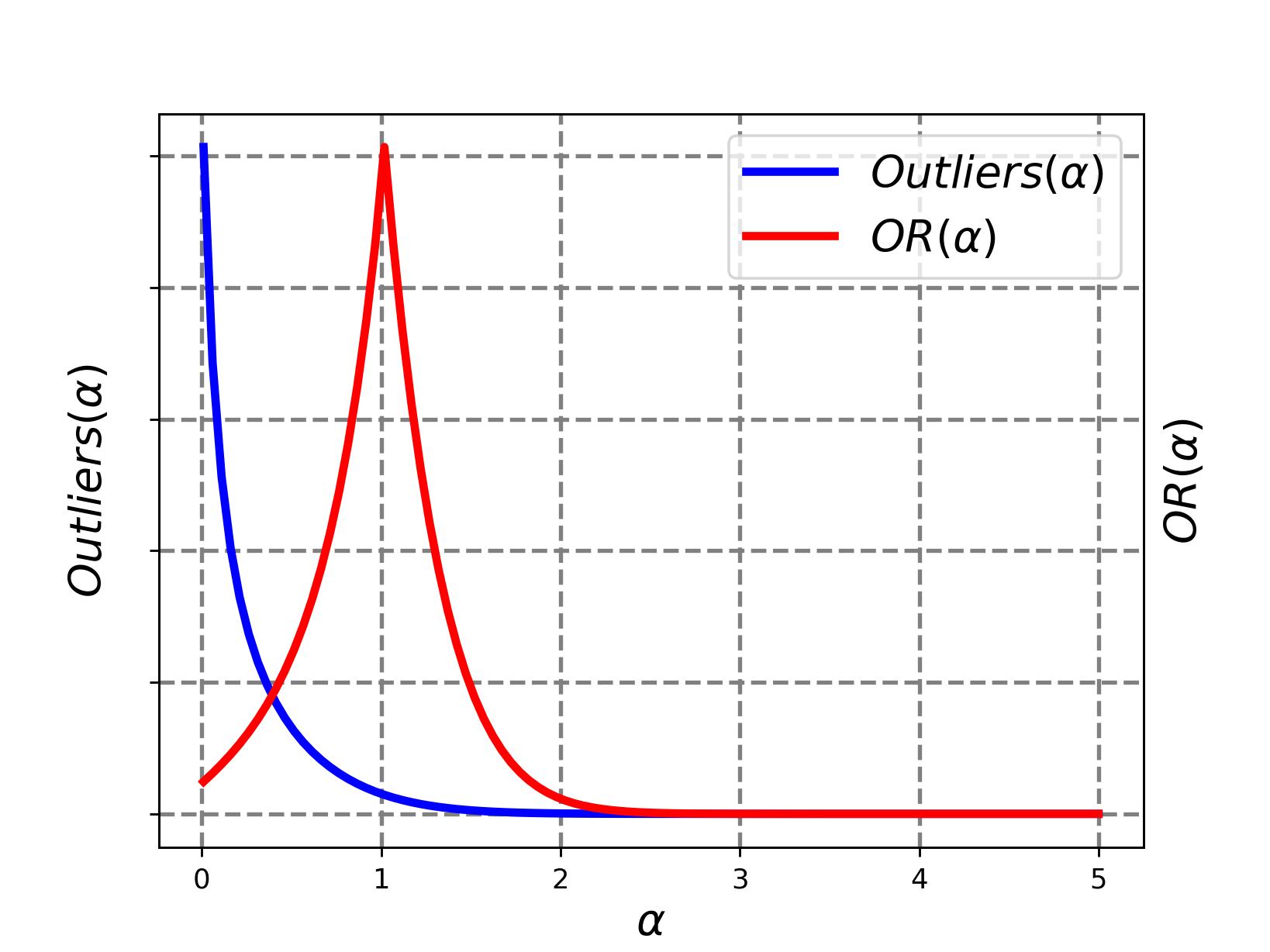}
    \caption{Graphical simulation of $Outliers(\alpha)$ and $OR(\alpha)$.}
    \label{fig_graphical_simulation} 
\end{figure}

To provide a clearer illustration of the relationship among $Outliers$, $OR$, and $\alpha$, a graphical simulation was conducted based on Equation \eqref{eq_outliers_result} – Equation \eqref{eq_or_2}, as shown in Figure \ref{fig_graphical_simulation}.

\section{Theoretical Analysis of Contrastive Loss on Convergence}
\label{appendix_contrastive_loss_on_convergence}
An important aspect of neural networks is understanding the non-Euclidean geometry of the parameter space of neural structures \cite{amari1998natural}. Our interest does not lie in the parameters themselves but in the mappings they represent. The non-Euclidean geometry of the parameter space, combined with the manifold of models exhibiting equivalent behavior under observation, allows movement from one region of the parameter space to another. This enables changes in the curvature of the model without altering the actual calculation of the function \cite{dinh2017sharp}.

\textbf{\textit{DEFINITION 4.}} \textit{Given a parameter $\theta^*$ as a local minimum point of the loss function $\mathcal{L}(\theta)$, such that $\nabla \mathcal{L}(\theta^*) = 0$ and $\nabla^2 \mathcal{L}(\theta^*)>0$.}

\textbf{\textit{THEOREM 3.}} \textit{Given a parameter transformation method $\eta^*= g^{-1}(\theta^*)$, the parameter $\eta^*$ is also a local minimum point of the loss function $\mathcal{L}(\theta)$.}

\textit{\textbf{PROOF.}} We define $\mathcal{L}_{\eta^*} = \mathcal{L} \circ g$ as the loss function with respect to parameters $\eta^*$. We extend the derivation as follows:
\begin{equation}
    \mathcal{L}_{\eta^*}(\eta^*) = \mathcal{L}(g(\eta^*))
\end{equation}
\begin{equation}
\begin{split}
    (\nabla \mathcal{L}_{\eta^*})(\eta^*) &= (\nabla \mathcal{L})(g(\eta^*))(\nabla g)(\eta^*) \\
                                          &= \nabla \mathcal{L}(\theta^*)(\nabla g)(\eta^*) 
\end{split}
\end{equation}
\begin{equation}
\begin{split}
    (\nabla^2 \mathcal{L}_{\eta^*})(\eta^*) &= (\nabla g)(\eta^*)^T(\nabla^2 \mathcal{L})(g(\eta^*))(\nabla g)(\eta^*) \\
                                            & + (\nabla \mathcal{L})(g(\eta^*))(\nabla^2 g)(\eta^*) \\
                                            &= (\nabla g)(\eta^*)^T \nabla^2 \mathcal{L}(\theta^*)(\nabla g)(\eta^*) \\
                                            & + (\nabla \mathcal{L})(\theta^*)(\nabla^2 g)(\eta^*) 
\end{split}
\end{equation}

By \textbf{\textit{DEFINITION 4}}, $\nabla \mathcal{L}(\theta^*) = 0$ and $\nabla^2 \mathcal{L}(\theta^*)>0$, we can thus derive:
\begin{equation}
    (\nabla \mathcal{L}_{\eta^*})(\eta^*) = 0
\end{equation}
\begin{equation}
    (\nabla^2 \mathcal{L}_{\eta^*})(\eta^*) = (\nabla g)(\eta^*)^T \nabla^2 \mathcal{L}(\theta^*)(\nabla g)(\eta^*) >0
\end{equation}

Thus, $\eta^*$ is also a local minimum point of the loss function $\mathcal{L}(\theta)$. It means that if the drifted data can converge on the classification model, then by employing contrastive learning to shift the model parameters within the non-Euclidean geometry to another parameter space, the new model will still converge on the drifted data.

\section{More Details of Experiments}
\subsection{Features Used in Our Experiments}
We use CICFlowMeter \cite{lashkari2017characterization} to extract approximately 80 relevant features and remove features such as source IP, destination IP, source port and destination port that could introduce "cheating effects." \cite{engelen2021troubleshooting} After experimental comparisons, we also eliminated features with numerous outliers (e.g. extremely large values or negative numbers) and features that had no impact on the experimental results. Finally, we selected 50 features used in our experiments, as shown in Table \ref{tab_used_features}.
\begin{table}[htbp]
    \centering
    \caption{Features Used in Our Experiments.}
    \begin{tabularx}{0.48\textwidth}{>{\centering\arraybackslash}p{0.14\textwidth}|>{\centering\arraybackslash}X}
    \toprule
    \textbf{Type} & \textbf{Features} \\

    \midrule
    Time-Realted & Flow Duration, Flow IAT Mean, Flow IAT Std, Flow IAT Max, Flow IAT Min, Fwd IAT Tot, Fwd IAT Mean, Fwd IAT Std, Fwd IAT Max, Fwd IAT Min, Bwd IAT Tot, Bwd IAT Mean, Bwd IAT Std, Bwd IAT Max, Bwd IAT Min \\
    \midrule
    Packet-Realted & Tot Fwd Pkts, Tot Bwd Pkts, Flow Byts/s, Flow Pkts/s, Fwd Pkts/s, Bwd Pkts/s, Subflow Fwd Pkts, Subflow Fwd Byts, Subflow Bwd Pkts, Subflow Bwd Byts, Init Fwd Win Byts, Init Bwd Win Byts, Fwd Act Data Pkts \\ 
    \midrule
    Length-Realted & TotLen Fwd Pkts, TotLen Bwd Pkts, Fwd Pkt Len Max, Fwd Pkt Len Min, Fwd Pkt Len Mean, Fwd Pkt Len Std, Bwd Pkt Len Max, Bwd Pkt Len Min, Bwd Pkt Len Mean, Bwd Pkt Len Std, Fwd Header Len, Bwd Header Len, Pkt Len Min, Pkt Len Max, Pkt Len Mean, Pkt Len Std, Pkt Len Var, Pkt Size Avg, Fwd Seg Size Avg, Bwd Seg Size Avg, Fwd Seg Size Min \\ 
    \midrule
    Protocol-Related & Fwd PSH Flags \\
    \bottomrule
    \end{tabularx}
    \label{tab_used_features}
\end{table}

\subsection{Settings of Drift Scenario}
\label{subsec_drift_scenario_settings}
Table \ref{tab_dataset_settings} illustrates details of two drift scenario settings in CICDS-2017 and CICIDS-2018 datasets.
\begin{table}[htbp]
    \centering
    \caption{Datasets Settings}
    \begin{tabularx}{0.48\textwidth}{>{\centering\arraybackslash}p{0.09\textwidth}|
    >{\centering\arraybackslash}p{0.09\textwidth}| >{\centering\arraybackslash}X| >{\centering\arraybackslash}X}
        \toprule
        Dataset & Type & Set1 & Set2 \\
        \toprule
        \multirow{8}{*}{CICIDS-2017} & \multirow{2}{*}{Local Drift} & Benign, DoS-GoldenEye & Benign, DoS-GoldenEye \\
        \cmidrule{2-4}
        & \multirow{6}{*}{Global Drift} & Benign, DoS-GoldenEye, Dos-Hulk, FTP-Patator, SSH-Patator. & Benign, Botnet, Portscan, DoS-Slowhttptest, DoS-Slowloris, XSS, DDoS, Bruteforce.\\
        \midrule
        \multirow{11}{*}{CICIDS-2018} & \multirow{2}{*}{Local Drift} & Benign, DoS-GoldenEye & Benign, DoS-GoldenEye \\
        \cmidrule{2-4}
        & \multirow{9}{*}{Global Drift} & Benign, DoS-Slowhttptest, DoS-Hulk, Dos-GoldenEye, DoS-Slowloris. & Benign, Botnet, FTP-Patator, SSH-Patator, DDoS-LOIC-HTTP, DDoS-LOIC-UDP, XSS, Web-Attacks, SQL-Injection.\\
        \bottomrule
    \end{tabularx}
    \label{tab_dataset_settings}
\end{table}

\subsection{Details of Model and Training}
Table \ref{tab_model_training_details} presents details of our model architecture and parameter counts, and Table \ref{tab_recommended_hyperparameters} provides the hyperparameters used in our experiments.
\begin{table}[htbp]
    \centering
    \caption{Details of Model Structure.}
    \begin{tabularx}{0.48\textwidth}{
    >{\centering\arraybackslash}p{0.08\textwidth}|>{\centering\arraybackslash}p{0.07\textwidth}|
    >{\centering\arraybackslash}X| >{\centering\arraybackslash}X|
    >{\centering\arraybackslash}X 
    }
        \toprule
        \multirow{2}{*}{\textbf{\makecell{Blocks}}} & \multirow{2}{*}{\textbf{Layers}} & \multicolumn{2}{c|}{\textbf{Dims}} & \multirow{2}{*}{\textbf{Parameters}} \\
        \cmidrule(lr){3-4}
        & & Input & Output \\
        \midrule
        \multirow{5}{*}{Encoder} & Linear & 50 & 96 & 4896 \\
        & ReLU & 96 & 96 & 0 \\
        & Linear & 96 & 64 & 6208 \\
        & ReLU & 64 & 64 & 0 \\
        & Linear & 64 & 32 & 2080 \\
        \midrule
        Classifier & Linear & 32 & 2 & 66 \\
        \bottomrule
    \end{tabularx}
    \begin{tablenotes}
        \footnotesize \item[*] * Total parameters: 11550.
    \end{tablenotes}
    \label{tab_model_training_details}
\end{table}

\begin{table}[!ht]
\centering
\caption{Recommended Hyperparameters.}
\begin{tabularx}{0.48\textwidth}{
    >{\centering\arraybackslash}X| >{\centering\arraybackslash}X|
    >{\centering\arraybackslash}X| >{\centering\arraybackslash}X
    }
    \toprule
    Type & Datasets & Hyper-Paras & Value \\
    \midrule
    \multirow{12}{*}{Global Drift} & \multirow{6}{*}{CICIDS-2017} & learning rate & 1e-4 \\ 
    & & batch size & 256 \\
    & & epoch & 500 \\
    & & $\mathcal{T}$ & 2.0 \\
    & & $\beta$ & 0.5 \\
    & & $\gamma$ & 1.0 \\
    \cmidrule(lr){2-4}
    & \multirow{6}{*}{CICIDS-2018} & learning rate & 1e-4 \\
    & & batch size & 256 \\
    & & epoch & 120 \\
    & & $\mathcal{T}$ & 2.0 \\
    & & $\beta$ & 0.3 \\
    & & $\gamma$ & 0.1 \\
    \midrule
    \multirow{12}{*}{Local Drift} & \multirow{6}{*}{CICIDS-2017} & learning rate & 1e-4 \\ 
    & & batch size & 256 \\
    & & epoch & 200 \\
    & & $\mathcal{T}$ & 2.0 \\
    & & $\beta$ & 0.5 \\
    & & $\gamma$ & 1.0 \\
    \cmidrule(lr){2-4}
    & \multirow{6}{*}{CICIDS-2018} & learning rate & 1e-5 \\
    & & batch size & 256 \\
    & & epoch & 100 \\
    & & $\mathcal{T}$ & 3.0 \\
    & & $\beta$ & 0.5 \\
    & & $\gamma$ & 0.1 \\
    \bottomrule
\end{tabularx}
\label{tab_recommended_hyperparameters}
\end{table}

\section{Supplementary Experiment}
\label{sec_supplementary_experiment}
\subsection{Feature Importance Analysis under Concept Drift}
\begin{figure}[htbp]
\centering
\subfloat[Historical Model.]{\includegraphics[width=0.23\textwidth]{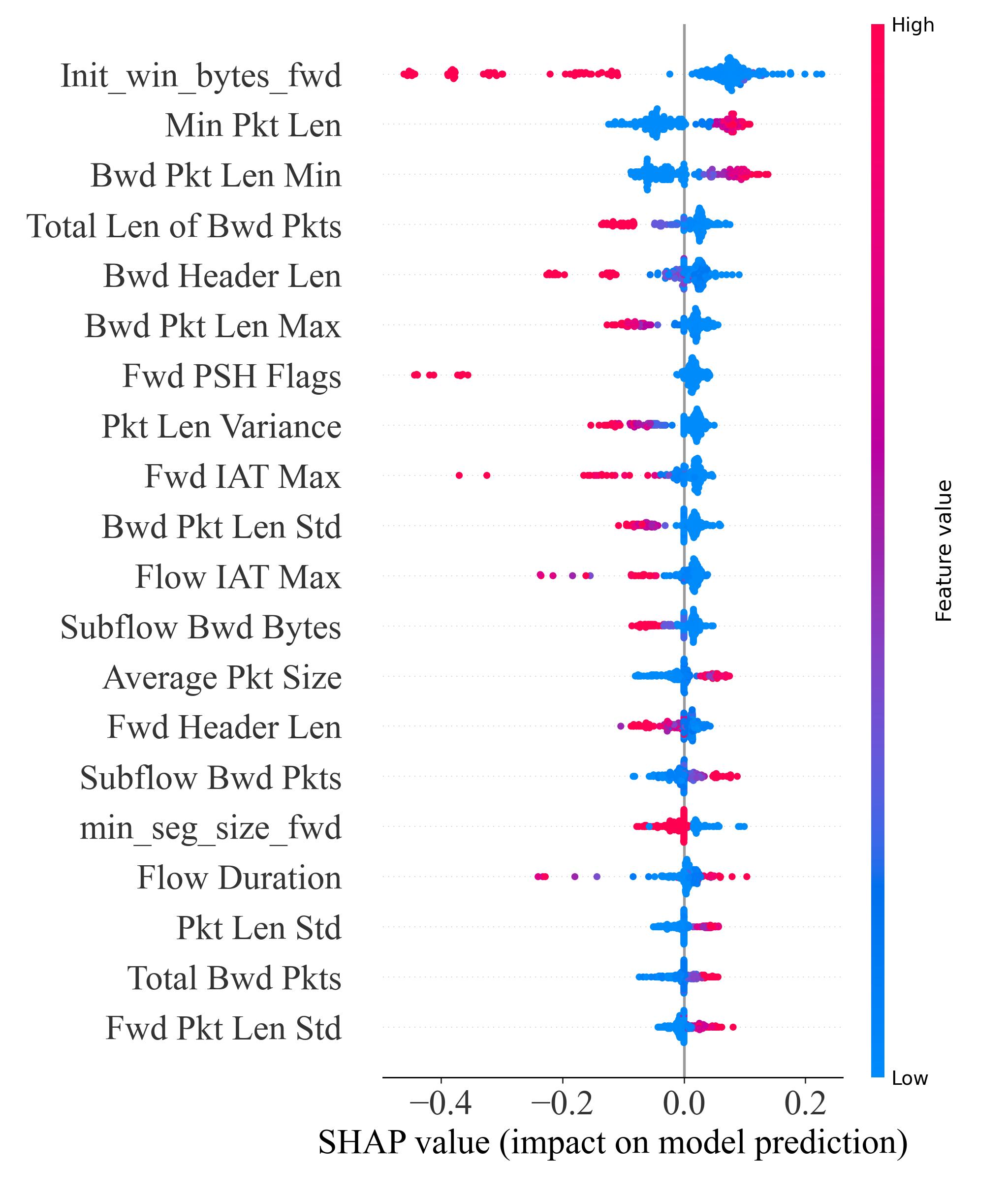}
\label{fig_shap_a}}
\subfloat[Drifted Model.]{\includegraphics[width=0.23\textwidth]{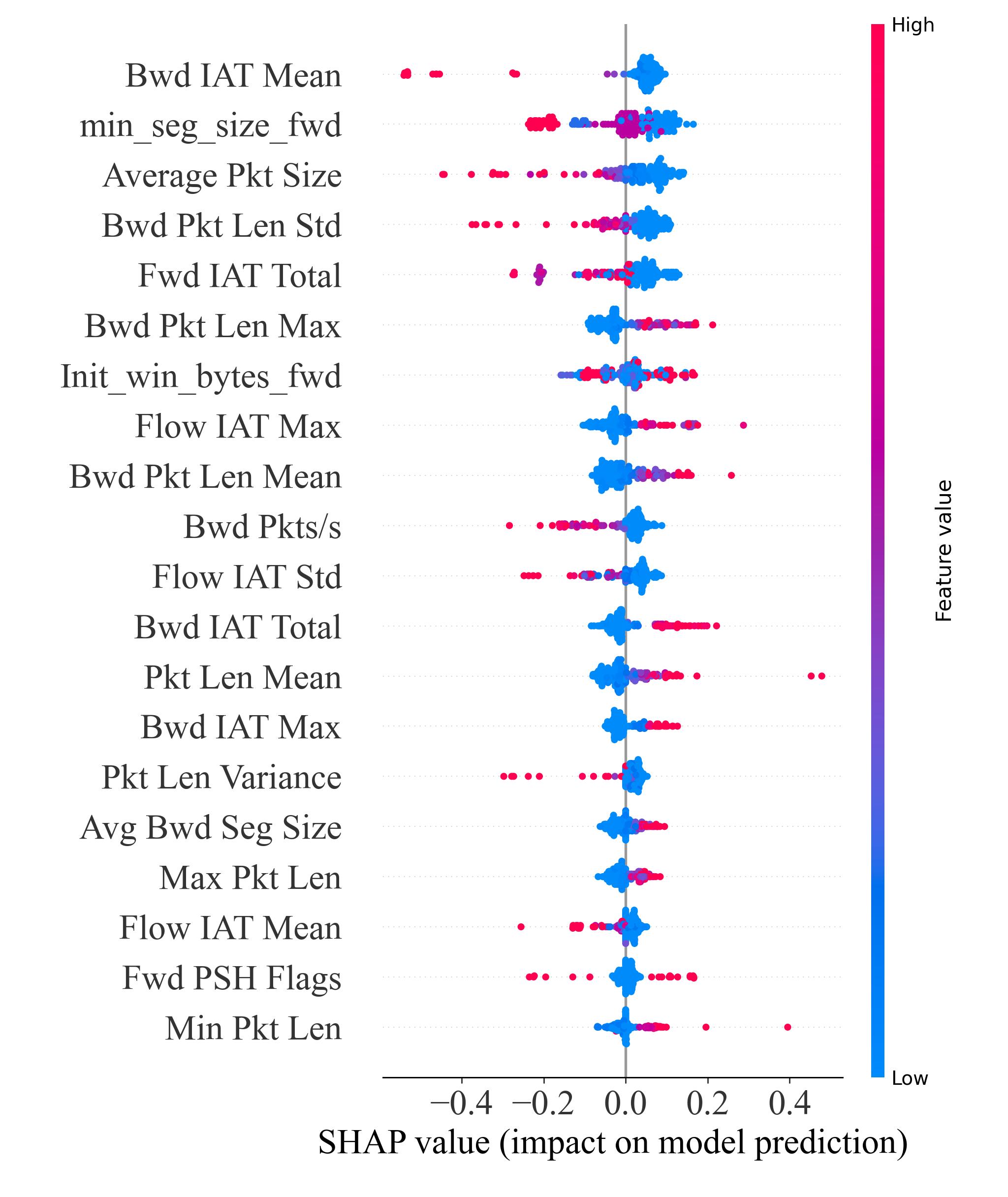}
\label{fig_shap_b}}
\caption{SHAP summary plots of the historical and drifted models. Higher-ranked features on the left indicate greater impact on predictions.}
\label{fig_shap}
\end{figure}

We analyze feature influence using SHAP (SHapley Additive Explanations) \cite{lundberg2017unified} for both historical and drifted models (Figure \ref{fig_shap}). The results show that key features differ, indicating drift beyond latent distributions. While initialization from the historical model alleviates catastrophic forgetting, SHAP reveals evolving feature importance. Therefore, freezing the first layer of the encoder preserves the initial feature influence while allowing subsequent fully connected layers to adapt, providing a well-justified balance between stability and adaptability.

\subsection{Analysis of Contrastive Loss}
\begin{figure}[htbp]
\centering
\subfloat[Local Drift.]{\includegraphics[width=0.23\textwidth]{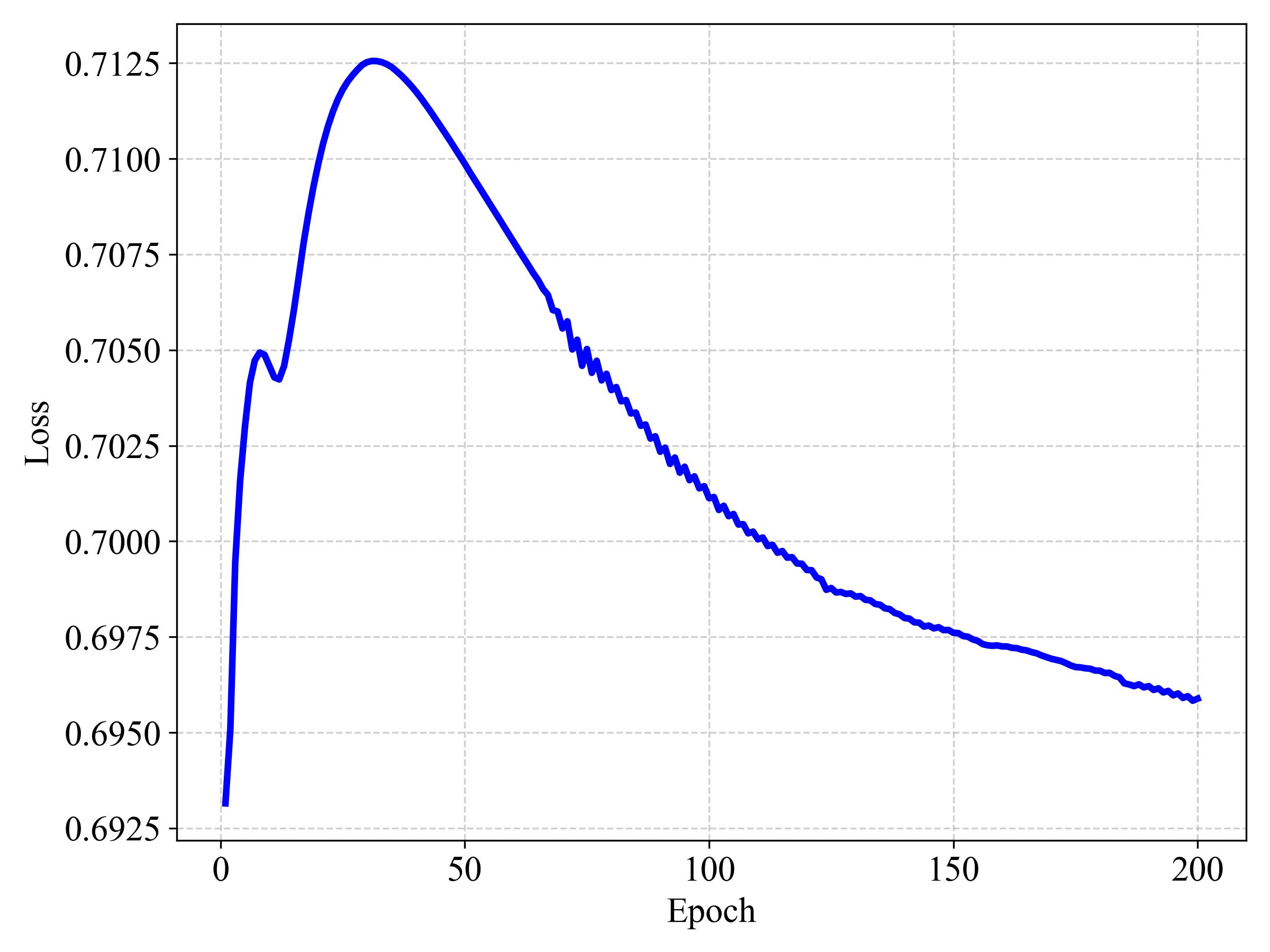}
\label{fig_con_loss_plot_a}}
\subfloat[Global Drift.]{\includegraphics[width=0.23\textwidth]{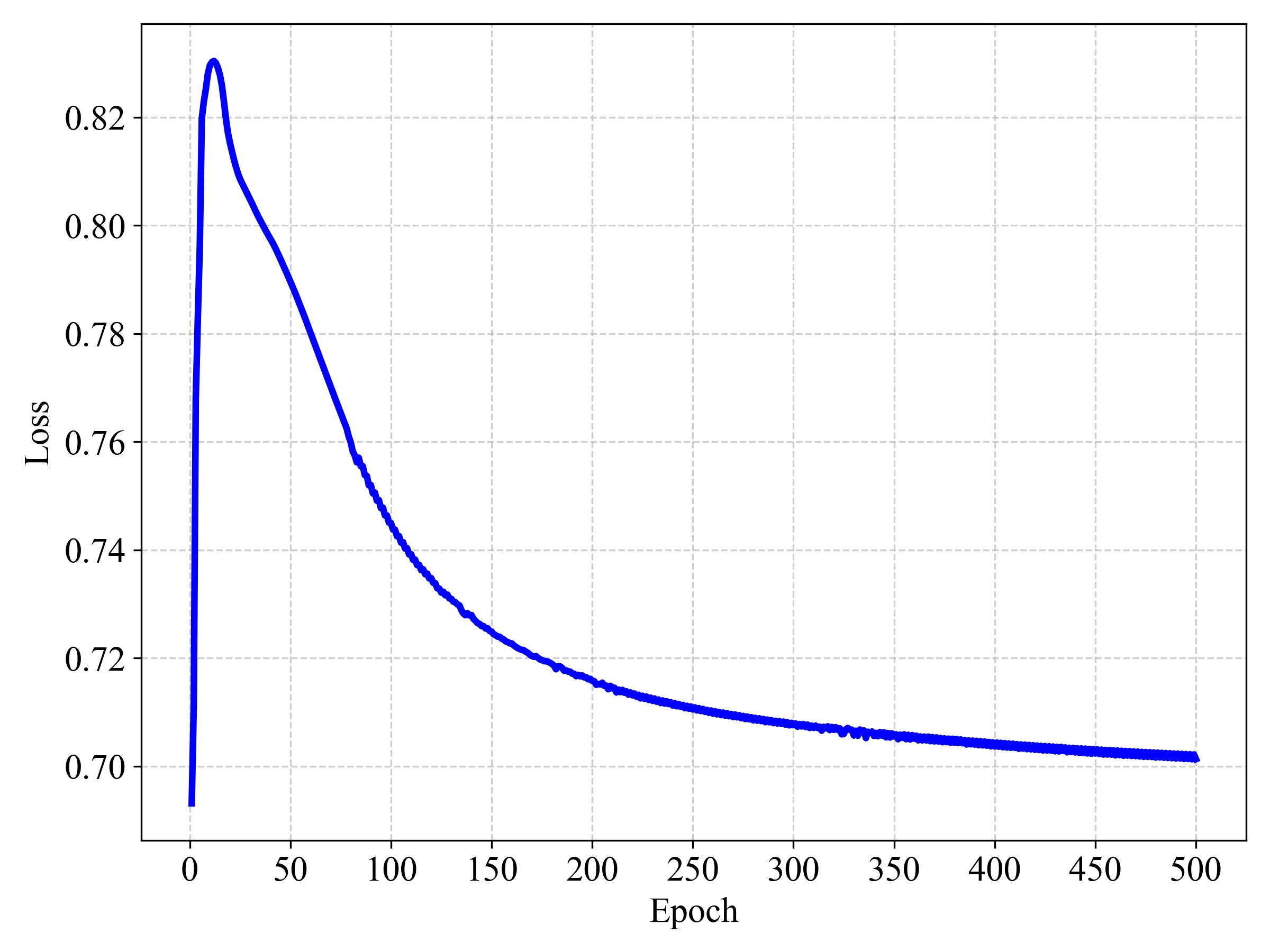}
\label{fig_con_loss_plot_b}}
\caption{Contrastive Loss During Training.}
\label{fig_con_loss_plot}
\end{figure}

During the model update process, both the aggregation model and the training model are initialized using the historical model parameters. According to contrastive loss (as shown in Equation \eqref{con_loss}, when both $\text{sim}(z, z_{his})$ and $\text{sim}(z, z_{prev})$ are $1.0$ (completely similar), contrastive loss should be $0.693$, indicating the optimal state of distribution alignment. However, as the model is trained on drifted data, it tends to fit the new data, causing parameter changes that shift the latent variable distribution away from the historical model's latent variable distribution. As a result, the contrastive loss gradually increases, meaning that during the initial epochs, the contrastive loss does not contribute to aligning the model. As training progresses, the contrastive loss begins to align the latent variable distribution of the new model with that of the aggregated model, and the contrastive loss should decrease steadily, but it should not fall below 0.693. The graph of contrastive loss during our experiments is shown in Figure \ref{fig_con_loss_plot}.

\subsection{Sensitivity Analysis of Hyperparameter on the CICIDS-2018 Dataset}
\label{sec_appendix_sa_hyperparameter_cicids_2018}
\begin{figure}[htbp]
\centering
\subfloat[Local Drift (Historical).]{\includegraphics[width=0.23\textwidth]{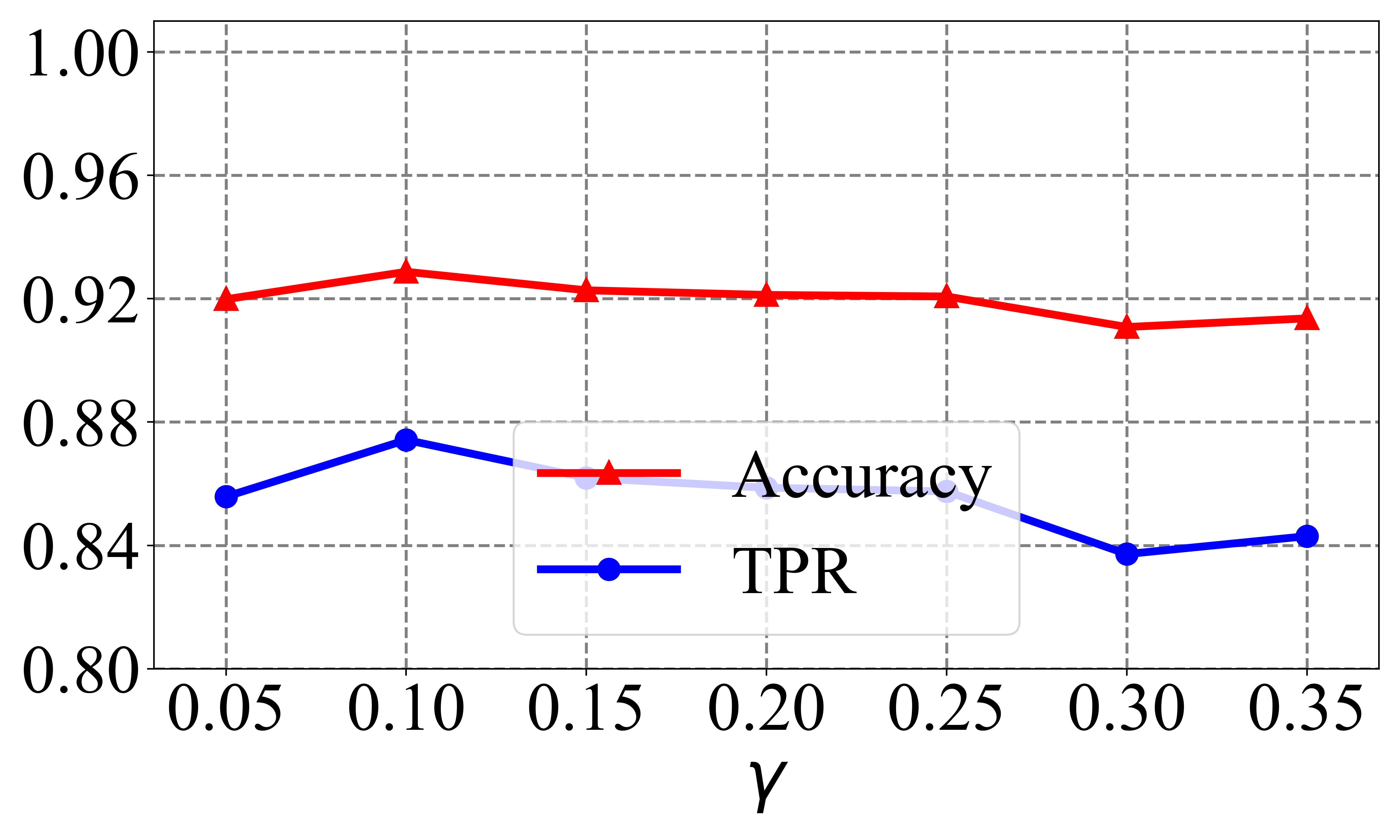}
\label{fig_appendix_sa_gamma_cicids2018_local_historical}}
\subfloat[Local Drift (Drifted).]{\includegraphics[width=0.23\textwidth]{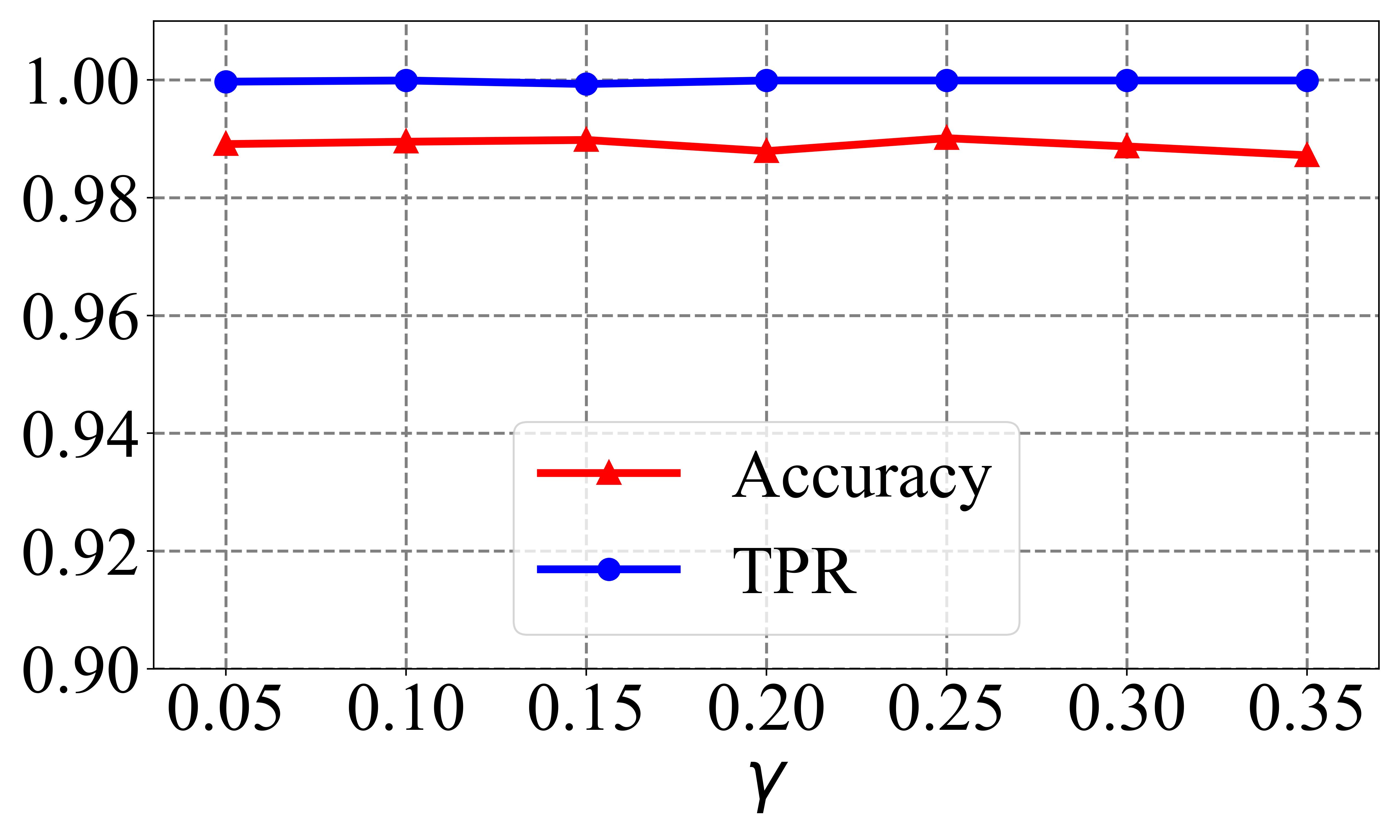}
\label{fig_appendix_sa_gamma_cicids2018_local_drifted}}

\subfloat[Global Drift (Historical).]{\includegraphics[width=0.23\textwidth]{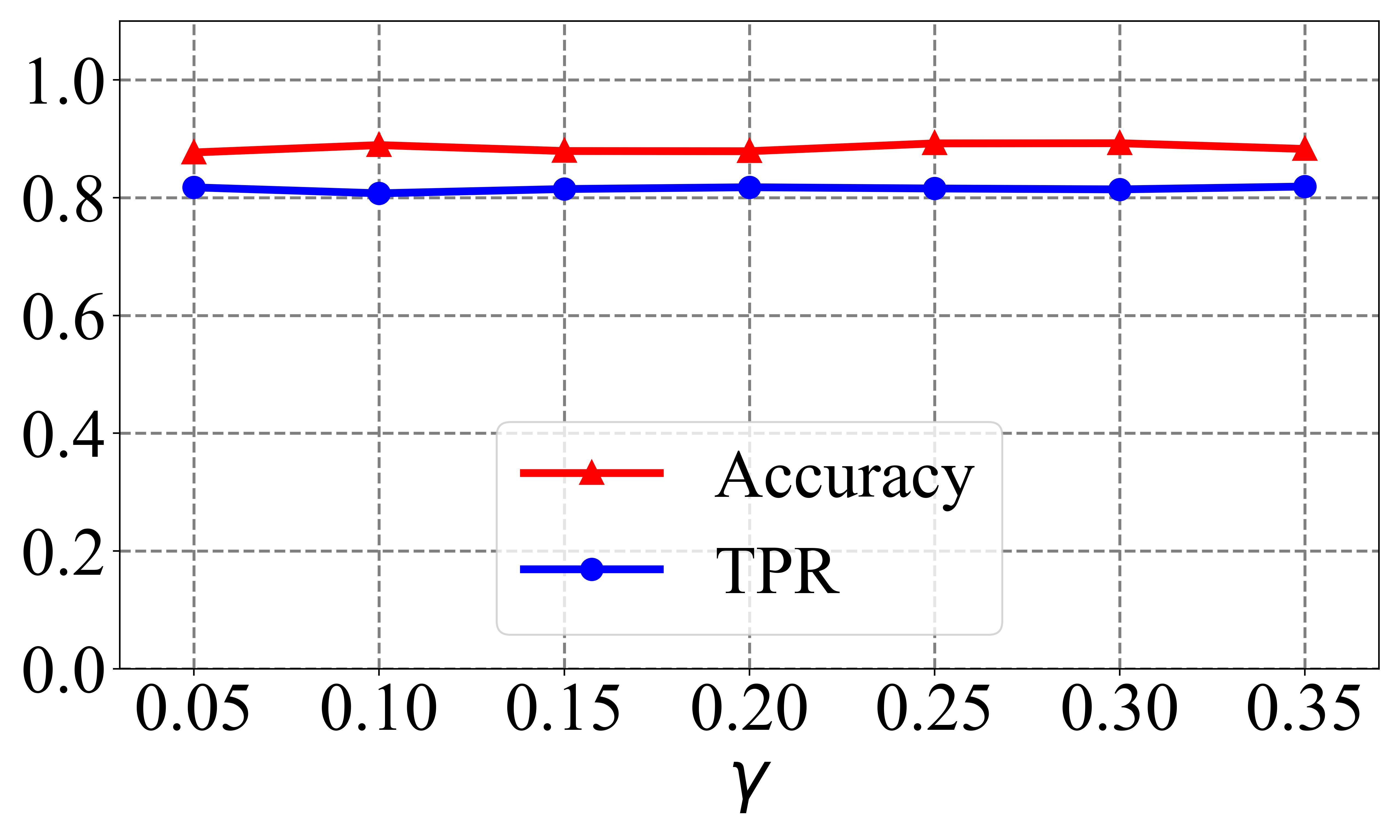}
\label{fig_appendix_sa_gamma_cicids2018_global_historical}}
\subfloat[Global Drift (Drifted).]{\includegraphics[width=0.23\textwidth]{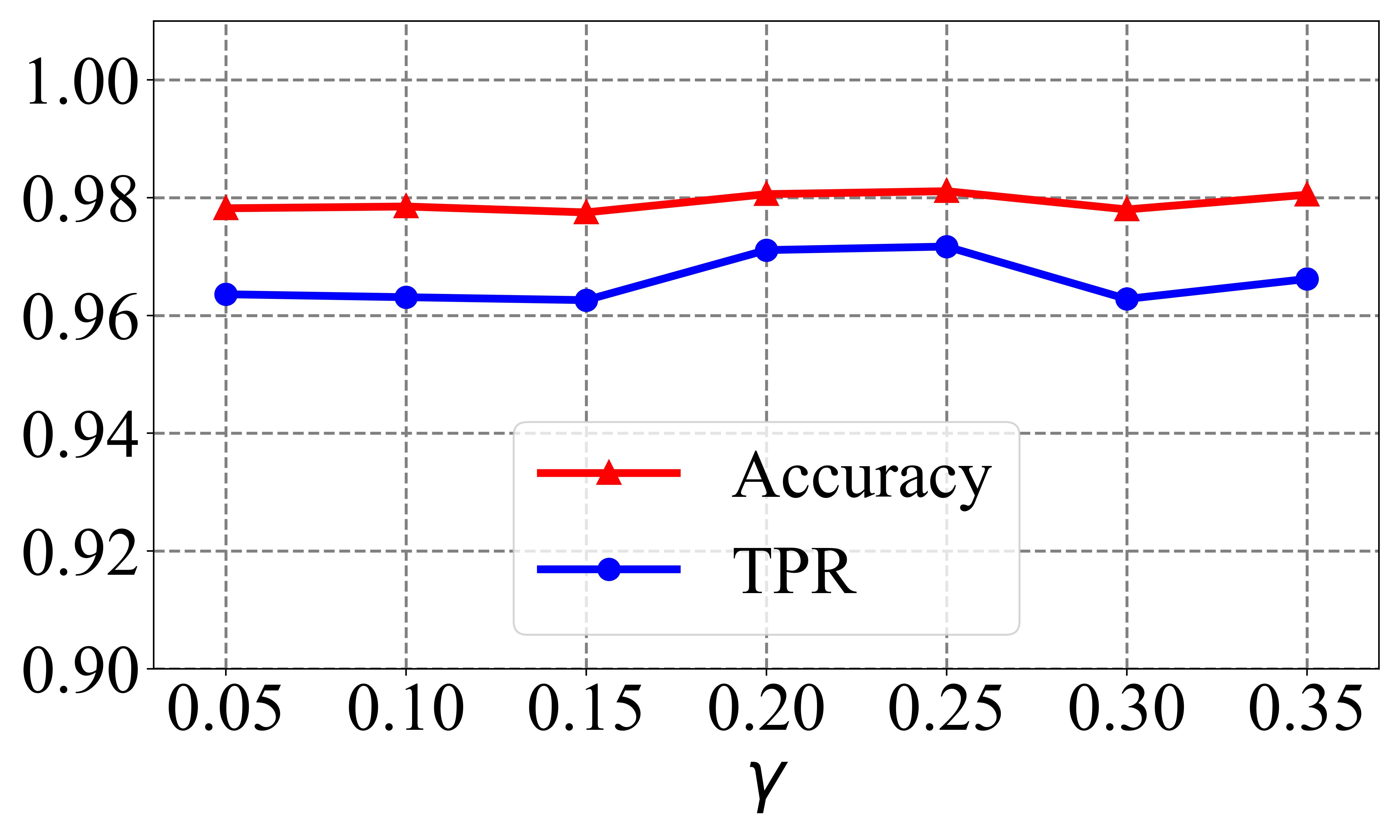}
\label{fig_appendix_sa_gamma_cicids2018_global_drifted}}
\caption{Sensitivity Analysis of Hyperparameter $\gamma$.}
\label{fig_appendix_sa_gamma_cicids2018}
\end{figure}

We conducted sensitivity analyses of the hyperparameters $\beta$ and $\gamma$ on the CICIDS-2018 dataset. The results are presented in Figure \ref{fig_appendix_sa_beta_cicids2018} and Figure \ref{fig_appendix_sa_gamma_cicids2018}, with the detailed analysis provided as follows:
\begin{figure}[htbp]
\centering
\subfloat[Local Drift (Historical).]{\includegraphics[width=0.23\textwidth]{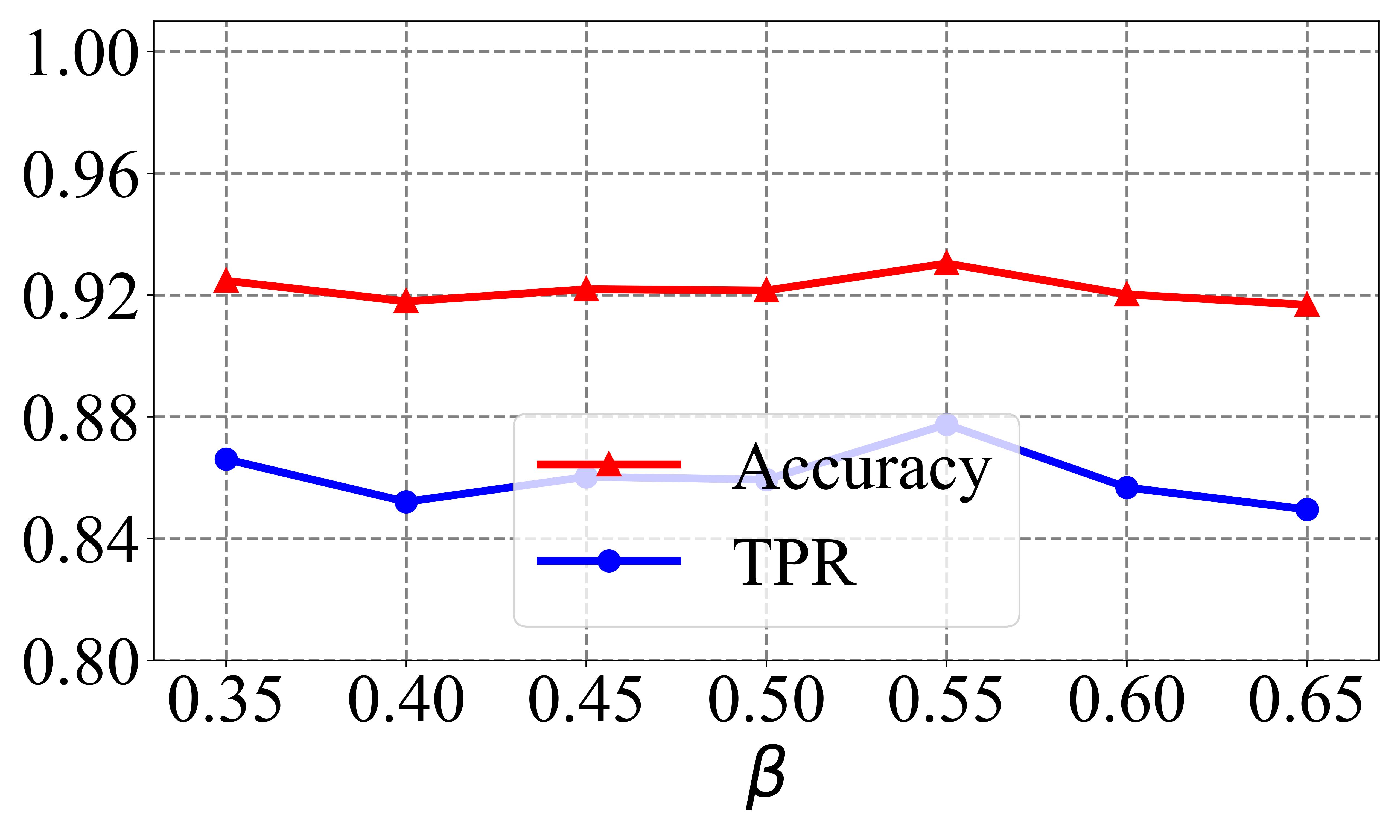}
\label{fig_appendix_sa_beta_cicids2018_local_historical}}
\subfloat[Local Drift (Drifted).]{\includegraphics[width=0.23\textwidth]{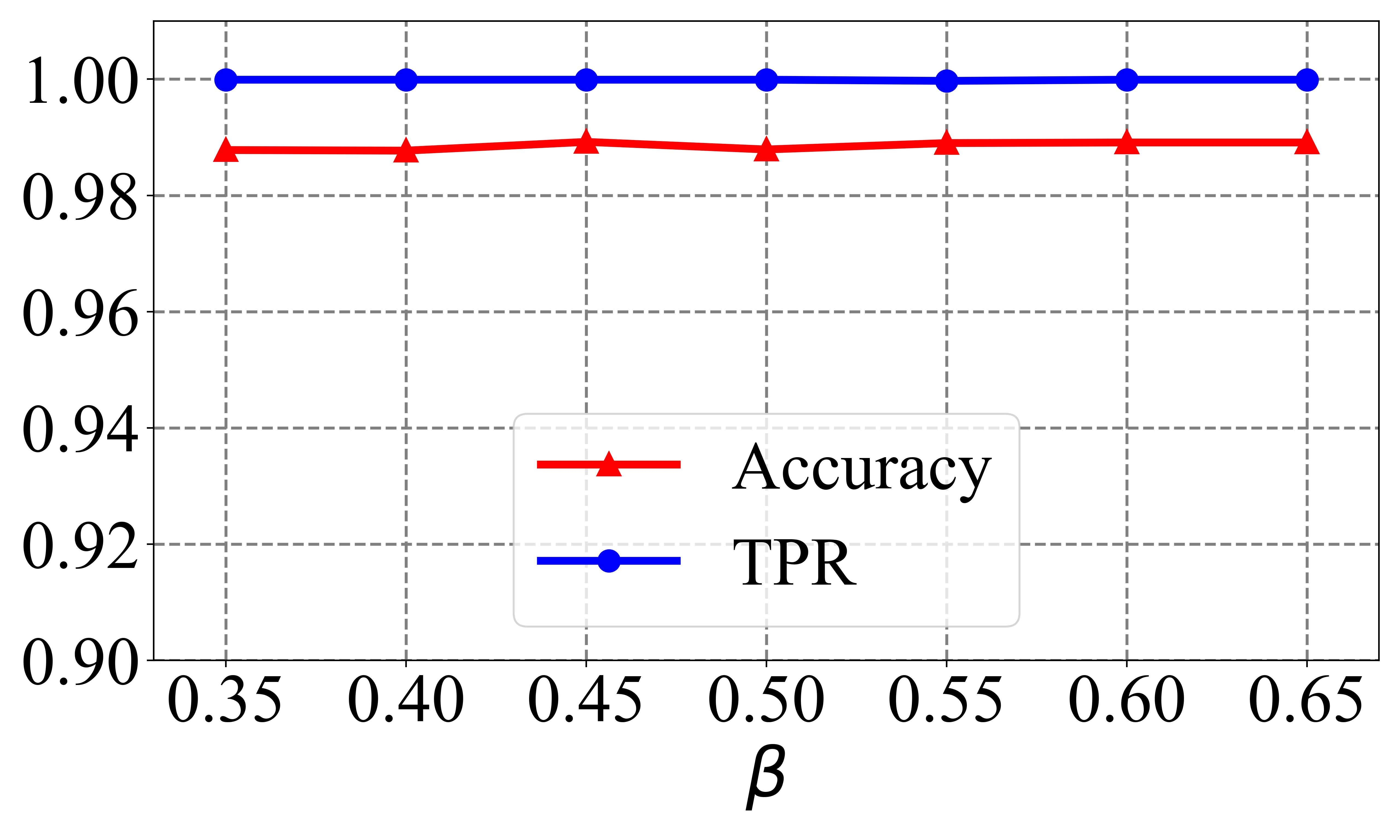}
\label{fig_appendix_sa_beta_cicids2018_local_drifted}}

\subfloat[Global Drift (Historical).]{\includegraphics[width=0.23\textwidth]{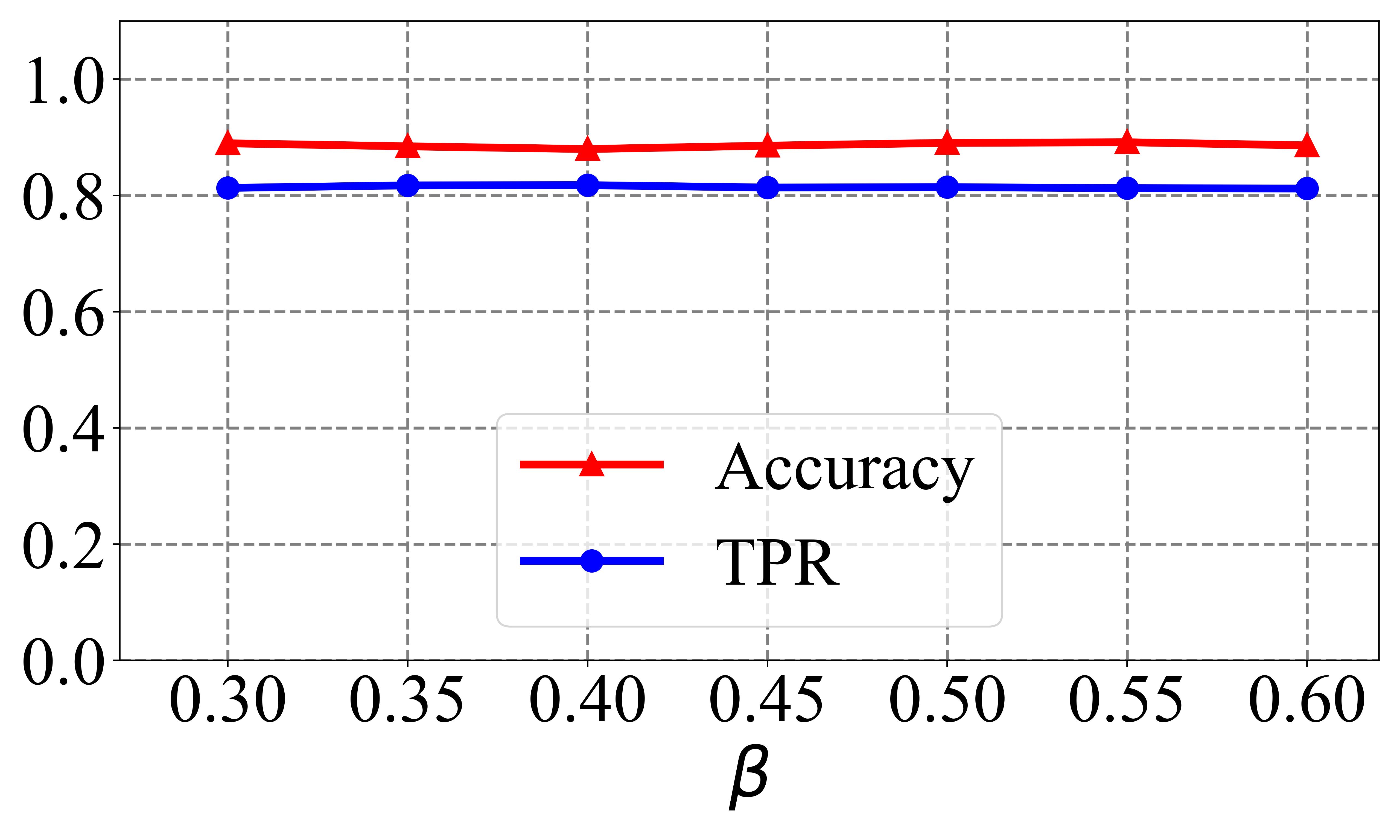}
\label{fig_appendix_sa_beta_cicids2018_global_historical}}
\subfloat[Global Drift (Drifted).]{\includegraphics[width=0.23\textwidth]{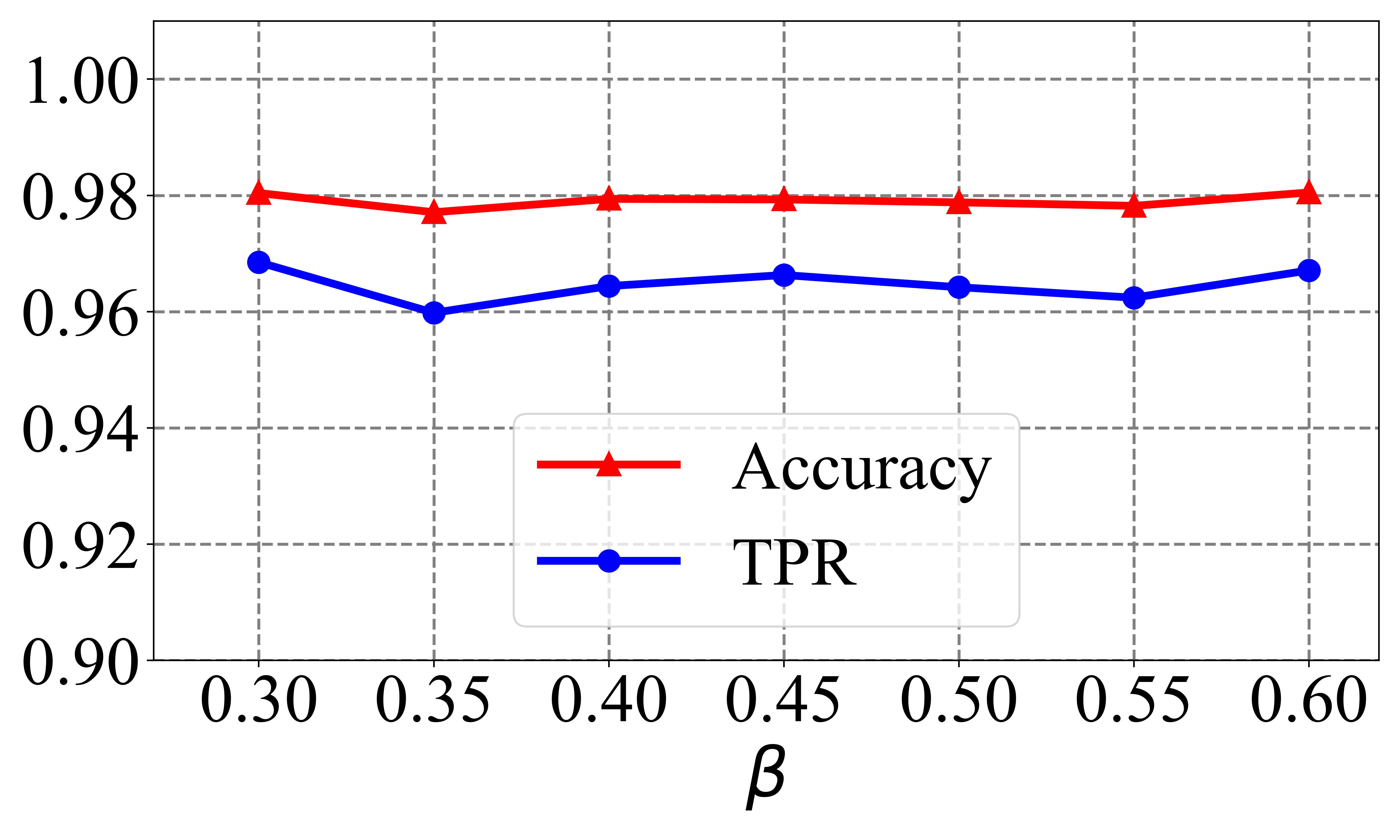}
\label{fig_appendix_sa_beta_cicids2018_global_drifted}}
\caption{Sensitivity Analysis of Hyperparameter $\beta$.}
\label{fig_appendix_sa_beta_cicids2018}
\end{figure}

\subsubsection{Sensiticity Analysis of $\beta$} We investigated the sensitivity of the hyperparameter $\beta$ on the CICIDS-2018 dataset under both local drift and global drift scenarios. DriftXpert maintains stable performance across different values of $\beta$ in both scenarios, with only marginal performance variations. This indicates that the proposed framework is not sensitive to the selection of $\beta$, and a wide range of values can effectively preserve the balance between catastrophic forgetting and model adaptation.

\subsubsection{Sensitivity Analysis of $\gamma$} We further analyzed the impact of the hyperparameter $\gamma$ under local and global drift scenarios. The performance remains relatively stable as $\gamma$ varies, without significant degradation in either scenario. This observation suggests that DriftXpert is insensitive to the choice of $\gamma$ and can achieve reliable adaptation across different parameter settings.

\subsection{Sensitivity Analysis of Hyperparameter on the Real-World Dataset}
\label{sec_appendix_sa_hyperparameter_realworld}
\begin{figure}[htbp]
\centering
\subfloat[Historical.]{\includegraphics[width=0.23\textwidth]{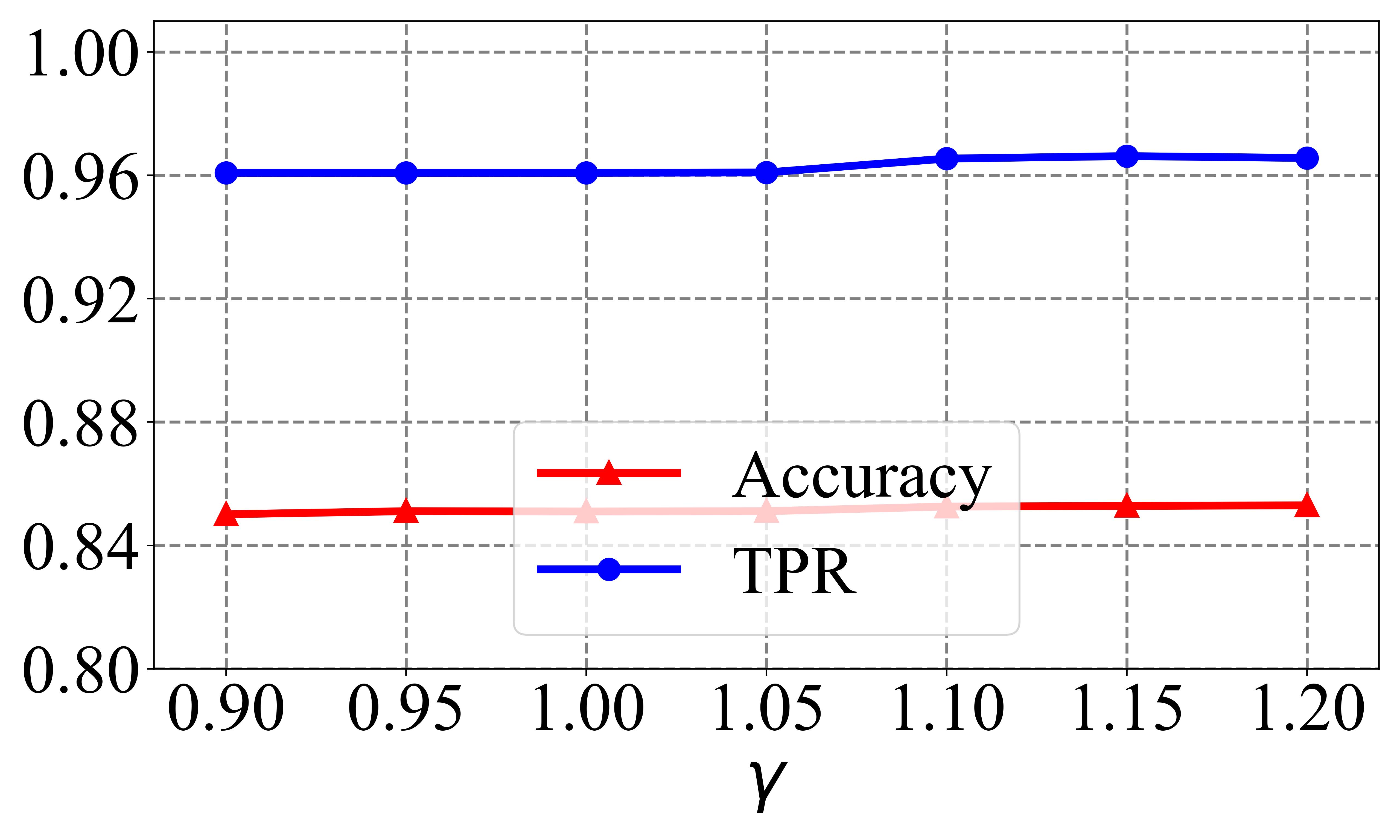}
\label{fig_appendix_sa_gamma_realworld_historical}}
\subfloat[Drifted.]{\includegraphics[width=0.23\textwidth]{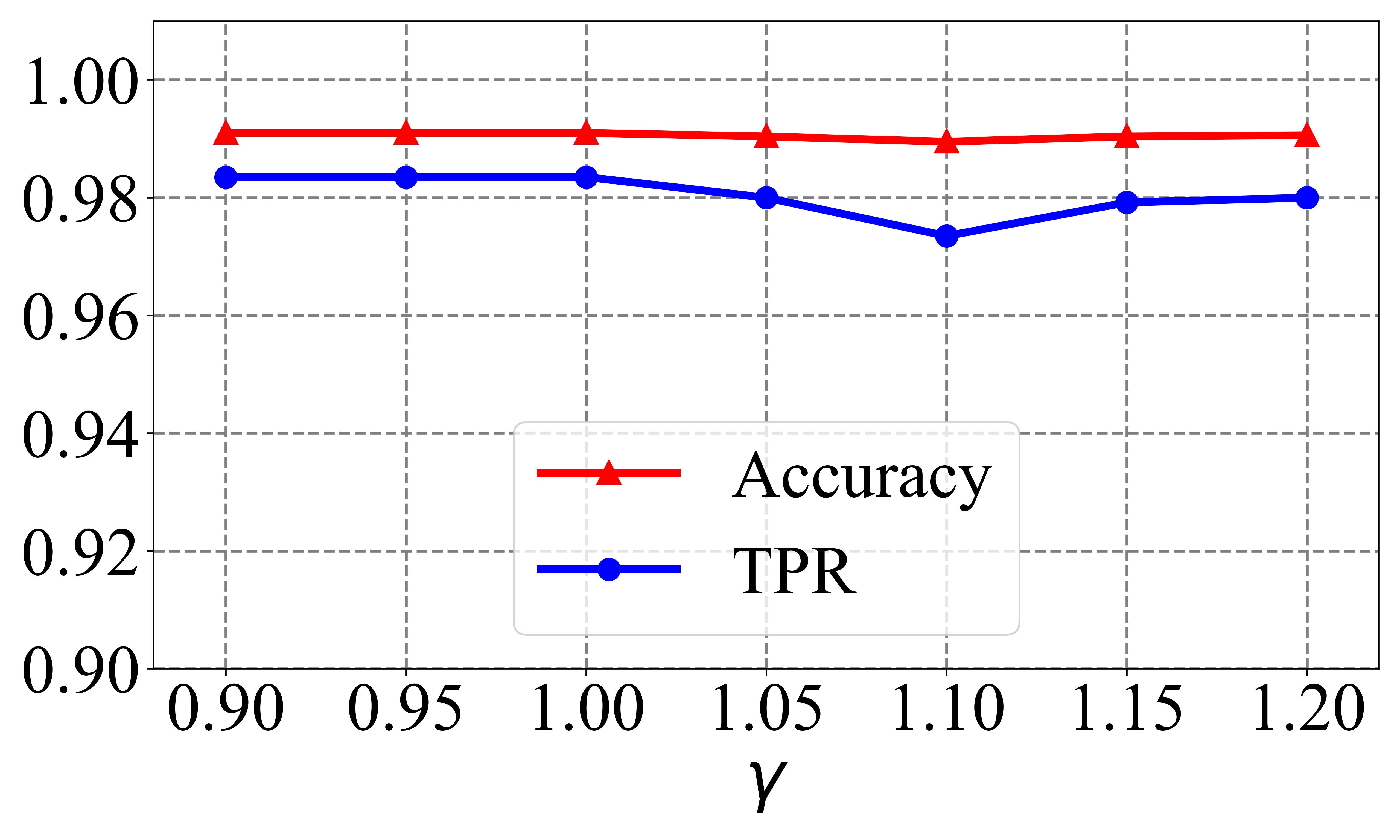}
\label{fig_appendix_sa_gamma_realworld_drifted}}
\caption{Sensitivity Analysis of Hyperparameter $\gamma$.}
\label{fig_appendix_sa_gamma_realworld}
\end{figure}

We further conducted sensitivity analyses of the hyperparameters $a$ and $b$ on the real-world dataset, with the results presented in Figure \ref{fig_appendix_sa_beta_realworld} and Figure \ref{fig_appendix_sa_gamma_realworld}. The detailed analysis is provided as follows:
\begin{figure}[htbp]
\centering
\subfloat[Historical.]{\includegraphics[width=0.23\textwidth]{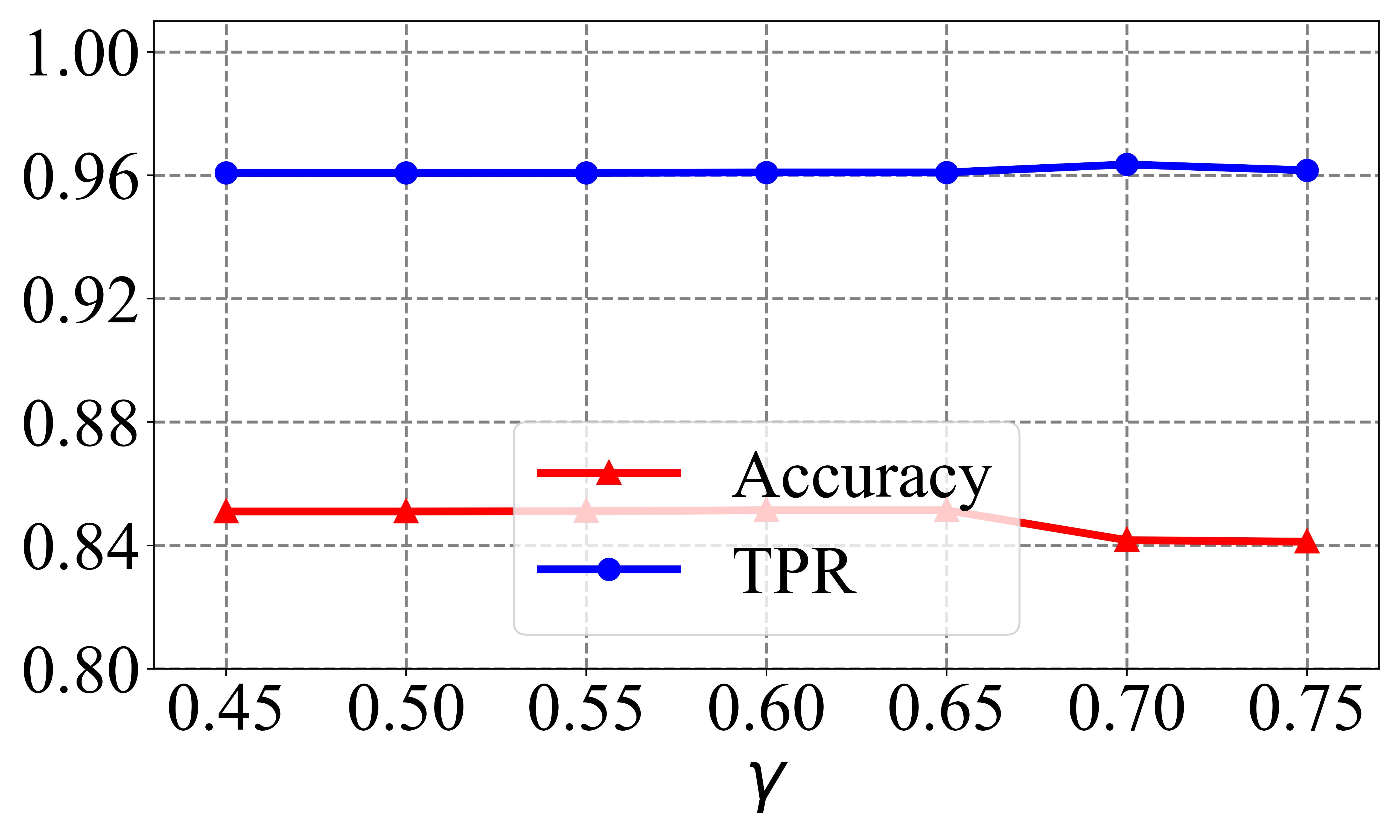}
\label{fig_appendix_sa_beta_realworld_historical}}
\subfloat[Drifted.]{\includegraphics[width=0.23\textwidth]{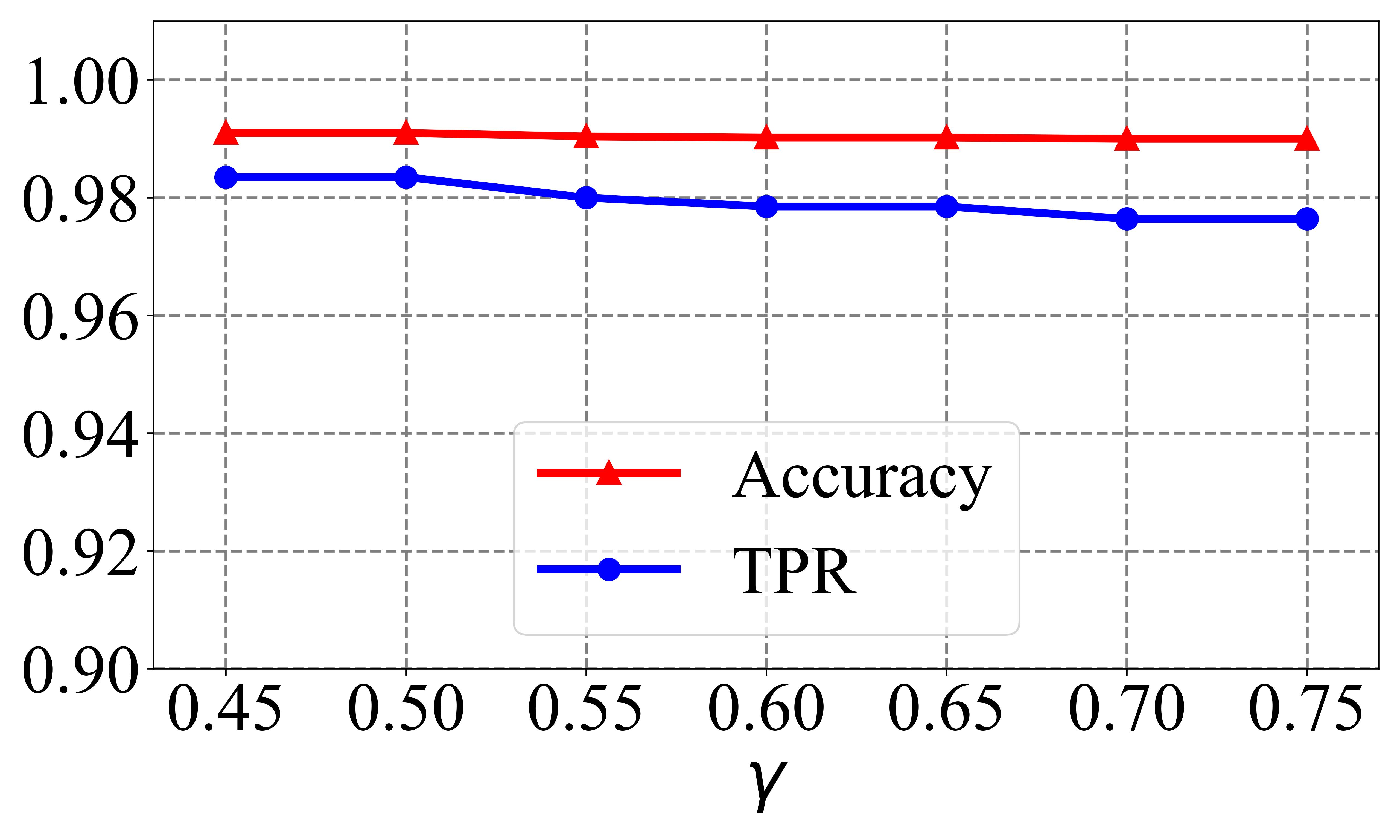}
\label{fig_appendix_sa_beta_realworld_drifted}}
\caption{Sensitivity Analysis of Hyperparameter $\beta$.}
\label{fig_appendix_sa_beta_realworld}
\end{figure}

\subsubsection{Sensiticity Analysis of $\beta$} Firstly, we evaluated the sensitivity of the hyperparameter $\beta$ on the real-world enterprise traffic dataset. DriftXpert achieves consistently stable performance across different values, with only minor fluctuations. This indicates that the proposed drift detection mechanism is robust to the variation of $\beta$ in practical deployment scenarios. Although $\beta$ controls the sensitivity of drift identification, DriftXpert can effectively distinguish evolving traffic patterns from normal variations across a wide range of settings, demonstrating its robustness in real-world environments.

\subsubsection{Sensitivity Analysis of $\gamma$} Secondly, we investigated the impact of $\gamma$ on the real-world enterprise traffic dataset. The results show that DriftXpert maintains stable performance under different configurations, without noticeable performance degradation. This demonstrates that the proposed adaptation strategy is not highly dependent on the exact choice of $\gamma$, and can consistently preserve previous knowledge while adapting to emerging traffic patterns. These results further validate the practicality and robustness of DriftXpert for long-term deployment in dynamic network environments.

\subsection{Performance Test}
\subsubsection{Memory Usage} We estimated the memory usage of the model. DriftXpert is lightweight, with 11,550 parameters (see Appendix D), which requires approximately 46 KB. Taking into account the storage of gradient and momentum, the optimizer adds an estimated 92 KB. Contrastive learning further requires storing the previous and aggregated models, adding another 92 KB. In total, the training memory footprint of DriftXpert is around 230 KB, making it suitable for deployment on devices with limited resources. 

\subsubsection{Latency} We measured per sample detection latency of the \textit{Encoder} and \textit{Classifier}, as illustrated in Figure \ref{fig_performance_test_a} and Figure \ref{fig_performance_test_b}. The \textit{Encoder} ranged from 0.1 to 2.1 µs (average 0.7 µs), while the simpler \textit{Classifier} ranged from 0.0 to 0.24 µs (average 0.1 µs). Overall latency did not exceed 2.3 µs, demonstrating the strong real-time performance of DriftXpert.

\subsubsection{Throughput} 
\begin{figure}[htbp]
\centering
\subfloat[Latency of Encoder.]{\includegraphics[width=0.23\textwidth]{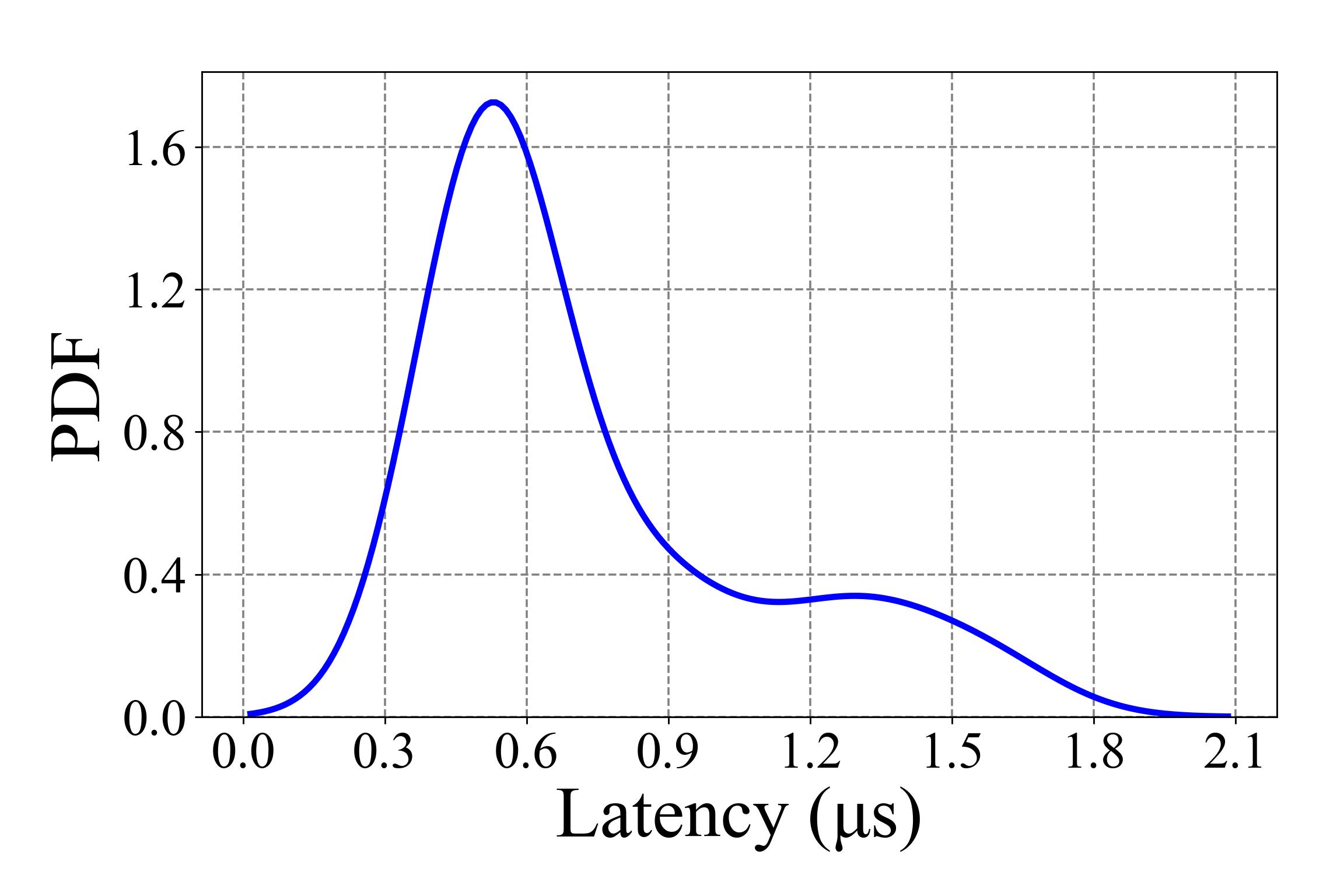}
\label{fig_performance_test_a}}
\subfloat[Latency of Classifier.]{\includegraphics[width=0.23\textwidth]{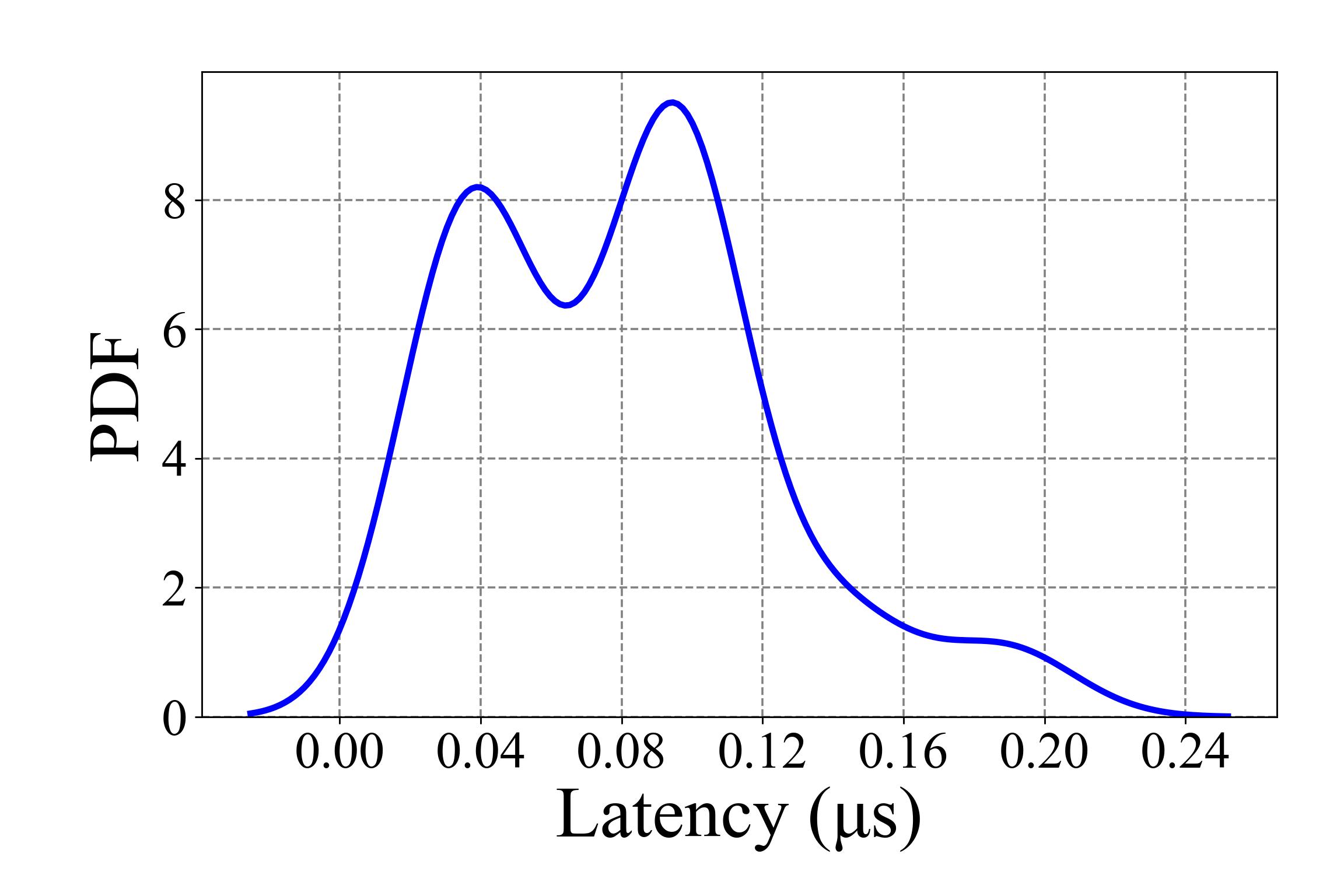}
\label{fig_performance_test_b}}

\subfloat[Throughput of Encoder.]{\includegraphics[width=0.23\textwidth]{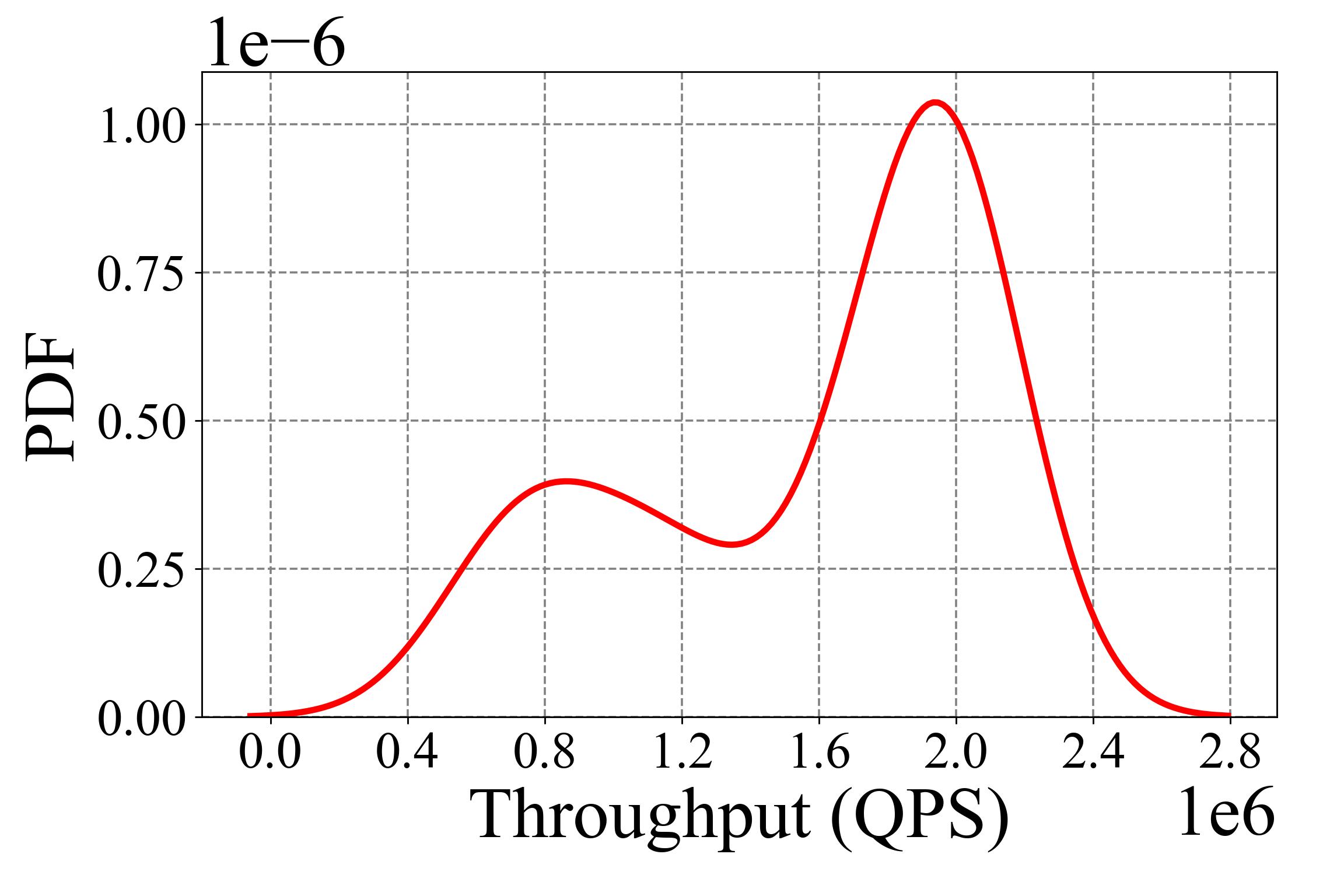}
\label{fig_performance_test_c}}
\subfloat[Throughput of Classifier.]{\includegraphics[width=0.23\textwidth]{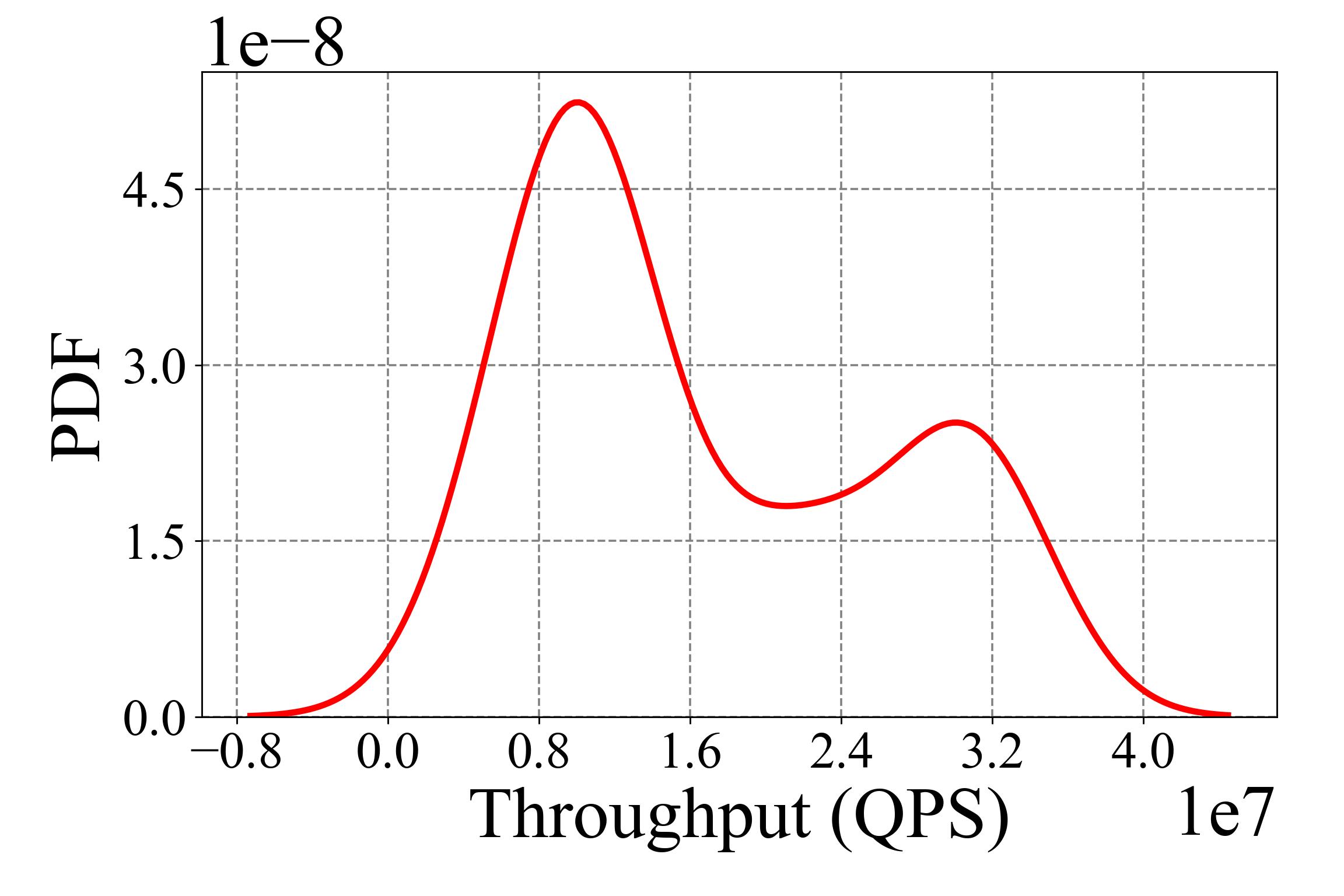}
\label{fig_performance_test_d}}
\caption{Performance results of DriftXpert.}
\label{fig_performance_test}
\end{figure}

We evaluated the throughput of the \textit{Encoder} and \textit{Classifier}, as illustrated in Figure \ref{fig_performance_test_c} and Figure \ref{fig_performance_test_d}. The \textit{Encoder} ranged from 0.4 to 2.8 MQPS, averaging 1.6 MQPS, while the \textit{Classifier} reached up to 40 MQPS, averaging 16 MQPS. In general, DriftXpert achieves an average throughput of ~1.6 MQPS, highlighting its strong real-time capability.

\end{document}